\documentclass[useamsfonts]{pasj01}
\usepackage[dvips]{graphicx}
\usepackage{subfigure}
\usepackage{natbib,color}
\usepackage{url}
\usepackage{bm}
\usepackage{xcolor}

\newcommand{\simgt}{\lower.5ex\hbox{$\; \buildrel > \over \sim \;$}}
\newcommand{\simlt}{\lower.5ex\hbox{$\; \buildrel < \over \sim \;$}}
\newcommand{\cmodel}{{\tt{cmodel}}}

\begin{document}

\title{Cosmology from cosmic shear power spectra with Subaru Hyper Suprime-Cam first-year data}
\author{Chiaki~\textsc{Hikage}\altaffilmark{1}}%
\author{Masamune~\textsc{Oguri}\altaffilmark{2,3,1}}%
\author{Takashi~\textsc{Hamana}\altaffilmark{4}}%
\author{Surhud~\textsc{More}\altaffilmark{1,5}}
\author{Rachel~\textsc{Mandelbaum}\altaffilmark{6}}%
\author{Masahiro~\textsc{Takada}\altaffilmark{1}}%
\author{Fabian~\textsc{K{\"o}hlinger}\altaffilmark{1}}%
\author{Hironao~\textsc{Miyatake}\altaffilmark{7,8,9,1}}%
\author{Atsushi~J.~\textsc{Nishizawa}\altaffilmark{7,8}}
\author{Hiroaki~\textsc{Aihara}\altaffilmark{3,1}}
\author{Robert~\textsc{Armstrong}\altaffilmark{10}}
\author{James~\textsc{Bosch}\altaffilmark{11}}
\author{Jean~\textsc{Coupon}\altaffilmark{12}}
\author{Anne~\textsc{Ducout}\altaffilmark{1}}
\author{Paul~\textsc{Ho}\altaffilmark{13}}
\author{Bau-Ching~\textsc{Hsieh}\altaffilmark{13}}
\author{Yutaka~\textsc{Komiyama}\altaffilmark{4,14}}
\author{Fran\c{c}ois~\textsc{Lanusse}\altaffilmark{6}}
\author{Alexie~\textsc{Leauthaud}\altaffilmark{15}}
\author{Robert~H.~\textsc{Lupton}\altaffilmark{11}}
\author{Elinor~\textsc{Medezinski}\altaffilmark{11}}
\author{Sogo~\textsc{Mineo}\altaffilmark{4}}
\author{Shoken~\textsc{Miyama}\altaffilmark{4,16}}
\author{Satoshi~\textsc{Miyazaki}\altaffilmark{4,14}}
\author{Ryoma~\textsc{Murata}\altaffilmark{1,3}}
\author{Hitoshi~\textsc{Murayama}\altaffilmark{1,17,18}}
\author{Masato~\textsc{Shirasaki}\altaffilmark{4}}
\author{Crist{\'o}bal~\textsc{Sif{\'o}n}\altaffilmark{11}}
\author{Melanie~\textsc{Simet}\altaffilmark{19,9}}
\author{Joshua~\textsc{Speagle}\altaffilmark{20}}
\author{David~N.~\textsc{Spergel}\altaffilmark{11,21}}
\author{Michael~A.~\textsc{Strauss}\altaffilmark{11}}
\author{Naoshi~\textsc{Sugiyama}\altaffilmark{8,22,1}}
\author{Masayuki~\textsc{Tanaka}\altaffilmark{4}}
\author{Yousuke~\textsc{Utsumi}\altaffilmark{23}}
\author{Shiang-Yu~\textsc{Wang}\altaffilmark{13}}
\author{Yoshihiko~\textsc{Yamada}\altaffilmark{4}}
\altaffiltext{1}{Kavli Institute for the Physics and Mathematics of the Universe (Kavli IPMU, WPI), University of Tokyo, Chiba 277-8582, Japan}
\altaffiltext{2}{Research Center for the Early Universe, University of Tokyo, Tokyo 113-0033, Japan}
\altaffiltext{3}{Department of Physics, University of Tokyo, Tokyo 113-0033, Japan}
\altaffiltext{4}{National Astronomical Observatory of Japan, Mitaka, Tokyo 181-8588, Japan}
\altaffiltext{5}{The Inter-University Center for Astronomy and Astrophysics, Post bag 4, Ganeshkhind, Pune, 411007, India}
\altaffiltext{6}{McWilliams Center for Cosmology, Department of Physics, Carnegie Mellon University, Pittsburgh, PA 15213, USA}
\altaffiltext{7}{Institute for Advanced Research, Nagoya University, Nagoya, Aichi 464-8602, Japan}
\altaffiltext{8}{Division of Particle and Astrophysical Science, Graduate School of Science, Nagoya University, Nagoya, Aichi 464-8602, Japan}
\altaffiltext{9}{Jet Propulsion Laboratory, California Institute of Technology, Pasadena, CA 91109, USA}
\altaffiltext{10}{Lawrence Livermore National Laboratory, Livermore, CA 94551, USA}
\altaffiltext{11}{Department of Astrophysical Sciences, Princeton University, 4 Ivy Lane, Princeton, NJ 08544, USA}
\altaffiltext{12}{Department of Astronomy, University of Geneva, ch. d'{\'E}cogia 16, 1290 Versoix, Switzerland}
\altaffiltext{13}{Academia Sinica Institute of Astronomy and Astrophysics, P.O. Box 23-141, Taipei 10617, Taiwan}
\altaffiltext{14}{SOKENDAI (The Graduate University for Advanced Studies), Mitaka, Tokyo 181-8588, Japan}
\altaffiltext{15}{Department of Astronomy and Astrophysics, University of California Santa Cruz, 1156 High St., Santa Cruz, CA 95064, USA}
\altaffiltext{16}{Hiroshima University, Higashi-Hiroshima, Hiroshima 739-8526, Japan}
\altaffiltext{17}{Department of Physics and Center for Japanese Studies, University of California, Berkeley, CA 94720, USA}
\altaffiltext{18}{Theoretical Physics Group, Lawrence Berkeley National Laboratory, MS 50A-5104, Berkeley, CA 94720, USA}
\altaffiltext{19}{University of California, Riverside, 900 University Avenue, Riverside, CA 92521, USA}
\altaffiltext{20}{Harvard University, 60 Garden St., Cambridge, MA 02138, USA}
\altaffiltext{21}{Center for Computational Astrophysics, Flatiron Institute, New York, NY 10010, USA}
\altaffiltext{22}{Kobayashi-Maskawa Institute for the Origin of Particles and the Universe, Nagoya University, Nagoya, 464-8602, Aichi, Japan}
\altaffiltext{23}{Kavli Institute for Particle Astrophysics and Cosmology, SLAC National Accelerator Laboratory, Stanford University, 2575 Sand Hill Road, Menlo Park, CA 94025, USA}

\email{chiaki.hikage@ipmu.jp}


\maketitle

\begin{abstract}
We measure cosmic weak lensing shear power spectra with the Subaru
Hyper Suprime-Cam (HSC) survey first-year shear catalog covering
137~deg$^2$ of the sky. Thanks to the high effective galaxy number
density of $\sim 17$~arcmin$^{-2}$ even after conservative cuts such
as magnitude cut of $i<24.5$ and photometric redshift cut of $0.3\leq
z \leq 1.5$, we obtain a high significance measurement of the cosmic
shear power spectra in 4 tomographic redshift bins, achieving a total
signal-to-noise ratio of 16 in the multipole range $300 \leq \ell \leq
1900$. We carefully account for various uncertainties in our analysis
including the intrinsic alignment of galaxies, scatters and biases in
photometric redshifts, residual uncertainties in the shear
measurement, and modeling of the matter power spectrum. The accuracy
of our power spectrum measurement method as well as our analytic model
of the covariance matrix are tested against realistic mock shear
catalogs.  For a flat $\Lambda$ cold dark matter ($\Lambda$CDM) model,
we find $S_8\equiv \sigma_8(\Omega_{\rm
  m}/0.3)^\alpha=0.800^{+0.029}_{-0.028}$ for $\alpha=0.45$
($S_8=0.780^{+0.030}_{-0.033}$ for $\alpha=0.5$) from our HSC
tomographic cosmic shear analysis alone. In comparison with {\it
  Planck} cosmic microwave background constraints, our results prefer
slightly lower values of $S_8$, although metrics such as the Bayesian
evidence ratio test do not show significant evidence for discordance
between these results. We study the effect of possible additional
systematic errors that are unaccounted in our fiducial cosmic shear
analysis, and find that they can shift the best-fit values of $S_8$ by
up to $\sim 0.6\sigma$ in both directions.  The full HSC survey data
will contain several times more area, and will lead to significantly
improved cosmological constraints.
\end{abstract}

\section{Introduction}
The $\Lambda$ Cold Dark Matter ($\Lambda$CDM) model has been
established as the standard cosmological model to describe the
expansion history and the growth of the large-scale structure of the
Universe.  Assuming the $\Lambda$CDM model, cosmological parameters
have been measured within percent-level uncertainties by a combination
of observations such as the cosmic microwave background (CMB)
experiments \citep[e.g.,][]{Hinshaw13,Planck16,Planck18_cosmology},
type-Ia supernovae \citep[e.g.,][]{Suzuki12,JLA}, and baryon acoustic
oscillations \citep[BAO; e.g.,][]{Anderson14,Alam17}. Despite the
success of the model, we are challenged by a fundamental lack of
physical understanding of the main components of the Universe, dark
matter and cosmological constant $\Lambda$ or more generally dark
energy. In order to understand these dark components, it is of great
importance to test the $\Lambda$CDM model at high precision using a
variety of cosmological probes.

Weak gravitational lensing provides an important means of studying the
mass distribution of the Universe including dark matter, because it is
a purely gravitational effect. In particular, the coherent distorted
pattern of distant galaxy images by gravitational lensing of
large-scale structure, commonly referred to as the cosmic shear
signal, is a powerful probe of the matter distribution in the Universe
\citep{Blandford91,MiraldaEscude91,Kaiser92}.  Cosmic shear, the
two-point correlation function or power spectrum of the weak lensing
signal, depends on both the growth of the matter density field and the
expansion history of the Universe, and serves as a unique cosmological
probe. It allows us to test a range of cosmological models including
dynamical dark energy and modified gravity \citep[see][for
  reviews]{BartelmannSchneider01,Kilbinger15}. Since the first
detections of cosmic shear around 2000
\citep{Bacon00,vanWaerbeke00,Wittman00,Kaiser00,Maoli01,Rhodes01,vanWaerbeke01,Hoekstra02,Bacon03,Jarvis03,Brown03,Hamana03},
cosmic shear studies have progressed in their precision thanks to the
progress of wide-field imaging surveys. For instance, the
Canada-France-Hawaii Telescope Lensing Survey (CFHTLS) survey observed
$\sim 10^7$ galaxies over 154 square degrees of the sky
\citep[CFHTLenS;][]{Heymans12} to conduct tomographic analyses
\citep{Hu99} of cosmic shear
\citep{Heymans13,Kilbinger13,Kitching14}. The Deep Lens survey (DLS)
conducted a deep cosmic shear analysis using galaxies with a limiting
magnitude 27 mag in $R$-band over 20 square degrees of the sky
\citep{Jee16}. Galaxy imaging surveys for even wider areas, which are
known as ``Stage III'' surveys, are on-going \citep{DETF06}. These
Stage III surveys, which include the Kilo-Degree survey
\citep[KiDS;][]{Kuijken15}, the Dark Energy Survey
\citep[DES;][]{Abbott16,Becker16}, and the Hyper Suprime-Cam (HSC)
survey \citep{HSCSSP,HSCDR1}, are expected to yield constraints on
cosmological parameters from the cosmic shear analyses that are
competitive with other dark energy probes. Cosmic shear is especially
sensitive to the combination of the matter density parameter
$\Omega_{\rm m}$ and the amplitude parameter of matter fluctuations
$\sigma_8$, i.e., $S_8(\alpha)\equiv\sigma_8 (\Omega_{\rm
  m}/0.3)^{\alpha}$ with $\alpha\sim 0.5$. In the next decade, we
expect that ``Stage IV'' galaxy surveys such as the Large Synoptic
Survey Telescope \citep[LSST;][]{LSST09}, the Wide Field Infrared
Survey Telescope \citep[WFIRST;][]{Spergel15} and Euclid
\citep{Euclid11} will provide even more accurate measurements of
cosmic shear from observations of $\sim 10^9$ galaxies over thousands
of square degrees.

Accurate cosmic shear measurements are needed in order to test the
concordance between the cosmological parameters obtained from the {\it
  Planck} CMB experiment, which is based on high redshift linear
physics, and lensing surveys which are based on much lower redshifts
and non-linear physics. In the flat $\Lambda$CDM model, the {\it
  Planck} temperature and polarization power spectra (without CMB
lensing) constrain $S_8 (\alpha=0.5)$ to be $0.848\pm 0.024$
\citep{Planck16}, whereas several lensing surveys infer values of
$S_8$ lower by about 2-3$\sigma$, e.g., $0.757^{+0.033}_{-0.038}$ from
the KiDS-450 correlation function analysis \citep{Hildebrandt17},
$0.651\pm0.058$ from the KiDS-450 power spectrum analysis
\citep{Kohlinger17}, $0.732^{+0.029}_{-0.031}$
from CFHTLenS \citep[][for the fiducial case where systematics
 are not included]{Joudaki17}, and $0.782^{+0.027}_{-0.027}$ from 
DES year one (Y1) data \citep{Troxel18}. While the original
  constraints on $S_8$ from DLS is consistent with Planck,
  $0.818^{+0.034}_{-0.026}$ \citep{Jee16}, \citet{Chang19} shows that
  the $S_8$ value decreases to $0.795\pm 0.032$ when the fitting
  formula of the nonlinear matter power spectrum is updated from
  \citet{Smith03} to \citet{Takahashi12}. 
The tension may indicate physics beyond
the $\Lambda$CDM model such as dynamical dark energy or modified
gravity \citep[e.g.,][]{Amendola16}, and therefore the possible
systematic effects should be carefully examined \citep[see
  also][]{Troxel18b,Chang19}.

The Hyper Suprime-Cam Subaru Strategic Program (HSC-SSP, hereafter the
HSC survey) is a wide-field imaging survey using a 1.77 deg$^2$
field-of-view imaging camera on the 8.2-meter Subaru telescope
\citep{Miyazaki12,Miyazaki15,HSC_Camera,Komiyama18,Furusawa18,Kawanomoto18}.
The HSC survey is unique due to the combination of its depth
($5\sigma$ point-source depth of the Wide layer of $i\sim 26$) and
excellent image quality (typical $i$-band seeing of $\sim 0\farcs
58$), which enable us to measure cosmic shear signals up to higher
redshifts with lower shape noise than KiDS and DES. The data from the
first 1.7 years (61.5 nights) was publicly released in Feb 2017
\citep{HSCDR1}. \citet{Mandelbaum18a} present the first-year shear
catalog (Y1) for weak lensing science, and carry out intensive null
tests of the catalog against various possible systematics such as
errors in the point-spread function (PSF) modeling and biases in the
shear estimation. These null tests demonstrated that the shear catalog
meets the requirements for carrying out science from this data without
being significantly affected by systematics. Here the requirements we
set are that residual systematic errors identified from the data are
sufficiently smaller than the overall statistical error in a
measurement of the cosmic shear correlation function, where the
overall statistical error indicates a total signal-to-noise ratio of
the correlation function measurement estimated using the HSC mock
shear catalogs. \citet{Oguri18} have reconstructed two- and
three-dimensional mass maps from the first-year shear catalog. They
found significant correlations between the mass maps and projected
galaxy maps, and no statistically significant correlations between the
mass maps and the maps of potential sources of systematics, further
demonstrating that the first-year shear catalog is ready for science
analyses.

In this paper, we present results from a tomographic cosmic shear
analysis using the HSC first-year shear catalog. We adopt a
pseudo-spectrum (hereafter pseudo-$C_\ell$) approach to obtain
unbiased cosmic shear power spectra from incomplete sky data
\citep{Hikage11,HikageOguri16}.  We perform a nested sampling analysis
of the HSC cosmic shear power spectra to constrain cosmological
parameters, especially focusing on $S_8$, in the context of the flat
$\Lambda$CDM model. In order to obtain robust cosmological constraints
from cosmic shear measurements, we take into account various
systematic errors, and perform a blind analysis to avoid confirmation
biases affecting our results.  One of the systematic errors we
consider is the measurement error of galaxy images due to imperfect
modeling of the PSF and the deconvolution error of the PSF model from
galaxy images \citep{Mandelbaum18a}. We account for additive and
multiplicative biases in our shape measurement method quantified by
\citet{Mandelbaum18b} using image simulations of the HSC
survey. Another source of systematic errors is related to the
photometric redshift (photo-$z$) uncertainties. Since it is not
feasible to measure the spectroscopic redshifts of all galaxies used
for the weak lensing analysis, the redshift distribution of source
galaxies is inferred from just their photometric information
\citep{Tanaka18}. Intrinsic shape correlations due to tidal
interactions also result in systematics in cosmic shear measurements
\citep{HirataSeljak04,Joachimi15,Kirk15}. There are also uncertainties
in modeling the matter power spectrum on small scales due to baryonic
effects such as star formation, supernovae, and AGN feedback
\citep{White04,Zhan04,Huterer05, Jing06,Bernstein09, Semboloni11}.  In
addition to testing for these systematics, we conduct various internal
consistency tests among different photo-$z$ bins, fields, and ranges
of angular scales, as well as null tests of B-modes, to check the
robustness of our results. We present tests for our cosmic shear
measurement as well as analysis methods using realistic mock
catalogs. We discuss the consistency of our constraints with {\it
  Planck} CMB data and other lensing surveys such as DES and KiDS, and
also explore effects of the dark energy equation of state and non-zero
neutrino mass.

This paper is organized as follows. In Section~\ref{sec:data}, we
briefly describe the HSC first-year shear catalog that is used in our
cosmic shear analysis. In Section~\ref{sec:method}, we describe and
validate the pseudo-$C_\ell$ method to estimate unbiased cosmic shear
spectra from finite-sky non-uniform data.
In Section~\ref{sec:measurement}, we also show our
measurements of tomographic cosmic shear spectra using the HSC first-year
shear catalog. Section~\ref{sec:model} summarizes model ingredients
for our cosmological analysis, including predictions of cosmic shear
signals and covariance and our methods to take account of various
systematics in cosmic shear analysis.  Our cosmological constraints
and their robustness to different systematics are presented in
Section~\ref{sec:cosmology}.  Finally we give our conclusions in
Section~\ref{sec:summary}.

Since the cosmological likelihoods for the final {\it Planck} data release
\citep{Planck18_cosmology} are not yet available at the time of
writing this paper, throughout this paper we use {\it Planck} 2015 CMB results
\citep{Planck16} for the comparison and the joint analysis with our
HSC first-year cosmic shear measurement. We use
the joint TT, EE, BB, and TE likelihoods for $\ell$ between 2 and 29
and the TT likelihood for $\ell$ between 30 and 2508,
commonly referred to as {\it Planck} TT + lowP \citep{Planck16}. We do not
use CMB lensing results, which contain information on the growth of
structure and the expansion history of the Universe at late stages,
except when we combine our joint analysis result with
distance measurements using baryonic acoustic oscillations and Type Ia
supernovae (Section~\ref{subsec:jointconst}).

Throughout this paper we quote 68\% credible intervals for
parameter uncertainties unless otherwise stated.

\section{HSC first-year shear catalog}
\label{sec:data}
Hyper Suprime-Cam (HSC) is a wide-field imaging camera with 1.77 deg$^2$
field-of-view mounted on the prime focus of the 8.2-meter
Subaru telescope \citep{Miyazaki12,Miyazaki15,HSC_Camera}. The HSC
survey is using 300 nights of Subaru time over 6 years to conduct a
multi-band wide-field imaging survey with HSC. The HSC survey consists
of three layers; Wide, Deep and UltraDeep. The Wide layer, which is
specifically designed for weak lensing cosmology, aims at covering
1400 square degrees of the sky with five broadbands, $grizy$, with a
$5\sigma$ point-source depth of $r\approx 26$ \citep{HSCSSP}. Since
$i$-band images are used for galaxy shape measurements for weak
lensing analysis, $i$-band images are preferentially taken when the
seeing is better. As a result, we achieve a median PSF FWHM of
$\sim 0\farcs58$ for the $i$-band images used to construct the HSC
first-year shear catalog. The details of the software pipeline used to
reduce the data are given in \citet{Bosch18}, and particulars about the
accuracy of the photometry and the performance of the deblender are
characterized using a synthetic imaging pipeline in \citet{Huang18}
and Murata et al.\ ({\em in prep.}), respectively. The HSC Subaru Strategic
Program (SSP) Data Release 1 (DR1), based on data taken using 61.5 nights
between March 2014 and November 2015, has been made public \citep{HSCDR1}.

The HSC first-year shear catalog \citep{Mandelbaum18a} is based on
about 90 nights of HSC Wide data taken from March 2014 to April 2016,
which is larger than the public HSC DR1 data. We apply a
number of cuts to construct a shape catalog for weak lensing analysis
which satisfies the requirements for carrying out first year key
science \citep[see][for more details]{Mandelbaum18a}. For instance, we
restrict our analysis to the regions of sky  with approximately full
depth in all 5 filters to ensure the homogeneity of the sample. We
also adopt a \cmodel\ magnitude cut of $i<24.5$ (see \citealt{Bosch18}
for definition of \cmodel\ magnitude in the context of HSC), which
is conservative given that the magnitude limit of the HSC is $i\sim 26.4$
\citep[$5\sigma$ for point
  sources;][]{HSCDR1}. We remove galaxies with PSF modeling failures
and those located in disconnected regions. Regions of sky around bright
stars ($\sim 16\%$ of the total area) are masked \citep{Mandelbaum18a}.
As a result, the final weak lensing shear catalog covers 136.9~deg$^2$
that consists of 6 disjoint patches: XMM, GAMA09H, GAMA15H, HECTOMAP,
VVDS, and WIDE12H. \citet{Mandelbaum18a} and \citet{Oguri18} performed
extensive null tests of the shear catalog to show that the shear
catalog satisfies the requirements of HSC first-year science for both
cosmic shear and galaxy-galaxy lensing.

The shapes of galaxies are estimated on the $i$-band coadded images
using the re-Gaussianization PSF correction method
\citep{HirataSeljak03}. An advantage of this method is that it has
been applied extensively to Sloan Digital Sky Survey data, and thus  the
systematics of the method are well understood
\citep{Mandelbaum05,Mandelbaum13}.
In this method, the shape of a galaxy image is defined as
\begin{equation}
\label{eq:distortion}
\bm{e}=(e_1,e_2)=\frac{1-(b/a)^2}{1+(b/a)^2}(\cos 2\phi, \sin 2\phi),
\end{equation}
where $b/a$ is the observed minor-to-major axis ratio and $\phi$ is the
position angle of the major axis with respect to the equatorial
coordinate system.  The shear of each galaxy, $\bm{\gamma}^{\rm
  (obs)}$, is estimated from the measured ellipticity $\bm{e}$ as
follows:
\begin{equation}
\label{eq:obsshear}
\bm{\gamma}^{\rm (obs)}=\frac{1}{1+\langle m\rangle}\left(\frac{\bm{e}}{2{\cal
    R}}-\bm{c}\right),
\end{equation}
where ${\cal R}$ represents the {\it responsivity} that describes the
response of our ellipticity definition to a small shear
\citep{Kaiser95,BernsteinJarvis02} and is given by
\begin{equation}
{\cal R}=1-\langle e_{\rm rms}^2\rangle\,.
\end{equation}
Here $e_{{\rm rms}}$ is the intrinsic root mean square (RMS) ellipticity  per
component. The symbols $\langle \cdot\cdot\cdot\rangle$ denote a weighted
average where each galaxy carries a weight $w$ defined as the inverse variance
of the shape noise
\begin{equation}
\label{eq:weight}
w=(\sigma_{e}^2+e_{\rm rms}^2)^{-1},
\label{eq:wlweight}
\end{equation}
where $\sigma_{e}$ represents the shape measurement error for each
galaxy. The $e_{\rm rms}$ values are also defined per-galaxy based on
the signal-to-noise ratio (SNR) and resolution factor calibrated by
using an ensemble of galaxies with SNR and resolution values similar
to the given galaxy. The values $m$ and $\bm{c}$ represent the
multiplicative and additive biases of galaxy shapes
\citep{Mandelbaum18}. Both shape errors and biases are
estimated per object using simulations of HSC images of the {\it
Hubble Space Telescope} COSMOS galaxy sample. The higher resolution of
this space-based input galaxy catalog makes it ideal for calibrating
the shape errors and biases \citep[see][for the details of the image
simulations]{Mandelbaum18b}. The multiplicative bias of the individual
shear estimates is corrected using the weighted average $\langle
m\rangle$ over the ensemble of galaxies in each tomographic sample,
whereas the additive bias is corrected per object.

The redshift distribution of source galaxies is estimated from the HSC
five broadband photometry. In the HSC survey, photometric redshifts
(photo-$z$'s) are measured using several different codes
\citep[see][for details]{Tanaka18}, including a classical
template-fitting code ({\tt Mizuki}), a machine-learning code based on
self-organizing map ({\tt MLZ}), a neural network code using the
PSF-matched aperture (afterburner) photometry ({\tt Ephor AB}), an
empirical polynomial fitting code ({\tt DEmP}) \citep{HsiehYee14}, a
hybrid code combining machine learning with template fitting ({\tt
  FRANKEN-Z}), and an extended (re)weighting method to find the
nearest neighbors in color/magnitude space from a reference
spectroscopic redshift sample ({\tt NNPZ}). Each code is trained with
spectroscopic and grism redshifts, as well as COSMOS 30-band photo-$z$
data \citep[see][]{Tanaka18}.

In addition, we estimate the redshift distribution by reweighting the
COSMOS 30-band photo-$z$ sample \citep{Ilbert2009,Laigle2016} such
that the distributions of the HSC magnitudes in all the five bands
match those of source galaxies we use for our analysis (More
et al.\ {\em in prep.}). In this paper, we adopt the COSMOS-reweighted
redshift distribution as our fiducial choice. However, in our analysis
we also take into account the difference between the COSMOS-reweighted
redshift distribution and redshift distributions obtained by stacking
the probability distribution functions (PDFs) of the HSC photo-$z$'s
from the various methods mentioned above in order to quantify our
systematic uncertainty in our knowledge of the redshift distribution
of our source galaxies. We explain  how we include the uncertainty due
to photometric redshift errors in cosmic shear analysis in
Section~\ref{subsec:photoz}. We use the sample of
galaxies with their {\tt best} estimates \citep[see][]{Tanaka18} of
their photo-$z$'s ($z_{\rm best}$) in the redshift range from 0.3 to
1.5 as determined by {\tt Ephor AB}. As the HSC filter set straddles
the 4000{\AA} break, the performance of the photo-$z$ estimation is
best in this redshift range \citep{Tanaka18}. After this cut in the
redshift range, the shear catalog contains a total of about 9.0
million galaxies with a mean redshift of $\langle z\rangle \simeq 0.81$.
The resulting total number density of source galaxies $n_{\rm g}$ is $\sim
18.5$~arcmin$^{-2}$. We estimate the effective number density using
two different definitions. One is the definition adopted in
\citet{Heymans12} 
\begin{equation}
\label{eq:neff_H12}
n_{\rm g,eff}^{\rm (H12)}=\frac{1}{\Omega_{\rm sky}}\frac{\left\{\sum_i w_i\right\}^2}{\sum_i w_i^2},
\end{equation}
where $\Omega_{\rm sky}$ is the sky area and $w_i$ is the weight of
each galaxy defined by equation~(\ref{eq:weight}). The other is the
definition used in \citet{Chang13} 
\begin{equation}
\label{eq:neff_C13}
n_{\rm g,eff}^{\rm (C13)}=\frac{1}{\Omega_{\rm sky}}\sum_i\frac{e_{{\rm rms},i}^2}{\sigma_{e,i}^2+e_{{\rm rms},i}^2}.
\end{equation}
We find $n_{\rm g,eff}^{\rm (H12)}=17.6$~arcmin$^{-2}$ and $n_{\rm
  g,eff}^{\rm (C13)}=16.5$~arcmin$^{-2}$, respectively. In our
tomographic analysis, we divide the galaxy sample into four photo-$z$
bins each 0.3 wide in redshift. Thus the redshift range of the
tomographic bins are ($0.3$, $0.6$), ($0.6$, $0.9$), ($0.9$, $1.2$),
and ($1.2$, $1.5$) for the binning number from 1 to 4
respectively. Table~\ref{tab:info} lists the mean redshift, number of
galaxies, (effective) number density, and the intrinsic RMS
ellipticity in each tomographic bin.  We note that the intrinsic
ellipticity is related to shear by equation~(\ref{eq:obsshear}). The
corresponding RMS dispersion of intrinsic shear becomes $\sim 0.28$,
which is comparable to the values in other surveys, 0.29 for KiDS
\citep{Hildebrandt17} and 0.27 for DES \citep{Troxel18}.

In Table~\ref{tab:info2}, we compare our setup of the tomographic bins
and the total number density of source galaxies with those in KiDS-450
\citep{Hildebrandt17} and DES Y1 \citep{Troxel18}. Although the survey
area is smaller than KiDS-450 and DES Y1, the effective source number
density of the HSC survey is 2--3 times higher than these other
surveys. In addition, the HSC survey reaches higher redshifts where
cosmic shear signals are also higher.

\begin{table*}
  \caption{Summary of properties of individual tomographic bins.$^*$}
\begin{center}
   \begin{tabular}{ccccccccc}
    \hline
    bin number & $z$ range & $z_{\rm med}$ & $N_{\rm g}$ & $n_{\rm g}$ [arcmin$^{-2}$] & $n_{\rm g,eff}^{\rm (H12)}$ [arcmin$^{-2}$] & $n_{\rm g,eff}^{\rm (C13)}$ [arcmin$^{-2}$] & $\langle e_{\rm rms}^2\rangle^{1/2}$ & $\langle e_{\rm rms}^2+\sigma_e^2\rangle^{1/2}$ \\
    \hline
    1 & 0.3 -- 0.6 & 0.446 & 2842635 & 5.9 & 5.5 & 5.4 & 0.394 & 0.411 \\
    2 & 0.6 -- 0.9 & 0.724 & 2848777 & 5.9 & 5.5 & 5.3 & 0.395 & 0.415 \\
    3 & 0.9 -- 1.2 & 1.010 & 2103995 & 4.3 & 4.2 & 3.8 & 0.404 & 0.430 \\
    4 & 1.2 -- 1.5 & 1.300 & 1185335 & 2.4 & 2.4 & 2.0 & 0.409 & 0.447 \\
    \hline
    All & 0.3 -- 1.5 & 0.809 & 8980742 & 18.5 & 17.6 & 16.5 & 0.398 & 0.423\\
    \hline
  \end{tabular}
\end{center}
  \begin{tabnote}
    $^*$We show redshift ranges ($z$ range), median redshifts ($z_{\rm
      med}$), total numbers of source galaxies ($N_{\rm g}$), raw
    number densities ($n_{\rm g}$), effective number densities
    ($n_{\rm g,eff}^{\rm (H12)}$; see equation~[\ref{eq:neff_H12}])
    defined in \citet{Heymans12}, effective number densities ($n_{\rm
      g,eff}^{\rm (C13)}$; see equation~[\ref{eq:neff_C13}]) defined
    in \citet{Chang13}, the mean intrinsic RMS ellipticity per
    component ($\langle e_{\rm rms}^2\rangle^{1/2}$) and the
      total RMS ellipticity per component ($\langle e_{\rm
        rms}^2+\sigma_e^2\rangle$), which are related to shear by
    equation~(\ref{eq:obsshear}), in our tomographic samples.  Source
    galaxies are assigned into four tomographic bins using photo-$z$
    {\tt best} estimates, $z_{\rm best}$, derived by the {\tt Ephor
      AB} photo-$z$ code (see text for details).  $z_{\rm med}$,
    $\langle e_{\rm rms}^2\rangle$ and $\langle e_{\rm
      rms}^2+\sigma_e^2\rangle$ are a weighted average
    [equation~(\ref{eq:weight})]
  \end{tabnote}
 \label{tab:info}
\end{table*}

\begin{table*}
   \caption{Comparison of lensing catalog properties of KiDS-450
    \citep{Hildebrandt17}, DES Y1 \citep{Troxel18}, and HSC Y1 (this
    paper) used for cosmic shear analyses.$^*$ }
\begin{center}
  \begin{tabular}{ccccccc}
    \hline
    survey catalog & area [deg$^2$] & No. of galaxies & $n_{\rm g,eff}^{\rm (H12)}$ [arcmin$^{-2}$] & $n_{\rm g,eff}^{\rm (C13)}$ [arcmin$^{-2}$] & $z$ range & tomography \\
    \hline
    KiDS-450 & 450  & 14.6M & 8.53 & 6.85 & 0.1 -- 0.9 & 4 bins \\
    DES Y1   & 1321 &  26M & 5.50 & 5.14 & 0.2 -- 1.3 & 4 bins \\
    HSC Y1  & 137  & 9.0M & 17.6 & 16.5  & 0.3 -- 1.5 & 4 bins \\
    \hline
  \end{tabular}
\end{center}
  \begin{tabnote}
    $^*$We compare the survey area, the number of galaxies after cuts for
    cosmic shear analysis, the effective number density, the redshift
    range, and the number of bins in tomographic analysis.
  \end{tabnote}
  \label{tab:info2}
\end{table*}

\begin{figure*}
\begin{center}
\includegraphics[width=14cm]{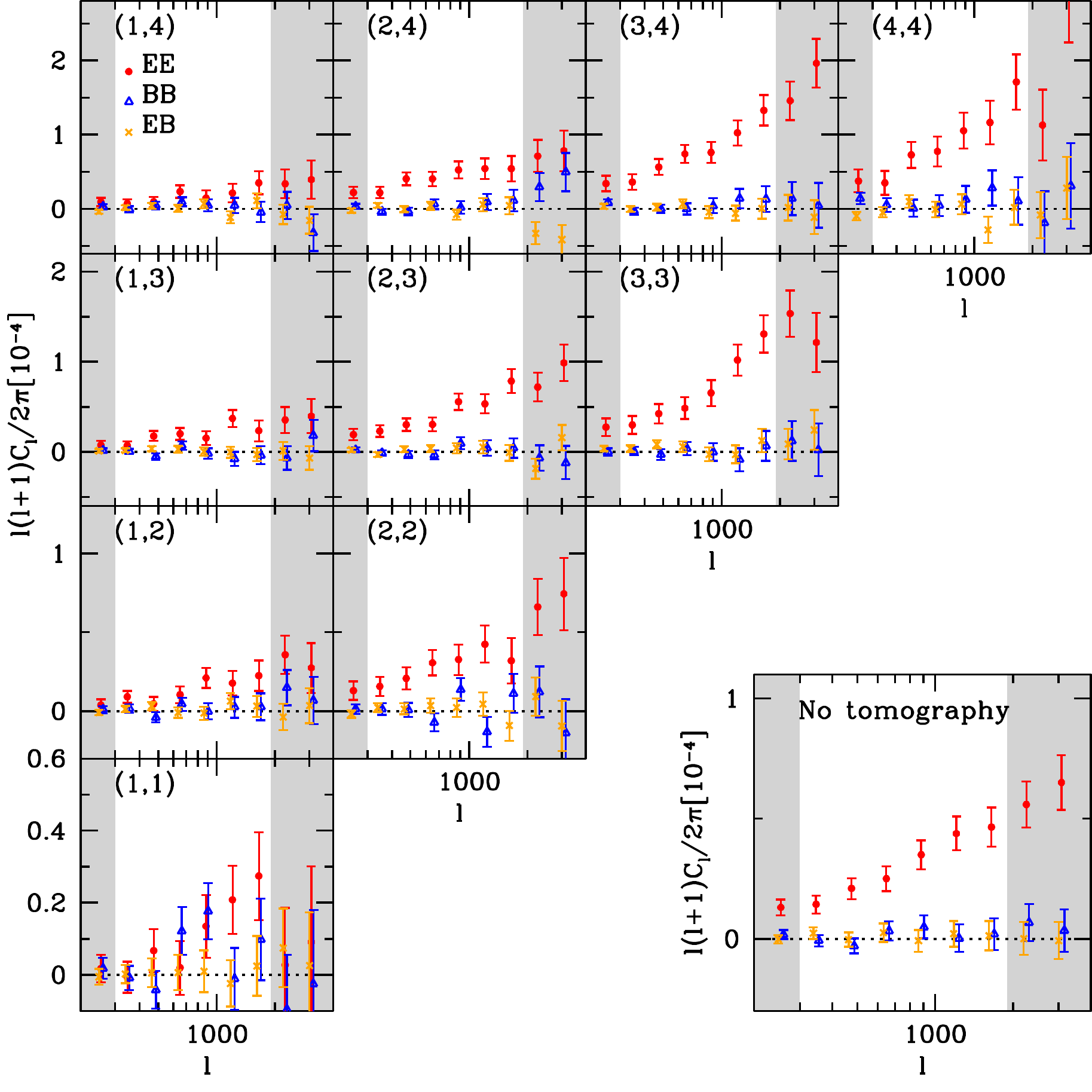}
\end{center}
\caption{Tomographic cosmic shear power spectra of EE ({\it red filled
    circles}), BB ({\it blue open triangles}), and EB ({\it yellow
    crosses}) modes. The galaxy samples are divided into four
  tomographic redshift bins using the {\tt Ephor AB} photo-$z$
  code. The redshift ranges of the four tomographic bins are set to
  [$0.3$, $0.6$], [$0.6$, $0.9$], [$0.9$, $1.2$], and [$1.2$,$1.5$],
  for binning number 1 to 4 (see also
  Table~\ref{tab:info}). The right-bottom panel shows the
  non-tomographic cosmic shear power spectrum. The multipole ranges of
  $\ell<300$ and $\ell>1900$ ({\it shaded regions}) are excluded in
  the cosmological analysis. The combined total detection significance
  of the tomographic EE-auto spectra is $16\sigma$ in the range of
  $300<\ell<1900$ ({\it unshaded regions}), whereas both BB and
  EB-mode spectra are consistent with zero.}
\label{fig:cl_obs_sum}
\end{figure*}

\section{Measurement methods}
\label{sec:method}
In this section, we summarize the measurement of cosmic shear power
spectra using the pseudo-$C_\ell$ method.  More details of the
formulation and validation tests using mock shear catalogs are given
in Appendix~\ref{sec:app1}. We also present the blinding methodology
adopted throughout our analysis.

\subsection{Pseudo-$C_\ell$ method}
\label{subsec:pseudocl}

We characterize cosmic shear signals using the power spectrum defined
in Fourier space. The power spectrum, which is the mean square of
fluctuation amplitudes as a function of wavenumber $k$, or multipole
$\ell$, is one of the most fundamental statistics to describe the
clustering properties of density fields \citep[e.g.,][]{Tegmark04}.
The power spectrum has been measured from different probes of the cosmic
density fields including CMB \citep[e.g.,][]{Hinshaw13,Planck16}, the
distribution of galaxies
\citep[e.g.,][]{Cole05,Yamamoto06,Percival10,Reid10,Blake11,Oka14,Alam17,Beutler17},
and the Lyman-$\alpha$ forest
\citep[e.g.,][]{McDonald06,Palanque-Delabrouille13,Viel13,Irsic17}.

However, it is non-trivial to measure the power spectrum in an
unbiased manner from data with incomplete sky coverage. In weak
lensing surveys, the sky coverage is usually very non-uniform due to
complicated survey geometry resulting from bright star masks, survey boundaries,
non-uniform survey depths, and
non-uniform galaxy shape weights. The observed shear field is given by the
weighted sum of shear values over galaxies in each sky pixel as
\begin{equation}
\bm{\gamma}^{\rm (obs)}(\bm{\theta})=W(\bm{\theta})\bm{\gamma}^{\rm (true)}(\bm{\theta}),
\end{equation}
where $W({\bm\theta})$ represents the survey window defined as the sum
of shear weights in each pixel. When a sky position ${\bm\theta}$ is
outside the survey area or masked due to a bright star,
$W({\bm\theta})$ is set to zero. We define a rectangular-shape region
enclosing each of the six HSC patches and then perform the Fourier
transformation of the observed shear field, $\bm{\gamma}^{\rm obs}$,
with typical pixel scale of about 0.88~arcmin, which is much smaller
than the scales we use in our cosmological analysis.
The power spectrum obtained simply from the amplitude of the
Fourier-transformed shear field is biased due to the convolution with
the mask field $W$. We apply the pseudo-$C_\ell$ method to obtain
unbiased estimates of the cosmic shear power spectrum by correcting
for the convolution with the survey window
\citep{Hikage11,Kitching12,HikageOguri16,Asgari18}. This method has also
been commonly used in CMB analyses \citep{Kogut03,Brown05}. The
details of the method may be found in Appendix~\ref{sec:app1}. In
short, the dimensionless binned lensing power spectrum ${\cal
  C}_b^{\rm (true)}$ corrected for the masking effect is
given by
\begin{equation}
{\cal C}_b^{\rm (true)}=\bm{M}_{bb'}^{-1}\sum_{\bm\ell}^{|{\bm\ell}|\in \ell_b'}P_{b'\ell}
(\bm{C}_{\bm\ell}^{\rm (obs)}-\langle \bm{N}_{\bm\ell}\rangle_{\rm MC}),
\label{eq:cl_measured}
\end{equation}
where $\bm{M}_{bb'}$ is the mode coupling matrix of binned spectra,
which is related to the survey window $W$ by
equation~(\ref{eq:convmat}), $\bm{C}_{\bm\ell}^{\rm (obs)}$ is the
pseudo-spectrum (masked spectrum) that we can directly measure from
the Fourier transform of ${\bf \gamma}^{\rm obs}$, and
$P_{b\ell}=\ell^2/2\pi$ is a conversion factor to the dimensionless
power spectrum.  The sum is over all Fourier modes in the given $\ell$
bin ($\ell_b'$). In order to
remove the shot noise, we randomly rotate orientations of individual
galaxies  to estimate the shot noise power spectrum
$\bm{N}_{\bm\ell}$, and subtract it from $\bm{C}_{\bm\ell}^{\rm
  (obs)}$. Specifically, we use 10000 Monte Carlo simulations with
random galaxy orientations to estimate the convolved noise spectrum
$\langle {\bm{N}}_\ell\rangle_{\rm MC}$. We use 15
logarithmically equal bins in the range $60\le \ell \le 6500$, although we
restrict ourselves to a narrower range for our cosmological inferences.

While the validity and accuracy of our pseudo-$C_\ell$ method have
been studied in depth in previous work \citep{Hikage11,HikageOguri16},
we explicitly check the accuracy of the pseudo-$C_\ell$ method for the
HSC first-year shear catalog by applying the method to the HSC mock
shear catalogs presented in \citet{Oguri18}. Note that this mock
  test is for verifying that our pseudo-$C_\ell$ method produces the unbiased measurement of lensing power spectra from inhomogeneous shear data, but not for
  verifying our modeling such as intrinsic alignment and baryon
  feedback. The mock shear catalogs have the same survey geometry
and spatial inhomogeneity as the real HSC first-year data, and include
random realizations of cosmic shear from the all-sky ray-tracing
simulation presented in \citet{Takahashi17}. These realistic mock
catalogs allow us to check the accuracy of the pseudo-$C_\ell$ method
in correcting for the masking effect, as well as the accuracy of our
analytic estimate of the covariance matrix as we will discuss
below. The results of the test with the HSC mock shear catalogs are
also presented in Appendix~\ref{sec:app1}. We find that our
pseudo-$C_\ell$ method recovers the input cosmic shear power spectrum
within 10\% of the current statistical errors at least over the range
of $\ell$ of interest, $80<\ell<2000$.  We also confirm that the input
values of $\Omega_{\rm m}$, $\sigma_8$, and $S_8$ are successfully
recovered from the mock catalogs. Specifically, from the analysis of
the mock catalogs we obtain $\Omega_{\rm m}=0.292\pm 0.014$,
$\sigma_8=0.801\pm 0.020$, and $S_8=0.791\pm 0.005$, which are
consistent with the input values, $\Omega_{\rm m}=0.279$,
$\sigma_8=0.82$, and $S_8=0.791$ to within the 68\% credible interval.
The credible intervals (error bars) are roughly $1/\sqrt{100}$ of the
accuracy we can achieve with the HSC first year shear catalog.

We note that the cosmic shear (E-mode) power spectrum is related to
the shear correlation functions $\xi_+$ and $\xi_-$ as
\begin{equation}
\xi_{\pm}(\theta)=\frac{1}{2\pi}\int d\ell\,\ell~ C_\ell J_{0,4}(\ell\theta),
\end{equation}
where $J_n(x)$ is the $n$-th order Bessel function of the first kind.
While mathematically the cosmic shear power spectrum carries the same
information as the shear two-point correlation functions for a
full-sky uniform survey, this is not
exactly true in finite-sky data. In addition, the covariance of
the power spectrum is diagonal in Gaussian fields, whereas the
covariance of the two-point correlation functions contains significant
non-diagonal elements even for Gaussian fields. Since the Gaussian
error  still  dominates in the current cosmic shear measurements,
the statistical independence is high among different $\ell$ modes.

\subsection{Blinding}
\label{subsec:blind}
We have entered an era of precision cosmology. With a growing number of
cosmological probes, one has to carefully guard against biases, including
confirmation bias, which may be particularly relevant when comparing results
with other experiments. To avoid confirmation bias, we perform our cosmological
analysis in a blind fashion. Within the HSC team, there are multiple projects
performing cosmological analysis on the weak lensing data, each with
separate individual timelines.  Therefore we pursue a two-tiered blinding
strategy such that unblinding one of the analysis teams does not automatically
unblind the others.  First, each analysis team is blinded separately at the
catalog level by preparing a set of three shear catalogs per analysis team with
different values of the multiplicative bias such that
\begin{equation}
  {\bf m}^{i}_{\rm cat} = {\bf m}_{\rm true} + \mathrm{d}{\bf m}^{i}_1
  + \mathrm{d}{\bf m}^{i}_2\,,
\end{equation}
where ${\bf m}_{\rm true}$ denotes the array of multiplicative bias values for
HSC galaxies as estimated in simulations and the index $i$ runs from 0 to 2 and
denotes the three different shear catalog versions.
The terms $\mathrm{d}{\bf m}^i_1$ and $\mathrm{d}{\bf m}^i_2$ are
different for each of three catalogs sent to each analysis team. The
values of each of these terms are stored in an encrypted manner in the
headers of the three shear catalogs. The term $\mathrm{d}{\bf m}^i_1$ can only be
decrypted by the analysis team lead, and this term is removed before
performing the analysis. The term $\mathrm{d}{\bf m}^i_2$ can only be decrypted by
the blinder-in-chief once the encrypted headers for $\mathrm{d}{\bf m}^i_2$ (stored
in the shear catalog) are passed on by the analysis team.  Exactly one
of the three values among $\mathrm{d}{\bf m}^i_2$ is zero, and can be
revealed by the blinder-in-chief once the analysis team is ready for unblinding.
The blinder-in-chief does not play any active role in the cosmological analysis
and is her/himself not aware of the values of $\mathrm{d}{\bf m}^i_2$ until the
end. The analysis group thus has to perform 3 analyses, a costly enterprise,
but then it avoids the need for reanalysis once the catalogs are unblinded.

The presence of $\mathrm{d}{\bf m}^i_1$ prevents accidental unblinding
by comparison of two sets of blinded catalogs sent out to two
different analysis teams. The presence of separate $\mathrm{d}{\bf
  m}^i_2$ allows each analysis team to remain blinded separately from
the other analysis teams. This constitutes the first tier of our
blinding strategy. The different multiplicative biases result in a
similar shape for the cosmic shear power spectra, but different
overall amplitudes, and thus different values of $S_8$. The values of
$\mathrm{d}{\bf m}^i_2$ are drawn randomly to allow
variations in $S_8$ at levels comparable to the differences between
the $S_8$ values inferred by {\it Planck} and other contemporary
lensing surveys.

We also guard ourselves against the possibility that the values of
$\mathrm{d}{\bf m}^i_2$ all come out close to each other by chance.  This would
automatically result in unblinding if we compared our cosmological constraints
to other surveys\footnote{Indeed it turned out that the values of
  $\mathrm{d}{\bf m}^i_2$ for our cosmic shear analysis happened to be
  close to each other by chance. We found out this fact after
  unblinding. Thanks to our strategy to adopt the analysis level
  blinding, the blinded nature of our analysis was not compromised.}.
Therefore as a second tier of protection, we also remain blinded at
the analysis level. We never compare the cosmic shear power spectra
obtained from any of our blinded catalogs with any theoretical
predictions on plots where the cosmological parameters of the
predictions were known beforehand. In addition, prior to unblinding we
always plot our cosmological constraints with the mean values
subtracted off and thus centered at zero.  Moreover, we do not compare
our cosmological constraints, even after shifting by the mean values,
with constraints from other surveys and experiments.

Prior to the start of the analysis we set down the conditions that must be
satisfied and systematics tests to be carried out before unblinding. The first
set of these conditions includes sanity checks about the satisfactory
convergence of posterior distributions on cosmological parameters for each of
the three shear catalogs given to the analysis team (see
Appendix~\ref{sec:convergence}). The second set concerns the analysis choices
for cosmic shear and study of their impact on the cosmological constraints.
These conditions were as follows:
\begin{itemize}
\item Make the code available to all collaboration members. Specific people
were assigned to review the code.
\item Test that the measurement code can recover the input power spectrum from
a mock data set within statistical uncertainties.
\item Test that the inference code can recover cosmological parameters, in
particular $S_8$, from the cosmic shear signal inferred from a mock dataset
within statistical uncertainties.
\item Estimate the systematic uncertainty due to the differences between various
photo-$z$ estimates, and quantify impact on the cosmological constraints.
\item Quantify the impact of the range of angular scales used, in particular
whether dropping the smaller angular scales results in a statistically
significant change to the inferred cosmological constraints.
\item Quantify the impact of removing individual photometric redshift bins from
the analysis and test whether any specific bins result in a statistically
significant shift of parameters.
\item Test the goodness of fit for the measured cosmic shear power spectra from
all of the blinded catalogs.
\item Quantify the impact of using different sets of matter power spectra
obtained from numerical simulations with a variation in the baryonic physics
recipes.
\end{itemize}

This paper describes the results of most of these tests.
Once a decision to unblind is reached, the second tier of blinding
(analysis level blinding) is removed by the analysis team just a few
hours prior to the final unblinding by the blinder-in-chief. Plots
with cosmological constraints and their comparison with the CMB
results are made for each of the three catalogs. The encrypted
headers with values of $\mathrm{d}{\bf m}^i_2$ are sent to the
blinder-in-chief so that the blinder-in-chief can decrypt these values
and  give them to the analysis team.

Lastly, we emphasize that we started our cosmological analysis
only after the HSC first year shear catalog \citep{Mandelbaum18a} was
finalized, i.e., the construction of the shear catalog was not
influenced by the cosmic shear signal, its shape or amplitude.
Although the shear catalog and the associated systematic
tests were finalized without blinding, the cosmological analysis does
not influence the decisions related to the shear catalog.

\section{Cosmic shear measurement}
\label{sec:measurement}
In this section, we present the measurement of tomographic cosmic
shear power spectra using the HSC first year shear catalog by
separating their E-mode and B-mode signals. We also estimate the
impact of PSF leakage and residual PSF model errors on our cosmic
shear measurement.

\subsection{Cosmic shear power spectra}
\label{subsec:shearspectra}
We use the pseudo-$C_\ell$ method described in
Section~\ref{subsec:pseudocl} to measure tomographic cosmic shear
power spectra of E-mode, B-mode auto, and EB-cross modes from the HSC
first-year shear catalog. The power spectra are shown in
Figure~\ref{fig:cl_obs_sum}. In deriving the spectra, we first
measured cosmic shear spectra in the six disjoint fields individually,
and then obtained a weighted average of the spectra using weights
computed from the sum of source weights of individual galaxies, $w_i$
(equation~\ref{eq:weight}).

We find that the B-mode signals appear qualitatively consistent with
zero, as expected. A possible exception is in the low multipole range
$\ell<300$, where the excess B-mode signals are significant.  As
\cite{Oguri18} found 2-3$\sigma$ B-mode residuals due to PSF modeling
errors, this is partly due to the PSF model ellipticity residuals, as
we will discuss in Section~\ref{subsec:resPSF}. Therefore, in our
cosmic shear analysis, we set the lower limit of the multipoles to
$\ell_{\rm min}=300$. We also set the upper limit to $\ell_{\rm
  max}=1900$ because of model uncertainties at such high multipole, as
we will discuss in Section~\ref{sec:model}. As shown by
\citet{Asgari18}, removing scales with significant B-modes does not
always ensure that the systematic error that causes those B-modes does
not impact the E-modes on other scales. To further mitigate the
systematic effect, we take account of both PSF leakage and residual
PSF model errors in our modeling, although their contribution is small
on the fiducial range of scales (see Section~\ref{subsec:PSF}).

We quantitatively check the consistency of the B-mode cosmic shear
power spectra with zero using the following chi-squared statistic,
\begin{equation}
\chi^2=\sum_{i,j,i',j'}\sum_{b,b'}^{\ell_{\rm min}\le\ell_b\le \ell_{\rm max}}
C_b^{BB (ij)}({\cal N}^{BB})^{-1}_{bb'}C_{b'}^{BB (i'j')}\,,
\label{eq:cl_bb}
\end{equation}
where the first summation runs over the four tomographic bins.  For
the covariance of the B-mode cosmic shear power spectra, we only use
the shape noise covariance\footnote{We estimate
the noise covariance matrix of B-mode power spectra from 10,000 Monte Carlo
realizations with random galaxy orientations as we described in
Section~\ref{subsec:shearspectra}.  However, note that the noise
covariance matrices for E and B modes are equivalent in the
statistical average sense.} ${\cal N}^{BB}$. While it is often
assumed that the shape noise covariance is given by
a simple analytic expression that depends only on the dispersion of galaxy
ellipticities and the number density of galaxies, in real observations
various effects such as the survey window function and the
inhomogeneous distribution of galaxies modify the shape noise
covariance \cite[e.g.,][]{Murata18,Troxel18b}. In order to obtain an
accurate shape noise covariance, we estimate the covariance directly
from the data, based on the estimate of the average shot noise power
spectrum discussed in Section~\ref{subsec:pseudocl} in which we
randomly rotate the orientations of source galaxies 10,000 times. From
this Monte Carlo sampling of shape noise power spectra, we can
directly construct the covariance matrix of shape noise power
spectra corrected for masking effects. We use this noise covariance matrix
throughout the paper. This noise covariance matrix is mostly diagonal, but
we find non-zero ($\lesssim 20\%$) off-diagonal components mostly
between neighboring multipole bins, which we also include in our analysis.

We find no significant B-mode signal for any of the auto- and
cross-power spectra measured between our fiducial four tomographic
bins. The most significant deviation from zero is found in the
lowest-redshift auto tomographic bin for which we find $\chi^2=12.1$
with $6$ data points, resulting in a $p$-value of 0.06.
The total $\chi^2$ over four bin tomographic B-mode auto spectra
becomes 60.7 with $60$ data points of the B-mode spectra (the
resulting $p$-value of 0.45) for our fiducial choice of
$300<\ell<1900$.  For the EB-cross mode, $\chi^2=59.7$ with the same
60 data points (with a resulting $p$-value of 0.49). We also confirm that
there are no significant B-mode signals even if we adopt other
photo-$z$ codes. We see no evidence either for systematics in the data
producing B-modes, or for leakage of E-mode power into B-mode power due to
the convolution of survey masks.  The latter indicates that our
pseudo-$C_\ell$ method successfully decomposes E- and B-modes as
expected from the analysis using HSC mock shear catalogs presented in
Appendix~\ref{sec:app1}.

\subsection{PSF leakage and residual PSF model errors}
\label{subsec:resPSF}

Systematics tests of the HSC first-year shear catalog presented in
\citet{Mandelbaum18a} and \citet{Oguri18} indicate that there are
small residual correlations between galaxy ellipticities and PSF
ellipticities resulting from imperfect PSF corrections. Such
residual PSF model errors could produce artificial two-point correlations
and hence bias our cosmic shear results. We check the impact of these
systematics in our cosmic shear measurements assuming that the
measured galaxy shapes have an additional additive bias given by
\begin{equation}
e^{\rm (sys)}=\tilde{\alpha} e^{\rm p}+\tilde{\beta} e^{\rm q}.
\label{eq:esys}
\end{equation}
The first term in the right hand side, referred to as PSF leakage,
represents the systematic error proportional to the PSF model
ellipticity $e^{\rm p}$ due to the deconvolution errors of the PSF
model. The second term represents the systematic error
associated with the difference between the model PSF ellipticity,
$e^{\rm  p}$, and the true PSF ellipticity that is estimated from
individual ``reserved'' stars,
$e^{\rm star}$, i.e., $e^{\rm q} \equiv e^{\rm p}-e^{\rm
  star}$ \citep{Troxel18}. The non-zero residual PSF ellipticity
$e^{\rm q}$ indicates an imperfect PSF estimate, which should also
propagate to shear estimates for galaxies. While the systematics tests carried out by
\citet{Mandelbaum18a} and \citet{Oguri18} suggest that these PSF
leakage and residual PSF model errors do not significantly affect our
cosmological analysis, it is of great importance to directly check the
potential impact of these errors on our measurements of the cosmic
shear power spectra.

When $e^{\rm (sys)}$ [equation~(\ref{eq:esys})] is added to the
observed galaxy ellipticity, these systematic terms change the
measured cosmic shear power spectrum as
\begin{eqnarray}
C_\ell & \rightarrow & C_\ell+\tilde{\alpha}^2C_\ell^{\rm pp}
+2\tilde{\alpha}\tilde{\beta} C_\ell^{\rm pq} +\tilde{\beta}^2C_\ell^{\rm qq}\,,
\label{eq:cl_psfsys}
\end{eqnarray}
where $C_\ell^{\rm pp}$, $C_\ell^{\rm qq}$, and $C_\ell^{\rm pq}$
represent the auto-spectrum of the model ellipticity $e_{\rm p}$, the
auto-spectrum of the residual PSF ellipticity $e_{\rm q}$, and the
cross-spectrum of $e_{\rm p}$ and $e_{\rm q}$, respectively.  We
subtract the shot noise term in the calculation of $C_\ell^{\rm pp}$
and $C_\ell^{\rm qq}$, which means that these power spectra would be
zero if there were no spatial correlation in $e_{\rm p}$ and $e_{\rm
  q}$, and that the value of the power spectrum shown on the plots
cannot be simply related to the typical PSF ellipticity value.  The
proportionality factors $\tilde{\alpha}$ and $\tilde{\beta}$ are
measured by cross-correlating $e^{\rm p}$ and $e^{\rm q}$ with the
observed galaxy ellipticities as
\begin{eqnarray}
C_\ell^{\rm gp}&=&\tilde{\alpha} C_\ell^{\rm pp}+\tilde{\beta} C_\ell^{\rm pq}\,,
\label{eq:cl_gp}\\
C_\ell^{\rm gq}&=&\tilde{\alpha} C_\ell^{\rm pq}+\tilde{\beta} C_\ell^{\rm qq}\,,
\label{eq:cl_gq}
\end{eqnarray}
where $C_\ell^{\rm gp}$ and $C_\ell^{\rm gq}$ denote the cross-spectra
between galaxy ellipticities used for the cosmic shear analysis and
$e_{\rm p}$ and $e_{\rm q}$, respectively. 

\begin{figure}
\begin{center}
\includegraphics[width=7.5cm]{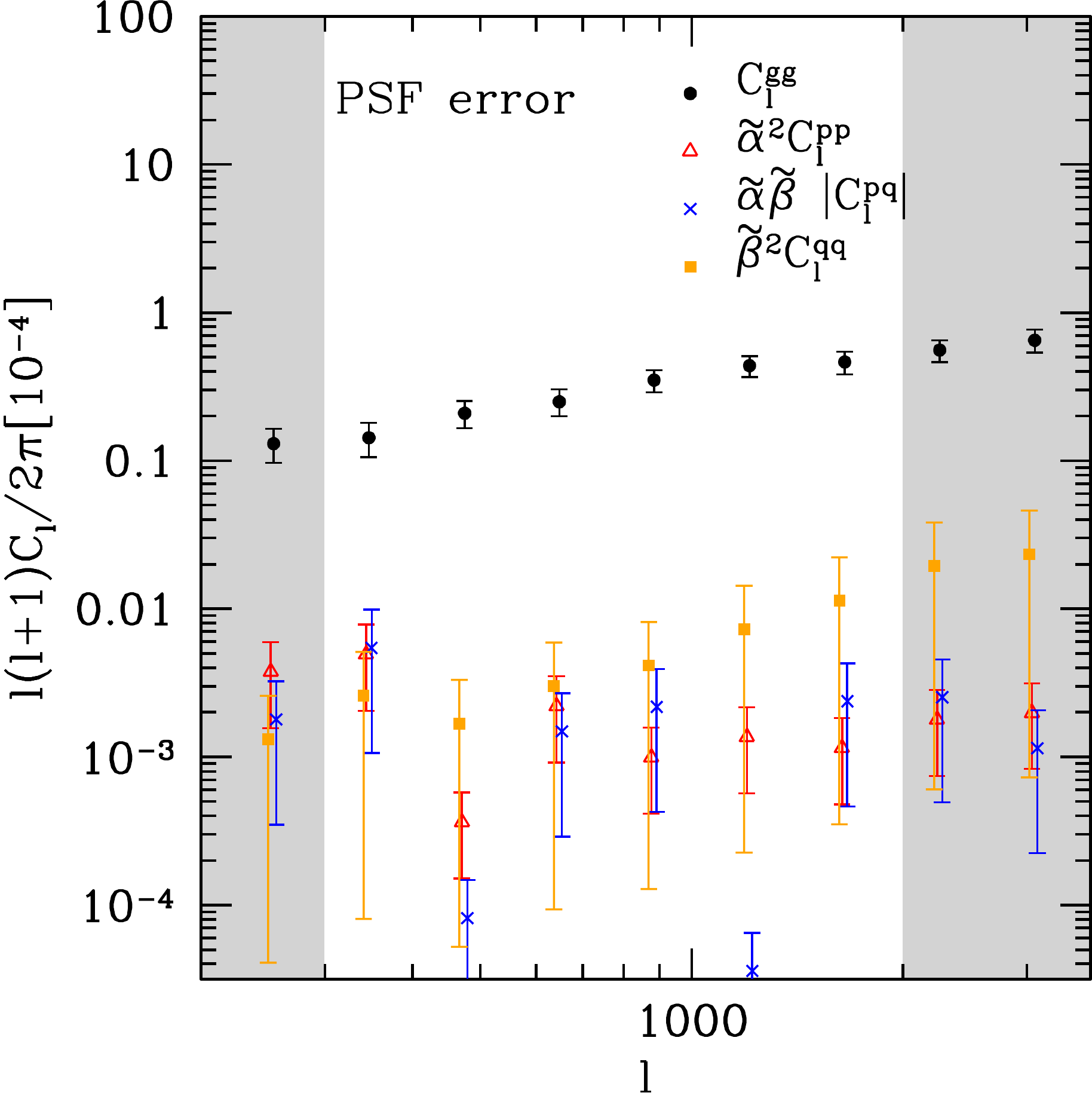}
\end{center}
\caption{The auto power spectrum of the PSF leakage
  $\tilde\alpha^2C_\ell^{pp}$ ({\it open triangles}), that of the
  residual PSF $\tilde\beta^2C_\ell^{qq}$ ({\it filled squares}), and
  their cross-spectrum $\tilde\alpha\tilde\beta |C_\ell^{pq}|$ ({\it
    crosses}) are compared with the non-tomographic cosmic shear power
  spectrum $C^{\rm gg}_\ell$ ({\it filled circles}) (see Section~\ref{subsec:resPSF} for
  details). The shaded region indicates the range of multipoles that
  is excluded in our fiducial cosmological analysis. It is found that
  both of the PSF systematics are subdominant in the fiducial
  multipole range.}
\label{fig:cl_psf}
\end{figure}

In the HSC software pipeline \citep{Bosch18}, PSF stars are selected
based on the distribution of high-$S/N$ objects with size.  However,
$\sim 20\%$ of such stars are not used for PSF modeling, so that they
can be used for cross-validation of PSF modeling. This $\sim 20\%$
sample of stars is referred to here as the reserved star sample. In
this paper, we use this subsample of stars for computing the auto- and
cross-spectra of $e_{\rm p}$ and $e_{\rm q}$
(Figure~\ref{fig:cl_psf}). Using equations~(\ref{eq:cl_gp}) and
(\ref{eq:cl_gq}), we find $\tilde{\alpha}=0.057\pm 0.018$ and
$\tilde{\beta}=-1.22\pm 0.74$, where the errors are estimated by
randomly rotating orientations of the stars.

Given that the systematics in galaxy shape measurements depend only
upon the shapes and brightness of galaxies and not directly upon their
redshifts, per se, as a first order approximation, it is reasonable to
assume that $\tilde{\alpha}$ and $\tilde{\beta}$ are common for all
tomographic bins. However, it is plausible that the values of
$\tilde{\alpha}$ and $\tilde{\beta}$ are slightly different for
different tomographic bins, reflecting the different distributions of
galaxy properties such as their sizes and $S/N$ ratio. In particular,
the impact of PSF model shape errors on the shear two-point
correlations depends on the size distribution of the galaxies compared
to the PSF, which we call ``resolution'' \citep[see, e.g., Section 3.4
  of][]{Jarvis16}. Therefore, we compare the distribution of
resolution factors among the four tomographic bins, and find that both
mean values and overall distributions of resolution factors are very
similar among the four tomographic bins. Specifically, the weighted
mean values of the resolution factor $R_2$ are 0.603, 0.592, 0.598,
and 0.596 from lowest to highest redshift bins \citep[see
  also][]{Mandelbaum18b}.  This suggests that the redshift dependence
of $\tilde{\beta}$ coming from the difference of galaxy sizes is
negligibly small. We also note that, even if small redshift
dependences of $\tilde{\alpha}$ and $\tilde{\beta}$ are present, the
values between the different tomographic bins would be highly
correlated. Given that the number of the reserved stars is modest, the
estimates of $\tilde{\alpha}$ and $\tilde{\beta}$ from the auto- and
cross-correlation analysis is noisy by nature, making it challenging
to estimate these values for individual tomographic bins as well as
their covariance between different tomographic bins. For these
reasons, we adopt the common values of $\tilde{\alpha}$ and
$\tilde{\beta}$ found above for all the tomographic bins as our
estimates of systematics from the PSF leakage and PSF model residuals.

Figure~\ref{fig:cl_psf} shows the auto spectra of the PSF leakage and
residual PSF model errors as well as their cross spectra. We find that
both the PSF systematics are subdominant and the amplitudes of the
power spectra of both the PSF systematics are less than 5\% of the
non-tomographic cosmic shear power spectrum over our fiducial range of
scales. Although the contribution of the PSF systematics to the total
cosmic shear power spectrum is not very significant, it could
represent a larger fraction of the tomographic cosmic shear power
spectra in low redshift bins.  Therefore, in our cosmological
analysis, we marginalize over the PSF systematics by introducing two
nuisance parameters $\tilde\alpha$ and $\tilde\beta$ with Gaussian
priors as obtained from our systematics analysis.

We note that the mean shear value of the HSC first-year shear catalog,
$\langle{\bf \gamma}^{\rm obs}\rangle$, is consistent with zero, as
can be seen from the comparison of the data with the HSC mock shear
catalogs that include cosmic variance
\citep{Mandelbaum18a,Oguri18}. We do not subtract the mean shear value
from the shear catalog because such subtraction would artificially
suppress the cosmic shear power especially at small $\ell$ where the
small sky coverage gives a limited number of modes.

\section{Model ingredients for cosmological analysis}
\label{sec:model}

In this section, we summarize our model for the measured cosmic shear
power spectra which we use for the cosmology analysis. In addition to
the cosmic shear signal, our model accounts for various astrophysical
effects such as the intrinsic alignment of galaxies and the impact of
baryon physics as well as systematics due to photo-$z$ and shape
measurement uncertainties. We also describe our analytical estimate of
the covariance matrix. We summarize the parameters of our fiducial
cosmological model as well as nuisance parameters and their priors
that we use in our analysis.

\subsection{Cosmic shear signals}
\label{subsec:signal}

The comparison of observed cosmic shear power spectra derived in
Section~\ref{subsec:shearspectra} with model-predicted power spectra
allows us to constrain cosmological parameters.
In particular adding redshift information of source galaxies into the
cosmic shear measurements, the so-called  {\it cosmic shear
  tomography} \citep{Hu99,Takada04}, enables us to improve the
cosmological constraints by lifting degeneracies among parameters.
We compute cosmic shear power spectra for arbitrary cosmological
models using the flat-sky and Limber approximations \citep[see][for
  the validity of these approximations in our
  study]{Kitching17,Kilbinger17} such that
\begin{eqnarray}
\label{eq:cl}
C_\ell^{(ij)} = \int_0^{\chi_H} d\chi
\frac{q^{(i)}(\chi)q^{(j)}(\chi)}{f_K^2(\chi)}P_{\rm mm}^{\rm
  NL}\left(k=\frac{\ell+1/2}{f_K(\chi)},z\right),
\label{eq:cl_model}
\end{eqnarray}
where $i$ and $j$ refer to tomographic bins, $\chi$ is the comoving
radial distance, $\chi_H$ is the comoving horizon distance and $f_{\rm
  K}(\chi)$ is the comoving angular diameter distance. As our fiducial
model, we use the fitting formula for the nonlinear matter power
spectrum $P_{\rm mm}^{\rm NL}(k,z)$ provided by \citet{Takahashi12},
which is an improved version of the {\tt halofit} model by
\citet{Smith03} (see also Section~\ref{subsec:baryon}). We use this
improved {\tt halofit} model implemented in {\tt Monte Python}
\citep{Audren2013} which adopts the Boltzmann code {\tt CLASS} to
compute the evolution of linear matter perturbations
\citep{Lesgourgues2011,Audren11,Blas11}. While we do not include
neutrino mass in our fiducial analysis, we also check the effect of a
non-zero neutrino mass by replacing the {\tt halofit} model of
\citet{Takahashi12} with that of \citet{Bird12}.

The lensing efficiency function $q^{(i)}(\chi)$ for the $i$-th
tomographic bin is defined as
\begin{eqnarray}
\label{eq:q_cl}
q^{(i)}(\chi) &\equiv &
\frac{3}{2}\Omega_{\rm m}H_0^2\int_{\chi}^{\chi_H} d\chi' p^{(i)}(\chi')
(1+z)\frac{f_K(\chi)f_K(\chi,\chi')}{f_K(\chi')}, \nonumber \\
\end{eqnarray}
where $p^{(i)}(\chi)$ denotes the redshift distribution of source
galaxies in the $i$-th tomographic bin and is normalized such that
$\int d\chi' p^{(i)}(\chi')=1$.

\subsection{Source redshift distributions}
\label{subsec:pofz}
We infer the source redshift distributions in individual tomographic bins,
$\bar{P}^{(i)}(z)\equiv(d\chi/dz)p^{(i)}(\chi)$, based on the broadband
photometry of the HSC survey. In order to estimate the redshift
distribution of the source galaxies, we would ideally obtain
spectroscopic redshifts for a representative subsample of galaxies in
our sample. Given the depth of the HSC survey ($i<24.5$), this is quite a
challenging task. There are a number of spectroscopic redshift (spec-$z$
hereafter) surveys that overlap with the HSC footprint, such as GAMA
\citep{Liske2015} and VVDS \citep{LeFevre2013}. The differences between the
populations of these existing spec-$z$ samples and the weak lensing
source galaxy sample could potentially be accounted for by using
clustering and reweighting techniques \citep[see][for assumptions and
  caveats of this method]{Bonnett2016,Gruen2017}. These methods place
galaxies in the source galaxy sample into groups with similar photometric
properties. Galaxies from the spectroscopic sample that belong to
the same groups are reweighted to mimic the distribution of the weak
lensing source galaxy sample \citep{Lima2008}. Unfortunately, the
number of galaxies in the spectroscopic samples is not large enough to
accurately represent the photometric properties of our source sample
even after reweighting.

Therefore, instead of the spec-$z$ sample, we use the 2016 version of
the COSMOS 30-band photo-$z$ catalog \citep{Ilbert2009, Laigle2016}
and accept it as the ground truth. There are a number of caveats that
come with this assumption. First, the COSMOS sample represents a small
area of the sky and could be affected by sample variance. Secondly,
the photometric redshift codes used in the HSC survey have been
trained on the COSMOS 30-band photo-$z$ sample
\citep[see][]{Tanaka18}, which could lead to some circularity in
logic. Thirdly, even though COSMOS photo-$z$'s use 30-band
information, they are not as good as having spectroscopic
redshifts. For example, COSMOS photo-$z$'s are known to  have
attractor solutions which could cause unnecessary  pile up of
photo-$z$'s at certain locations in photo-$z$ space \citep[see also
  discussions in][]{Tanaka18}.

The reweighting method can overcome sample variance to some extent, as it
determines appropriate weights to map the COSMOS 30-band photo-$z$
sample to the HSC weak lensing sample that we use in our analysis. This
relies on the assumption that the color-redshift relation does not vary with
environment \citep{Hogg2004, Gruen2017}. Based on the variance of the four
CFHTLS deep fields, \citet{Gruen2017} estimated the cosmic variance
contribution in the context of galaxy-galaxy lensing and found the effect
on the angular diameter distance ratios in lensing to be approximately
3\% of the lens redshifts. Unfortunately only two of these fields are
in the current HSC footprint, which makes it difficult to compute
similar estimates of cosmic variance for our cosmic shear results. The
other way to estimate the cosmic variance would be to populate mock
simulation catalogs with galaxies with appropriate spectral energy
distributions \citep{Hoyle2018}. The algorithms to achieve an
appropriate assignment of HSC colors to the galaxies based on
environment and its evolution with redshift is still a subject of
active research.

As a workaround to the second caveat, we reserved 20$\%$ of the
galaxies from the original COSMOS 30-band photo-$z$ catalog, which are
not used for training the HSC photo-$z$'s. We use this subsample for
testing purposes. In the future, to avoid the photo-$z$ issues, we
plan to  make use of the cross-correlation technique to obtain
clustering redshifts \citep{Newman2008, Menard2013, McQuinn2013}.
Unfortunately, the area covered in the current
data release is not large enough to apply this method. Therefore, we
use the COSMOS reweighted distribution as our fiducial choice, but use
the stacked photo-$z$ PDFs to propagate our uncertain knowledge of the redshift
distributions to our cosmological constraints\footnote{Stacking photometric
redshift distributions to infer the underlying redshift distributions of the
population of galaxies is not a mathematically sound way to infer the
underlying redshift distribution of the sources. It is expected to inflate the
scatter in the inferred redshift distribution and could potentially also result
in biases \citep[see][for difficulties in estimation of the underlying redshift
distribution]{Padmanabhan2005}.  Nevertheless, such techniques have been used
previously in the analysis of cosmic shear \citep[see e.g.,][]{Kitching14}.
In the simplest case that the photo-$z$ PDFs are symmetric with respect to
the best constrained photometric redshift of a galaxy, the mean of the stacked
photo-$z$ PDF is expected to be an unbiased estimator for the mean of the
redshift of the galaxy sample despite resulting in a wider distribution.
Therefore, we do not directly use these distributions in our fiducial analysis,
but only to gauge the potential systematic impact of difference in these
methods.}.

Here we describe our procedure to obtain the weights that map the
COSMOS 30-band photo-$z$ sample to the HSC weak lensing sample.
For this purpose, we employ the HSC Wide observation of the COSMOS
field, although it is not included in our HSC first year shear catalog
presented in \citet{Mandelbaum18a}. The HSC $i$-band images in the
COSMOS field, which were obtained in the same observing constraints as
the HSC Wide survey overall, allows us to obtain weak lensing weights
[equation~(\ref{eq:wlweight})] for each of the COSMOS galaxies, as well
as the photo-$z$ as inferred by our different pipelines exclusively
based on the HSC photometry. We only use those COSMOS galaxies which
also pass our weak lensing cuts.

We first sort the galaxies in our entire weak lensing shear
catalog using their $i$-band {\tt cmodel} magnitude and 4 colors
(based on the afterburner photometry) into cells of a self-organizing
map (SOM; More et al., {\em in prep.}). The self-organizing map is a
clustering technique which groups objects of similar properties
together \citep[see][for application to photometric
redshifts]{2015ApJ...813...53M}. Given the HSC photometry, we classify the
COSMOS galaxies into SOM cells defined by the source galaxy sample. We then
compute weights for galaxies that belong to each SOM cell ($w^i_{\rm SOM}$)
such that
\begin{equation}
    w^{i}_{\rm SOM} = \frac{N_{\rm wl}^{i}}{N_{\rm COSMOS}^{i}}
\end{equation}
where $N_{\rm wl}^i$ denotes the number of galaxies in the weak
lensing source galaxy sample in the $i^{\rm th}$ cell, and $N_{\rm
COSMOS}^{i}$ denotes the number of galaxies in the COSMOS galaxy
sample. The weighted COSMOS 30-band photo-$z$ sample thus mimics our
source galaxy sample in terms of the photometry. We do not account for
the errors in the HSC photometry as the errors are the same
for the HSC weak lensing sample and the COSMOS photo-$z$
sample \citep{Gruen2017}.

In order to compute the redshift distribution of the
sources in the four different redshift bins of our weak lensing shear
catalog, we mimic our selection criteria on the COSMOS 30-band sample,
using HSC photometry and the HSC-derived photometric redshifts
($z_{\rm best}$) from the {\tt Ephor AB} pipeline, which is the code
we used to define the tomographic bins. Given these samples, we infer
the redshift distribution as a histogram of COSMOS 30-band photo-$z$
weighted by the lensing source weights times the SOM weights. In order
to compute the statistical noise due to the limited number of COSMOS
galaxies lying in certain SOM bins, we also perform a jackknife
estimate of the statistical error on the $P(z)$ distribution using
10 jackknife samples of the COSMOS galaxies.

On the other hand, the stacked photo-$z$ PDFs from the different HSC
photo-$z$ codes are derived by stacking the full PDF of photo-$z$'s
for individual galaxies $P_j(z)$ with their weight $w_j$
[equation~(\ref{eq:weight})]
\begin{equation}
\bar{P}^{(i)}(z)=\frac{\sum_j^{z_{\rm min}^{(i)}<z_j^{\rm best}<z_{\rm
      max}^{(i)}} w_jP_j(z)}{\sum_j^{z_{\rm min}^{(i)}<z_j^{\rm
      best}<z_{\rm max}^{(i)}} w_j}.
\label{eq:pofz}
\end{equation}
We note that $\bar{P}^{(i)}(z)$ estimated by COSMOS-reweighted or
stacked PDF of photo-$z$s has tails that extend beyond the tomographic
bin range $z_{\rm min}^{(i)}<z<z_{\rm max}^{(i)}$ simply because of
the finite width in the $P^{(i)}(z)$ of all galaxies.

We adopt the COSMOS-reweighted redshift distribution as our fiducial choice for
the redshift distributions. To be conservative, we allow these redshift
distributions to shift in the redshift direction by an amount $\Delta z$ in
each bin, independently, which result in a corresponding shift to the means of
the redshift distribution. We use the differences between the COSMOS reweighted
photo-$z$'s and the stacked photo-$z$ PDFs to put priors on $\Delta z$ and
propagate our uncertain knowledge of the redshift distributions to our
cosmological constraints (see Section~\ref{subsec:photoz}).

\subsection{Covariance}
\label{subsec:cov}
Accurate covariance matrices are crucial to make a robust estimation
of cosmological parameters from the measured cosmic shear power
spectra.  Without loss of generality we can break down the covariance
matrix of cosmic shear power spectra into three parts:
\begin{equation}
{\rm Cov}^{\rm (tot)}=
{\rm Cov}^{\rm (G)} + {\rm Cov}^{\rm (NG)} + {\rm Cov}^{\rm (SSC)},
\label{eq:cov_tot}
\end{equation}
where ${\rm Cov}^{\rm (G)}$, ${\rm Cov}^{\rm (NG)}$, and ${\rm Cov}^{\rm
  (SSC)}$ denote the Gaussian (G), the non-Gaussian (NG), and the
super-sample covariance (SSC) contribution to the covariance,
respectively. As discussed in Appendix~\ref{sec:covariance},
we adopt an analytical halo model for computing the covariance
\citep{CooraySheth02}, except that we use the direct estimate of the
shape noise covariance, ${\cal N}^{BB}_{bb'}$ [see around
  equation~(\ref{eq:cl_bb})]. The shape noise covariance is one part
of ${\rm Cov}^{\rm (G)}$ [see equation~(\ref{eq:cov_gauss_nn}) in
  Appendix~\ref{sec:covariance}].

The Gaussian and non-Gaussian power spectrum covariances have been
well studied in previous work \citep{Scoccimarro99,TakadaJain09}.  The
excess covariance due to super-sample modes has also been studied
\citep{TakadaBridle07,TakadaJain09,TakadaHu13} and tested using
ray-tracing simulations \citep{Sato09}. Based on these
findings, we employ an analytical halo model to compute the sample
variance contribution. Our analytic model of covariance includes all
these components, as well as its cosmological dependence.
While the analytic model involves various approximations, we show in
Appendix~\ref{sec:covariance} that it agrees well with the covariance
matrix estimated using the HSC mock shear catalogs.

\subsection{Intrinsic alignment}
\label{subsec:IA}
One of the major astrophysical systematic effects in the cosmic shear
analysis is the intrinsic alignment (IA) of galaxy shapes
\citep[see][for recent reviews]{Joachimi15,Kirk15,Kiessling15}.
The intrinsic alignment comes from two contributions. One is the
correlation between the intrinsic shapes of two galaxies residing in
the same local field
\citep{Heavens00,CroftMetzler00,Lee00,Catelan01}. The
other is the correlation of the gravitational shear acting on one
galaxy and the intrinsic shape of another galaxy \citep{HirataSeljak04}.

In this paper we adopt a nonlinear alignment (NLA) model
\citep{BridleKing07}, which is commonly used in cosmic shear analysis,
to describe the IA contributions. The NLA model is based on the linear
alignment model \citep{HirataSeljak04}, but the linear matter power
spectrum is replaced with the nonlinear power spectrum. This
phenomenological model has been found to fit the galaxy-shear cross
correlations down to $\simgt 1-2h^{-1}$Mpc quite well
\citep{Singh15,Blazek15},  and has been used in various cosmic shear
analyses \citep{Heymans13,Kitching14,Abbott16,Hildebrandt17}. In this
model, the intrinsic-intrinsic (II) and shear-intrinsic (GI) power
spectra are given by
\begin{eqnarray}
P_{\rm II}(k,z)&=&F^2(z)P_{\rm mm}^{\rm NL}(k,z), \\
P_{\rm GI}(k,z)&=&F(z)P_{\rm mm}^{\rm NL}(k,z).
\end{eqnarray}
The redshift- and cosmology-dependent factor relating
the galaxy ellipticity and the gravitational tidal field
is often parametrized as
\begin{equation}
\label{eq:Fz}
F(z)=-A_{\rm IA}C_1\rho_{\rm crit,0}\frac{\Omega_{\rm m}}{D(z)}
\left(\frac{1+z}{1+z_0}\right)^\eta\left(\frac{\bar{L}(z)}{L_0}\right)^\beta,
\end{equation}
where $A_{\rm IA}$ is a dimensionless amplitude parameter, $\rho_{\rm
  crit,0}$ is the critical density of the Universe at $z=0$, and
$D(z)$ is the linear growth factor normalized to unity at $z=0$. In
the expression above, additional redshift ($z$) and $r$-band
luminosity ($\bar{L}(z)$) dependences are assumed to have a power-law
form, with indices $\eta$ and $\beta$ being the power-law indices of
the redshift and luminosity dependences, respectively.  The
normalization constant factor $C_1$ is set to be $5\times
10^{-14}h^{-2}M_\odot^{-1}{\rm Mpc}^3$ at $z_0=0.62$, which is
motivated by the observed ellipticity variance in SuperCOSMOS
\citep{Brown02} and also used in other lensing surveys such as DES
\citep{Troxel18} and KiDS \citep{Hildebrandt17}.

Previous studies have detected IA signals only for red galaxies.  The
index for the luminosity dependence of the IA signal for red galaxies has
been measured to be $\beta=1.13_{-0.20}^{+0.25}$ for the
MegaZ-LRG+SDSS LRG \citep{Joachimi11} and $\beta=1.30\pm 0.27$ for the
SDSS LOWZ samples \citep{Singh15}. So far there is no evidence of
additional redshift dependence, i.e., $\eta$ is consistent with
zero, although admittedly these tests have been carried out at $z\lesssim 0.7$,
below our median redshift. For the HSC first-year shear catalog, even
if $\eta=0$, we expect an apparent redshift evolution of IA amplitudes
from the difference of source galaxy luminosities at different redshifts.
Therefore, in the paper we adopt the following functional form for the
prefactor
\begin{equation}
\label{eq:Fz_eff}
F(z)=-A_{\rm IA}C_1\rho_{\rm crit}\frac{\Omega_{\rm m}}{D(z)}
\left(\frac{1+z}{1+z_0}\right)^{\eta_{\rm eff}},
\end{equation}
where $\eta_{\rm eff}$ represents the effective redshift evolution of the
IA amplitudes due to a possible intrinsic redshift evolution and/or
the change of the galaxy population as a function of redshift,
and therefore includes the effects of both $\eta$ and $\beta$ in
equation~(\ref{eq:Fz}).

Here we discuss plausible values of $\eta_{\rm eff}$ from available
observations. Since intrinsic alignments have only been observed for
red galaxies, we simply assume that only red galaxies have intrinsic
alignments. We assume that the IA signal of red galaxies is
proportional to $L^\beta$ with $\beta=1.2\pm 0.3$, which is consistent
with the observations we quoted above. We also assume that
there is no intrinsic redshift dependence ($\eta=0$). In this case,
the redshift evolution of the IA amplitude is given as $F(z)\propto
f_{\rm red}(z)\bar{L}^\beta(z)$, where $f_{\rm red}(z)$ is the
fraction of red galaxies in our source sample at redshift $z$ and
$\bar{L}$ is the average absolute $r$-band luminosity of red galaxies
in our source sample as a function of redshift. We divide the HSC
shear catalog into red and blue galaxies using the color-$M_*$ plane,
where we use intrinsic $M_u$-$M_g$ color and stellar mass $M_*$
estimated by the {\tt Mizuki} photo-$z$ code
\citep{Tanaka15}\footnote{Although our fiducial samples are defined
  using the best redshifts for {\tt Ephor AB}, we use the template
  fitting code {\tt Mizuki} for this purpose, since only {\tt Mizuki}
  provides stellar mass and specific star formation rate
  estimates. Since we never use the individual photometric redshift
  anywhere other than sample selection, the difference in choice of
  photometric redshift codes should not have any impact on our
  calculations.}.
Specifically, we divide red and blue galaxies by the
line $M_u-M_g=0.1\{\log(M_*/10^9M_\odot)\}+1.12$ in the color-$M_*$
plane.  We find that the red fraction of the HSC first-year shear
catalog is nearly constant of $\sim 20$\% for the redshift range of
$0.3<z<1.5$.  This is not surprising because our sample is
flux-limited. Even if the red galaxy fraction {\it at fixed
  luminosity} decreases with increasing redshift, our high redshift
sample only includes very bright galaxies among which red galaxies
dominate, which more or less compensates the intrinsic decrease of the
red fraction at fixed luminosity.  We note that our result does not
change much even if we use specific star formation rate (sSFR) values,
which are also derived by {\tt Mizuki}, to divide the catalog into red
and blue galaxies using the threshold ${\rm sSFR}=10^{-10.3}{\rm
  yr^{-1}}$.  On the other hand, the mean luminosity of red galaxies,
$\bar{L}$, is found to evolve as $\sim (1+z)^{2.5}$. This leads to an
effective power-law index of the redshift dependence $\eta_{\rm
  eff}=3\pm 0.75$, which is obtained by multiplying 2.5 with
$\beta=1.2\pm 0.3$. However, given a large uncertainty in this
estimate of plausible values of $\eta_{\rm eff}$, in the following
analysis we fit both the dimensionless amplitude $A_{\rm IA}$ and the
effective power-law index $\eta_{\rm eff}$ as free parameters with
flat priors, which are marginalized over when deriving cosmological
constraints.

Given the three-dimensional II and GI power spectra, the GI and II
angular power spectra are respectively given by
\begin{eqnarray}
C_{\rm GI}^{(ij)}(\ell)&=&\int_0^{\chi_H} d\chi
\frac{q^{(i)}(\chi)p^{(j)}(\chi)+p^{(i)}(\chi)q^{(j)}(\chi)}{f_K^2(\chi)} \nonumber \\
&&\times P_{\rm GI}\left(k=\frac{\ell+1/2}{f_K(\chi)},z\right),
\end{eqnarray}
and
\begin{eqnarray}
C_{\rm II}^{(ij)}(\ell)=
\int_0^{\chi_H} d\chi\frac{p^{(i)}(\chi)p^{(j)}(\chi)}{f_K^2(\chi)}P_{\rm II}\left(k=\frac{\ell+1/2}{f_K(\chi)},z\right),
\nonumber\\
\end{eqnarray}
where is $p^{(i)}(\chi)$ is the normalized redshift distribution
function of source galaxies in the $i$-th tomographic bin and
$q^{(i)}(\chi)$ is the lensing efficiency function in the $i$-th
tomographic bin defined in equation~(\ref{eq:q_cl}).
Note that the cross power spectra between intrinsic galaxy shapes in
the different tomographic bins, $C_{\rm II}^{(ij)}$ with $i\ne j$,
can be non-zero due to an overlapping between the redshift
distributions of the galaxies in the different bins (see
Figure~\ref{fig:fullpz}).

It has been argued that on small scales, $\simlt 1-2h^{-1}$Mpc
($\ell\simgt 2000$) in the redshift range of our sample, the NLA model
underestimates the IA signal \citep{Schneider13,Sifon15,Singh15,Blazek15} and an
additional one-halo term is needed to
accurately model the observed IA signal. However, since the one-halo
term is not well understood, adding the one-halo term in the IA model would
introduce additional model uncertainties. This is one of the reasons
why we limit our cosmic shear analysis to $\ell<1900$ where the
one-halo term contribution is subdominant.

\subsection{Effects of baryon feedback on the matter power spectrum}
\label{subsec:baryon}
Hydrodynamical simulations including baryonic physics such as supernova
and AGN feedback effects indicate that the matter power spectrum can be
significantly modified at $\simlt 1h^{-1}$Mpc scales \citep[e.g.,][]{Schaye10,
vanDaalen2011, Vogelsberger14, Mead15, Hellwing16, McCarthy17, Springel18, Chisari18}.
There is significant uncertainty about how to incorporate the effects of baryonic
processes on scales well below the resolution limit of cosmological
simulations. The resultant uncertainty in theoretical matter power spectra
could potentially bias the cosmological parameters derived from cosmic
shear if small scales are used in the analysis
\citep{White04,Zhan04, Jing06, Bernstein09, Semboloni11, Osato15}.

We evaluate the impact of baryons following the methodology used in
\citet{Kohlinger17} and use a fitting formula from \citet{Harnois2015}
that interpolates between the
result for the matter power spectrum in the collisionless case and a
model with extreme baryonic feedback, with the help of a single extra
parameter. This fitting formula is based on the ``AGN'' model from
cosmological hydrodynamical simulations, the OverWhelming Large Simulations
\citep[OWLS;][]{Schaye10,vanDaalen2011},
where the AGN model has the largest effect on the matter power spectrum.
The matter power spectrum is modeled as
\begin{equation}
\label{eq:baryon}
P_{\rm mm}^{\rm (baryon)}(k,z)=b_{\rm baryon}^2(k,z)P_{\rm mm}^{\rm (T12)}(k,z),
\end{equation}
with
\begin{equation}
\label{eq:baryon_harnois}
b_{\rm baryon}^2(k,z)=1-A_B\left[A_ze^{(B_zx-C_z)^3}-D_zx e^{E_zx}\right],
\end{equation}
where $x=\log(k/[h {\rm Mpc}^{-1}])$, and $P_{\rm mm}^{\rm (T12)}$ is
the matter power spectrum in the absence of baryonic effects.
The parameters $A_z$, $B_z$,
$C_z$, $D_z$, and $E_z$ are redshift-dependent, and we use the
functional forms and values of the parameters as given by
\citet{Harnois2015} for the AGN model in OWLS. The parameter $A_B$
controls the strength of the baryon feedback effect.  The case with
$A_B=1$ corresponds to the matter power spectrum in the AGN model
presented in \citet{Harnois2015}, whereas $A_B=0$ corresponds to the
matter power spectrum in the collisionless case, i.e., the revised
{\tt halofit} model of \citet{Takahashi12}). As a further check, we
also adopt another fitting formula derived by \citet{Mead15}, which
is based on the same OWLS simulations, whose result is shown in
Section~\ref{subsec:robustness}.

We note that the baryonic effect on the matter power spectrum has also
been investigated using other state-of-the-art simulations with
baryonic physics fully implemented, including the EAGLE simulation
\citep{Hellwing16}, the IllustrisTNG simulations \citep{Springel18},
and the Horizon set of simulations \citep{Chisari18}. Although their
predictions of the baryonic effects have significant variations, all
of these simulations predict that baryon effects have a smaller effect
on the matter power spectrum than the OWLS AGN feedback model we use
here. The BAHAMAS simulations are an extension of the OWLS AGN model
with the feedback parameters calibrated to reproduce a wider range of
observations such as the galaxy stellar mass function and the X-ray
gas fractions in groups and clusters \citep{McCarthy17}.
\citet{McCarthy18} further extend the BAHAMAS to include massive
neutrinos to show that non-minimal neutrino mass can resolve the
tension between {\it Planck} and large-scale structure observations.

As shown in Section~\ref{subsec:robustness}, the baryonic effects on
our cosmological constraints are less than $1\sigma$ even using
the most extreme model that we adopt here.  This is a result of our
conservative choice for the upper limit of the multipole in our cosmic
shear analysis, $\ell_{\rm max}=1900$. Therefore, in our fiducial
analysis we simply adopt the matter power spectrum in the DM-only
model, i.e., the revised {\tt halofit} model of \citet{Takahashi12},
which is equivalent to fixing $A_B=0$ in
equation~(\ref{eq:baryon_harnois}). We also examine the baryon effect
on cosmological constraints by varying $A_B$ in our robustness checks
presented in Section~\ref{subsec:robustness}.

\subsection{Effects of PSF leakage and residual PSF model errors}
\label{subsec:PSF}

In Section~\ref{subsec:resPSF}, we explored the impact of
PSF leakage and residual PSF model errors on our cosmic shear power
spectrum measurements. While the contribution from these errors to the
non-tomographic cosmic shear power spectrum was found to be
small, they could still make non-negligible contributions
to the tomographic cosmic shear power spectra in low redshift bins for
which the power spectrum amplitudes are smaller. Thus,
following equation~(\ref{eq:cl_psfsys}), we add contributions from PSF
leakage and residual PSF model errors to all the tomographic cosmic
shear power spectra, and include the proportional factors
$\tilde{\alpha}$ and $\tilde{\beta}$ as model parameters. Based on the
measurements from cross-correlations (see
Section~\ref{subsec:resPSF}), we include Gaussian priors of
$\tilde{\alpha}=0.057\pm 0.018$ and $\tilde{\beta}=-1.22\pm 0.74$ in
our nested sampling analysis.

\subsection{Multiplicative bias and selection bias}
\label{subsec:multip}

The multiplicative bias $m$ for each source galaxy is estimated by
performing image simulations with properties carefully matched to real
data \citep{Mandelbaum18b}. The simulations are analyzed using the HSC
pipeline, just like the real data, allowing us to impose the same set
of flag cuts and cuts on object properties as in the real shear
catalog (see Section~\ref{sec:data}). In that paper, it was shown that
the residual multiplicative bias $m$ in the HSC first-year shear
catalog is controlled at the 1\% level, and thus satisfies our
requirements for HSC first-year science. Given this, we include a 1
percent uncertainty on the residual multiplicative bias, $\Delta m$,
as
\begin{equation}
\label{eq:merror}
C_\ell^{(ij)}\rightarrow (1+\Delta m)^2C_\ell^{(ij)}.
\end{equation}
We include a Gaussian prior with zero mean and a standard deviation of $0.01$
to $\Delta m$ when performing our analysis. As in the case of PSF leakage and
residual PSF model errors (see Section~\ref{subsec:resPSF}), we assume that the
$\Delta m$ value is common for all tomographic bins, because the multiplicative
bias does not depend directly on galaxy redshifts and hence values of $\Delta
m$ between different tomographic bins are expected to be highly correlated with
each other.

\begin{figure}
\begin{center}
\includegraphics[width=7.5cm]{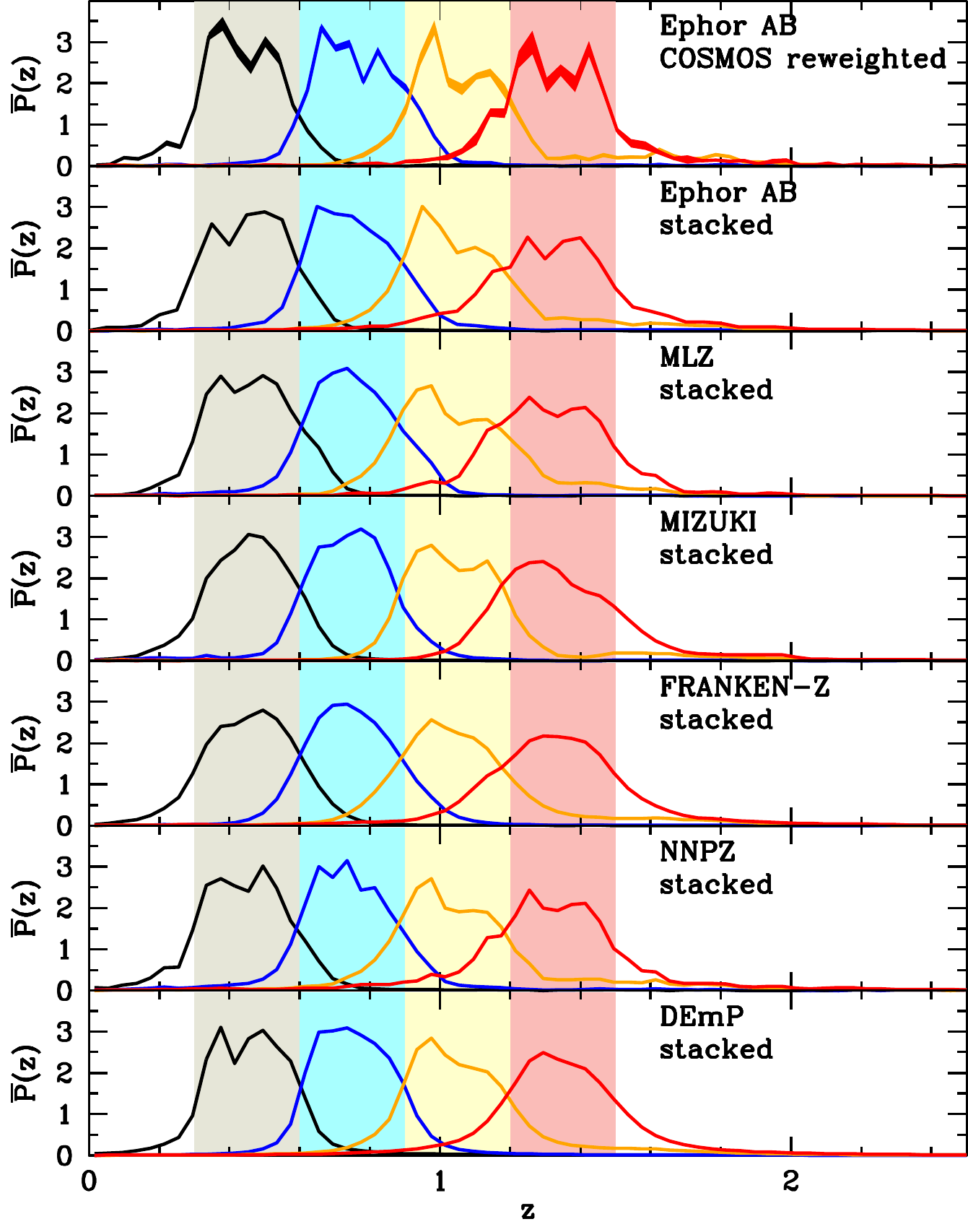}
\end{center}
\caption{Comparison of COSMOS-reweighted redshift distribution for the
  four tomographic bins with corresponding redshift distributions obtained by
  stacking the photo-$z$ PDFs using different photo-$z$ codes; {\tt Ephor
    AB} (photo-$z$'s used to define the tomographic bins), {\tt MLZ},
  {\tt Mizuki}, {\tt FRANKEN-Z}, {\tt NNPZ}, and {\tt DEmP}.  Shaded
  regions show our definition of 4 tomographic bins for {\tt best}
  photo-$z$'s, i.e., $0.3<z<0.6$, $0.6<z<0.9$, $0.9<z<1.2$, and
  $1.2<z<1.5$. The shaded region around each curve in the top panel shows
  the statistical error
  of the COSMOS-reweighted redshift distributions estimated by the
  bootstrap resampling technique. In our cosmology analysis, we take
  account of these differences of redshift distributions $\bar{P}(z)$
  using $z$-shift parameter $\Delta z$
  [equation~(\ref{eq:deltaz})]. We also discuss the impact of
  different $\bar{P}(z)$ on cosmological parameters in
  Section~\ref{subsec:robustness}.}
\label{fig:fullpz}
\end{figure}

\begin{table*}
  \caption{Summary of the photo-$z$ distribution for each method.}
\begin{center}
   \begin{tabular}{ccccc}
    \hline
    photo-$z$ method & $0.3<z<0.6$ & $0.6<z<0.9$ & $0.9<z<1.2$ & $1.2<z<1.5$ \\
    \hline
Fiducial & 0.44 ( 6.1\%), 0.43, 0.11 & 0.77 ( 3.6\%), 0.75, 0.12 & 1.05 (12.6\%), 1.03, 0.13 & 1.33 ( 5.9\%), 1.31, 0.15 \\
Ephor AB & 0.45 ( 7.4\%), 0.44, 0.12 & 0.75 ( 5.1\%), 0.76, 0.12 & 1.04 ( 7.2\%), 0.98, 0.16 & 1.32 ( 4.8\%), 1.29, 0.19 \\
MLZ      & 0.46 ( 2.8\%), 0.46, 0.12 & 0.75 ( 2.3\%), 0.74, 0.11 & 1.04 ( 3.2\%), 1.02, 0.13 & 1.32 ( 1.6\%), 1.31, 0.16 \\
Mizuki   & 0.45 ( 3.3\%), 0.46, 0.10 & 0.74 ( 1.8\%), 0.74, 0.10 & 1.04 ( 3.0\%), 1.04, 0.10 & 1.33 ( 1.2\%), 1.32, 0.12 \\
Franken-Z& 0.45 ( 5.9\%), 0.46, 0.12 & 0.75 ( 3.3\%), 0.74, 0.13 & 1.04 ( 7.3\%), 1.03, 0.14 & 1.34 ( 4.3\%), 1.33, 0.16 \\
NNPZ     & 0.44 ( 6.7\%), 0.44, 0.11 & 0.74 ( 4.0\%), 0.73, 0.12 & 1.03 ( 8.6\%), 1.01, 0.12 & 1.33 ( 7.9\%), 1.31, 0.14 \\
DEmP     & 0.45 ( 5.9\%), 0.45, 0.11 & 0.75 ( 3.6\%), 0.74, 0.11 & 1.04 ( 6.2\%), 1.02, 0.13 & 1.34 ( 6.5\%), 1.32, 0.16 \\
    \hline
  \end{tabular}
\end{center}
  \begin{tabnote}
    $^*$We show the mean, median, and $1\sigma$ dispersion of the
    photo-$z$ distribution for each method, for each of the 4-bin
    tomographic bins, as shown in Figure~\ref{fig:fullpz}. The mean
    value is estimated by clipping the distribution outside the
    $3\sigma$ range, where the clipping is repeated until the mean
    value converges. The values in parentheses denote the clipped
    fractions, which reflect outlier fractions of photo-$z$'s of
    individual galaxies.
  \end{tabnote}
 \label{tab:zinfo}
\end{table*}

In addition, we take account of the multiplicative selection bias
$m_{\rm sel}$ due to cuts in the resolution factor that characterize
galaxy size.  \citet{Mandelbaum18b} found that the selection bias
is proportional to the fraction of galaxies at the sharp
  boundary of galaxy size cut (the resolution factor $R_2$ in our
  terminology)
as $m_{\rm sel}=A\times p(R_2=0.3)$ with $A=0.0087\pm 0.0026$. We
use this formula to estimate the selection bias in each tomographic
bin. It is found that $m_{\rm sel}$ is at the level of 0.01, as listed
in Table~\ref{tab:msel}. Since the statistical errors in $m_{\rm sel}$ are $\sim
0.003$ and therefore are smaller than $\Delta m$ introduced above, we
ignore the statistical error of $m_{\rm sel}$.

Furthermore, we also include the responsivity correction due to the
dependence of the intrinsic ellipticity variations on redshift. In the
HSC first-year shear catalog, the intrinsic ellipticity was estimated
as a function of $S/N$ and resolution factor $R_2$ using image
simulations \citep{Mandelbaum18b}. Inferring the intrinsic ellipticity
dispersion required us to separate out the measurement error
contribution to the total shape variance.  This separation was carried
out using pairs of galaxies simulated at 90$^\circ$ with respect to
each other, for which we derived an analytic method of inferring the
measurement error contribution to the total shape variance.  Once we
have estimated the measurement error contribution as a function of
galaxy properties, it is possible to infer the intrinsic shape noise
dispersion (assuming independence of the measurement error and
intrinsic shape contributions for each galaxy). We recently found that
the intrinsic ellipticity varies with redshift such that the intrinsic
rms error $e_{\rm rms}$ is smaller between $1<z<1.5$ than that at
other redshifts\footnote{While this result may seem surprising, since
  galaxies at higher redshift generally have a more irregular
  morphology, this has been seen before, e.g., in Figure 18 of
  \citet{Leauthaud2007}, which used a different second moments-based
  shape estimator in higher-resolution {\it Hubble Space Telescope}
  images.}. This variation of $e_{\rm rms}$ affects our cosmic shear
signals via the responsivity factor. We include this correction in our
theoretical model by introducing an additional multiplicative bias
factor $m_{\cal R}$, which has the value $m_{\cal R}=0.015$ for
$0.9<z<1.2$, $m_{\cal R}=0.03$ for $1.2<z<1.5$, and zero otherwise
\citep[see Section~5.3 of][]{Mandelbaum18b}.

Together with the uncertainty of the original multiplicative bias
factor in equation~(\ref{eq:merror}), we correct the theoretical model
of tomographic lensing power spectra as
\begin{equation}
C_\ell^{(ij)}\rightarrow (1+\Delta m)^2(1+m_{\rm sel}^{(i)}+m_{\cal R}^{(i)})(1+m_{\rm
  sel}^{(j)}+m_{\cal R}^{(j)})C_\ell^{(ij)},
\end{equation}
where $\Delta m$ is a model parameter with Gaussian prior, and $m_{\rm
  sel}^{(i)}$ and $m_{\cal R}^{(i)}$ are fixed numbers listed in
Table~\ref{tab:msel}.

\begin{table}
  \caption{Multiplicative bias correction factors associated with the
    selection bias due to cuts in the resolution factor, $m_{\rm
      sel}$, and with the responsivity correction due to intrinsic
    ellipticity variations with redshifts, $m_{\cal R}$, for the four
    tomographic bins.}
\begin{center}
  \begin{tabular}{cccc}
    \hline
    $z$ range & $100m_{\rm sel}$ & $100m_{\cal R}$ \\
    \hline
    0.3 -- 0.6 & 0.86 & 0.0 \\
    0.6 -- 0.9 & 0.99 & 0.0 \\
    0.9 -- 1.2 & 0.91 & 1.5 \\
    1.2 -- 1.5 & 0.91 & 3.0 \\
    \hline
  \end{tabular}
\end{center}
\label{tab:msel}
\end{table}

\subsection{Redshift distribution uncertainty}
\label{subsec:photoz}

In our cosmological analysis, we take into account uncertainty in the redshift
distribution of our source galaxies by comparing redshift distributions
of source galaxies from the reweighting method to that of the stacked
photo-$z$ PDF method, as well as the difference of stacked photo-$z$
PDFs among different photo-$z$ codes. Figure~\ref{fig:fullpz} shows
the comparison of the COSMOS-reweighted redshift distribution with
stacked photo-$z$ PDFs [equation~(\ref{eq:pofz})] among different
photo-$z$ codes \citep{Tanaka18} (also see Section~\ref{sec:data})
for all four tomographic bins. Statistical uncertainties in the
COSMOS-reweighted redshift distributions are estimated by the
bootstrap resampling technique.

\begin{table}
  \caption{
    The methodological errors $\Delta z_i$ and the code uncertainties
    $\sigma^{\rm code}_{\Delta z_i}$, as well as the total photo-$z$
    uncertainties $\sigma_{\Delta z_i}^{\rm tot}$ from the quadrature
    sum of these two uncertainties, for the four tomographic bins.
     }
\begin{center}
  \begin{tabular}{cccc}
    \hline
    $z$ range & $100{\Delta z_i}^{\rm method}$ & $100\sigma_{\Delta z_i}^{\rm code}$ & $100\sigma_{\Delta z_i}^{\rm tot}$ \\
    \hline
    0.3 -- 0.6 & 2.66 & 1.01  & 2.85 \\
    0.6 -- 0.9 & $-1.07$ & 0.83 & 1.35 \\
    0.9 -- 1.2 & $-3.79$ & 0.55 & 3.83 \\
    1.2 -- 1.5 & $-3.20$ & 1.98 & 3.76 \\
    \hline
  \end{tabular}
\end{center}
\label{tab:zshift}
\end{table}

Model predictions of cosmic shear signals depend crucially on redshift
distributions of source galaxies (see Section~\ref{subsec:signal}),
suggesting that it is important to take proper account of photo-$z$
uncertainties and especially the effect of the photo-$z$ errors on the
mean of the redshift distribution in each tomographic bin. We quantify
the impact of the photo-$z$ error on the cosmic shear power spectrum
using the $z$-shift parameter $\Delta z_i$, which uniformly shifts the
redshift distribution of source galaxies in the $i$-th tomographic bin
as
\begin{equation}
\label{eq:deltaz}
\bar{P}^{(i)}(z)\rightarrow \bar{P}^{(i)}(z+\Delta z_i).
\end{equation}
For each estimate of the redshift distribution that is different from the
fiducial one, we derive a value of $\Delta z_i$ so that the cosmic shear power
spectrum amplitude computed using that $\bar{P}^{(i)}(z)$ matches our fiducial
cosmic shear power spectrum computed using the COSMOS-reweighted
$\bar{P}^{(i)}(z)$. We have verified that given the signal-to-noise ratio of
our cosmic shear measurements, the shifts that we consider here cannot be
distinguished from differences in the shape of the redshift distribution.

We estimate photo-$z$ uncertainties in two different ways in order to
avoid any double counting of photo-$z$ uncertainties. First, we
evaluate $\Delta z_i$ between the fiducial COSMOS-reweighted
$\bar{P}^{(i)}(z)$ and those obtained by using the stacked photo-$z$
PDFs using the {\tt Ephor AB} code. This $\Delta z_i$, which we denote
$\Delta z_i^{\rm method}$, represents the methodological
uncertainty. Next, we evaluate the photo-$z$ uncertainties due to the
photo-$z$ algorithm as the scatter of $\Delta z_i$ among the six
photo-$z$ codes (see Figure~\ref{fig:fullpz}). Specifically, for each
photo-$z$ code we estimate $\Delta z_i$ between the fiducial
COSMOS-reweighted $\bar{P}^{(i)}(z)$ and the stacked photo-$z$ PDF
from that photo-$z$ code by matching the amplitudes of the cosmic
shear power spectrum, and regard the standard deviation, $\sigma^{\rm
  code}_{\Delta z_i}$, among the six $\Delta z_i$ from the six
photo-$z$ codes as the photo-$z$ uncertainty due to the photo-$z$
algorithm.  We present more details of these photo-$z$ distributions
Table~\ref{tab:zinfo}, in which we list the mean, median, and
1$\sigma$ dispersion of each distribution.

We list values of $\Delta z_i^{\rm method}$ and
$\sigma_{\Delta z_i}^{\rm code}$ for the four tomographic bins in
Table~\ref{tab:zshift}.  We find that
$\left|\Delta z_i^{\rm method}\right|$ is at the
level of $0.03-0.04$ except for the lower-intermediate redshift
bin. In contrast, $\sigma_{\Delta z_i}^{\rm code}$ is $\lesssim 0.02$ and
therefore is smaller than $\Delta z_i^{\rm method}$. In our cosmic
shear analysis, we combine these two uncertainties in quadrature
\begin{equation}
  \sigma_{\Delta z_i}^{\rm tot}
  =\sqrt{(\Delta z_i^{\rm method})^2+(\sigma_{\Delta z_i}^{\rm code})^2},
\end{equation}
and for each tomographic bin we include $\Delta z_i$ as defined in
equation~(\ref{eq:deltaz}) as a model parameter with Gaussian prior,
and set the mean and standard deviation of the Gaussian prior to zero
and $\sigma_{\Delta z_i}^{\rm tot}$, respectively. The values of
$\sigma_{\Delta z_i}^{\rm tot}$ for the four tomographic bins are
listed in Table~\ref{tab:zshift}. We find that the statistical errors
of the COSMOS-reweighted redshift distributions estimated by bootstrap
resampling translate into values $\Delta z_i$ that are a factor of 5-10
times smaller than $\sigma_{\Delta z_i}^{\rm tot}$. Thus
this source of statistical errors is negligible, and we do not
explicitly account for it.

The procedure above assumes that uncertainties of photo-$z$'s in
individual bins are parametrized by single parameters $\Delta z_i$,
which might be too simplistic. To check the robustness of our results
to changes in redshift distributions such as outlier
fractions, in Section~\ref{subsubsec:photoztest} we will conduct a
robustness check in which we replace $\bar{P}(z)$ from the fiducial
COSMOS reweighted method with those from stacked photo-$z$ PDFs with
different photo-$z$ codes.

\begin{table*}
  \caption{Summary of parameters and priors used in our nested
    sampling analysis of tomographic cosmic shear power spectra.$^*$ }
\begin{center}
  \begin{tabular}{lcr}
    \hline
    Parameter & symbols &  prior \\
    \hline
    physical dark matter density & $\Omega_{\rm c}h^2$ & flat [0.03,0.7] \\
    physical baryon density & $\Omega_{\rm b}h^2$ & flat [0.019,0.026]  \\
    Hubble parameter & $h$ & flat [0.6,0.9] \\
    scalar amplitude on $k=0.05$Mpc$^{-1}$ & $\ln (10^{10}A_s)$ & flat [1.5,6] \\
    scalar spectral index & $n_{\rm s}$ & flat [0.87,1.07]  \\
    optical depth & $\tau$ & \it{flat [0.01,0.2]} \\
    neutrino mass & $\sum m_\nu$ [eV] & fixed (0)$^\dagger$, fixed (0.06) or flat [0,1] \\
    dark energy EoS parameter & $w$ & fixed ($-1$)$^\dagger$ or flat $[-2,-0.333]$ \\
    \hline
    amplitude of the intrinsic alignment & $A_{\rm IA}$ & flat $[-5,5]$  \\
    redshift dependence of the intrinsic alignment & $\eta_{\rm eff}$ & flat $[-5,5]$  \\
    baryonic feedback amplitude & $A_B$ & fixed (0)$^\dagger$ or flat $[-5,5]$ \\
    \hline
    PSF leakage                & $\tilde{\alpha}$   & Gauss $(0.057,0.018)$ \\
    residual PSF model error   & $\tilde{\beta}$    & Gauss $(-1.22,0.74)$ \\
    uncertainty of multiplicative bias $m$ & $100\Delta m$ & Gauss $(0,1)$ \\
    photo-$z$ shift in bin 1 & $100\Delta z_1$ & Gauss $(0,2.85)$ \\
    photo-$z$ shift in bin 2 & $100\Delta z_2$ & Gauss $(0,1.35)$ \\
    photo-$z$ shift in bin 3 & $100\Delta z_3$ & Gauss $(0,3.83)$ \\
    photo-$z$ shift in bin 4 & $100\Delta z_4$ & Gauss $(0,3.76)$ \\
    \hline
  \end{tabular}
\end{center}
  \tabnote{$^*$In our fiducial analysis, we vary five cosmological
    parameters ($\Omega_{\rm c}h^2$, $\Omega_{\rm b}^2$, $h$, $A_s$, and
    $n_s$) within
    the $\Lambda$CDM model and an additional parameter $w$ in
    the $w$CDM model, two parameters to model the IA (amplitude $A_{\rm
      IA}$ and the power-law index of the redshift dependence
    $\eta_{\rm eff}$), and seven additional nuisance parameters; PSF leakage
    $\tilde{\alpha}$, residual PSF model error $\tilde{\beta}$, the
    multiplicative bias error $\Delta m$, and photo-$z$ mean shift
    values $\Delta z_i$ for the four tomographic bins. Optical
    depth $\tau$ is not used when analyzing cosmic shear measurements only, but
    is included when CMB datasets are combined, and its prior is
    indicated in italics. There are three types of priors; fixed ($x$)
    means the value is fixed to be $x$; flat $[x,y]$ means a flat prior
    between $x$ and $y$, and Gauss $(x,y)$ means a Gaussian prior
    with  mean value $x$ and  standard deviation $y$. In our fiducial
    analysis, the neutrino mass $\sum m_\nu$ is fixed to be 0
    and the baryonic feedback amplitude $A_B$ is fixed to be 0 when
    the baryonic effect is not included [see equation~(\ref{eq:baryon_harnois})].
    The parameter $A_B$ is included as a model parameter only when we
    check the robustness of our results in Section~\ref{subsec:robustness}.
    When we have multiple priors for each parameter, priors with
    $^\dagger$ indicate those adopted in our fiducial $\Lambda$CDM
    model analysis.
    }
  \label{tab:params}
\end{table*}

\subsection{Posterior distribution of parameters using nested sampling}
We draw samples from the posterior distribution of parameters given the cosmic
shear signal with the help of nested sampling as implemented in the publicly
available {\tt Multinest} (Multi-Modal Nested Sampler) code \citep{Feroz2008,
Feroz2009, Feroz2013, Buchner14} together with the package {\tt Monte Python}
\citep{Audren2013}. The log-likelihood of the data given our parameters obeys
the following equation,
\begin{eqnarray}
-2\ln\mathcal{L}&=&\sum_{iji'j'}\sum_{b,b'}^{\ell_{\rm min}\le\ell_b,\ell_{b'}\le \ell_{\rm max}}
\Delta C_{b}^{(ij)}\left[{\rm Cov}\right]^{-1}\Delta C_{b'}^{(i'j')} \nonumber \\
&& +\ln |{\rm Cov}| + {\rm const},
\label{eq:likelihood}
\end{eqnarray}
where $\Delta C_{b}^{(ij)}$ is the difference between measured
(Section~\ref{sec:measurement}) and model (Section~\ref{sec:model})
cosmic shear power spectra, and $i$ and $j$ refer to different
tomographic bins. The latter is a function of cosmological
parameters as well as various nuisance parameters. Given that we measure the
cosmic shear power spectra in multipole bins [see
equation~(\ref{eq:cl_measured})], the model predictions must be computed as a
bin-average of the theoretical power spectra weighted by the number of modes
present at each multipole within the bin.
To be explicit, we compute the model power spectrum in each multipole
bin, $C^{\rm model}_b$, as
\begin{equation}
C^{\rm model}_{b}=\frac{\sum_{\bm\ell}^{\ell\in \ell_b}
  P_{b\ell}C_\ell^{\rm model}}{\sum_{\bm\ell}^{\ell\in \ell_b} P_{b\ell}},
    \label{eq:cl_model_binned}
\end{equation}
where $\ell=|{\bm\ell}|$, $P_{b\ell}=\ell^2/2\pi$ is the conversion factor to the
dimensionless power spectrum. When computing $C^{\rm model}_b$,
the summation over ${\bm\ell}$ in each multipole bin
is matched to the measured spectra as in equation~(\ref{eq:cl_measured})
The covariance matrix is estimated by an
analytic model and also depends on cosmological parameters (see
Section~\ref{subsec:cov}). Since we adopt a cosmology-dependent
covariance matrix, we include the determinant of the covariance,
$|{\rm Cov}|$, in the expression of our log-likelihood.

The cosmological parameter set we use is summarized in
Table~\ref{tab:params}. We include the standard $\Lambda$CDM
cosmological parameters, such as the dark matter density $\Omega_{\rm
  c}h^2$, the baryon density $\Omega_{\rm b}h^2$, the dimensionless
Hubble constant $h$, the scalar spectral index $n_{\rm s}$, the
amplitude of the primordial curvature perturbation $A_s$ with
reasonably broad flat priors. In the fiducial case, the neutrino mass
is fixed to be zero. As we will show later, our results do not change
significantly when we apply the lower bound of the absolute sum of
neutrino mass, $\sim 0.06$~eV, indicated from the neutrino oscillation
experiments \citep[e.g., see][for a review]{Lesgourgues:2013neco.book.....L}.
We also constrain models where the dark energy equation of state
parameter $w$ is allowed to vary, in our extended analysis.  When
combining with CMB datasets, we add the free-electron scattering
optical depth $\tau$ as an additional parameter. In addition to
these cosmological parameters, we constrain two parameters for IA, the
amplitude $A_{\rm IA}$ and the redshift-dependence power-law index
$\eta_{\rm eff}$, as well as seven additional nuisance parameters to model
uncertainties in shear and photo-$z$ measurements.

We check the convergence of nested sampling using {\tt nestcheck}
\citep{Higson2018}, which is a publicly available code to assess the
convergence of the posterior distribution and implementation-specific
error due to the correlated samples and missing parts of
the posterior. We find that each run terminates when the remaining
posterior mass is sufficiently small. We also find no significant
implementation-specific errors. Details of this convergence test are
given in Appendix~\ref{sec:convergence}. As a sanity
check, we also derive the posterior distributions using the
traditional Markov-Chain Monte-Carlo algorithm implemented in the
{\tt CosmoMC} code \citep{Lewis02} and find excellent agreement
with the posterior distribution of our parameters obtained from nested
sampling.

\begin{figure*}
\begin{center}
\includegraphics[width=14cm]{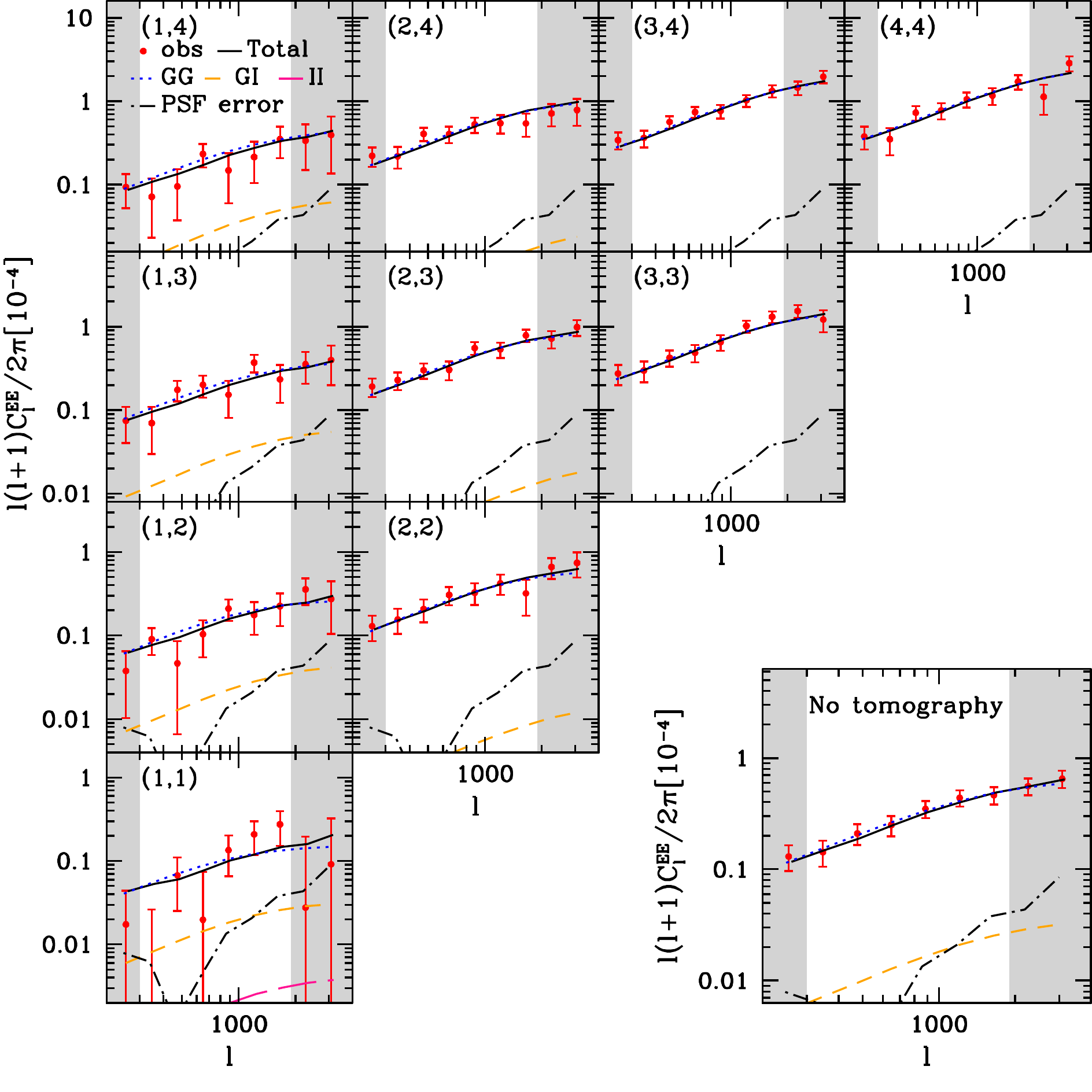}
\end{center}
\caption{Comparison of the measured tomographic shear power spectra
  with our theoretical model with best-fit values for the fiducial
  $\Lambda$CDM model. Best-fit IA power spectra of $C_{\rm GG}$ ({\it dotted}),
  $-C_{\rm GI}$ ({\it short dashed}), and $C_{\rm II}$ ({\it long
    dashed}) as well as power spectra arising from PSF leakage and PSF
  model error [equation~(\ref{eq:cl_psfsys})] ({\it dash-dotted}) are
  also plotted. The redshift range of $z_{\rm best}$
  in each tomographic bin is =$[0.3,0.6]$, $[0.6,0.9]$, $[0.9,1.2]$, and
  $[1.2,1.5]$ from 1 to 4. The right-bottom panel shows the measured
  non-tomographic cosmic shear power spectrum and the model spectra with the
  best-fit values from the tomographic analysis.
  The $C_{\rm II}$ term is so small that it is absent from all panels except for 11.}
\label{fig:cl_tomography}
\end{figure*}

\begin{figure*}
\begin{center}
\includegraphics[width=8cm]{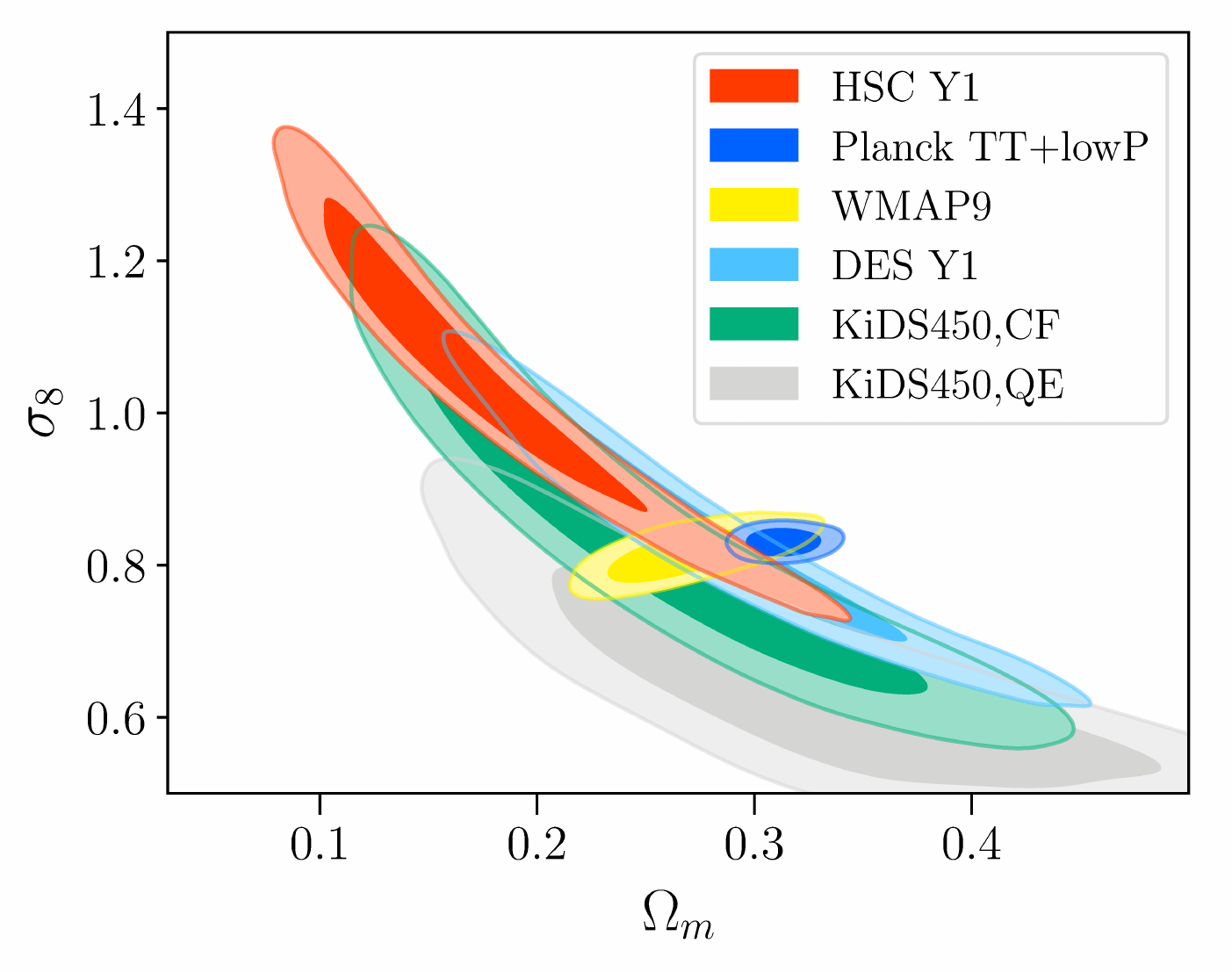}
\includegraphics[width=8cm]{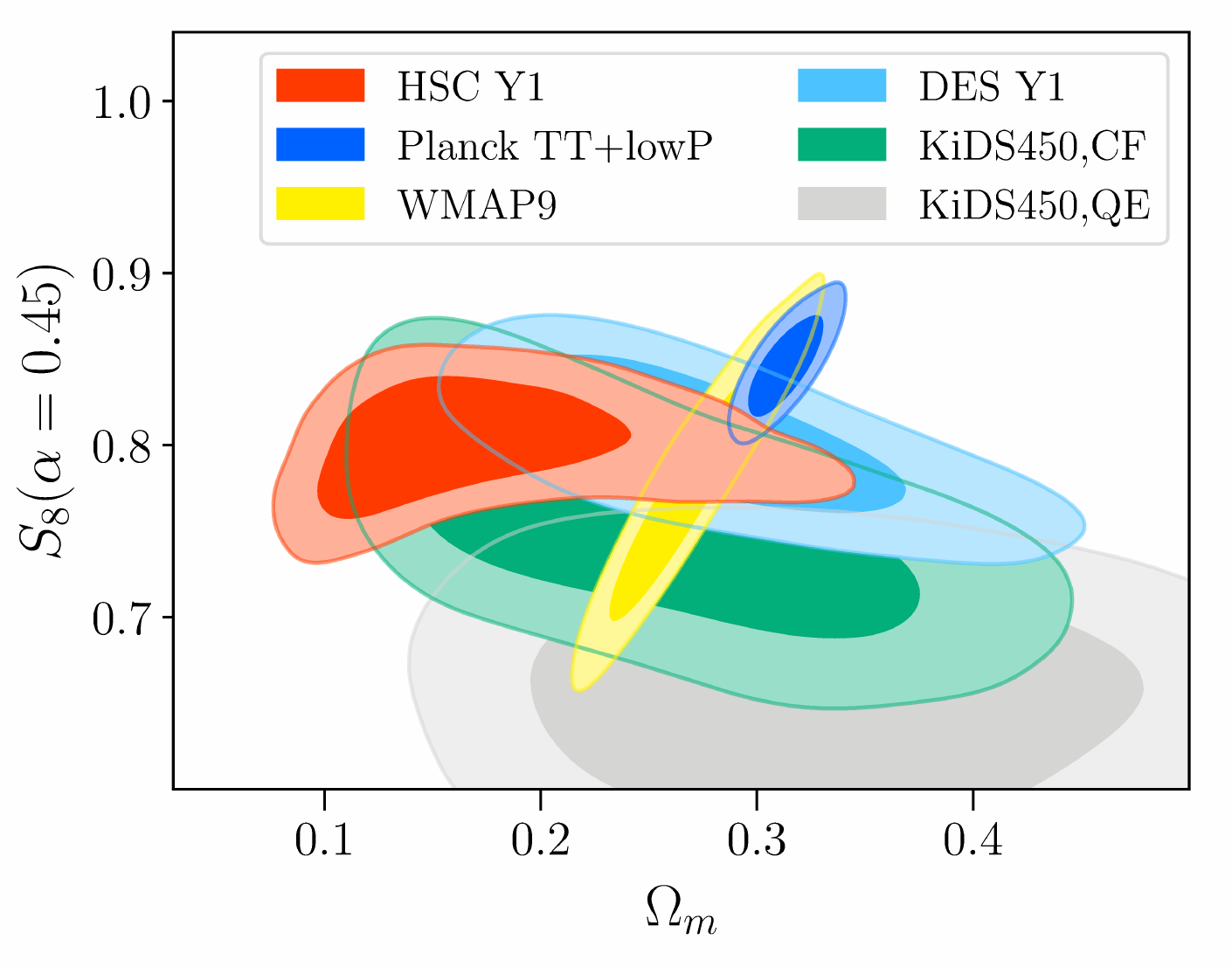}
\end{center}
\caption{Marginalized posterior contours in the $\Omega_{\rm
    m}$-$\sigma_8$ plane ({\it left}) and in the $\Omega_{\rm
    m}$-$S_8(\alpha=0.45)$ plane ({\it right}), where $S_8
  (\alpha)\equiv \sigma_8 (\Omega_{\rm m}/0.3)^\alpha$, in the
  fiducial $\Lambda$CDM model. Both 68\% and 95\% credible levels
  are shown.  For comparison, we plot cosmic shear
  results from KiDS-450 with correlation function (CF) estimators
  \citep{Hildebrandt17} and  with quadratic estimators (QE)
  \citep{Kohlinger17} and DES Y1 \citep{Troxel18b} with the same set of
  cosmological parameters and priors as adopted in this paper, as well
  as {\it WMAP9} \citep{Hinshaw13} ({\it yellow}) and {\it Planck}
  2015 CMB constraints without CMB lensing \citep{Planck16} ({\it purple}).}
\label{fig:Omm-sig8_LCDM}
\end{figure*}

\begin{figure*}
\begin{center}
\includegraphics[width=15cm]{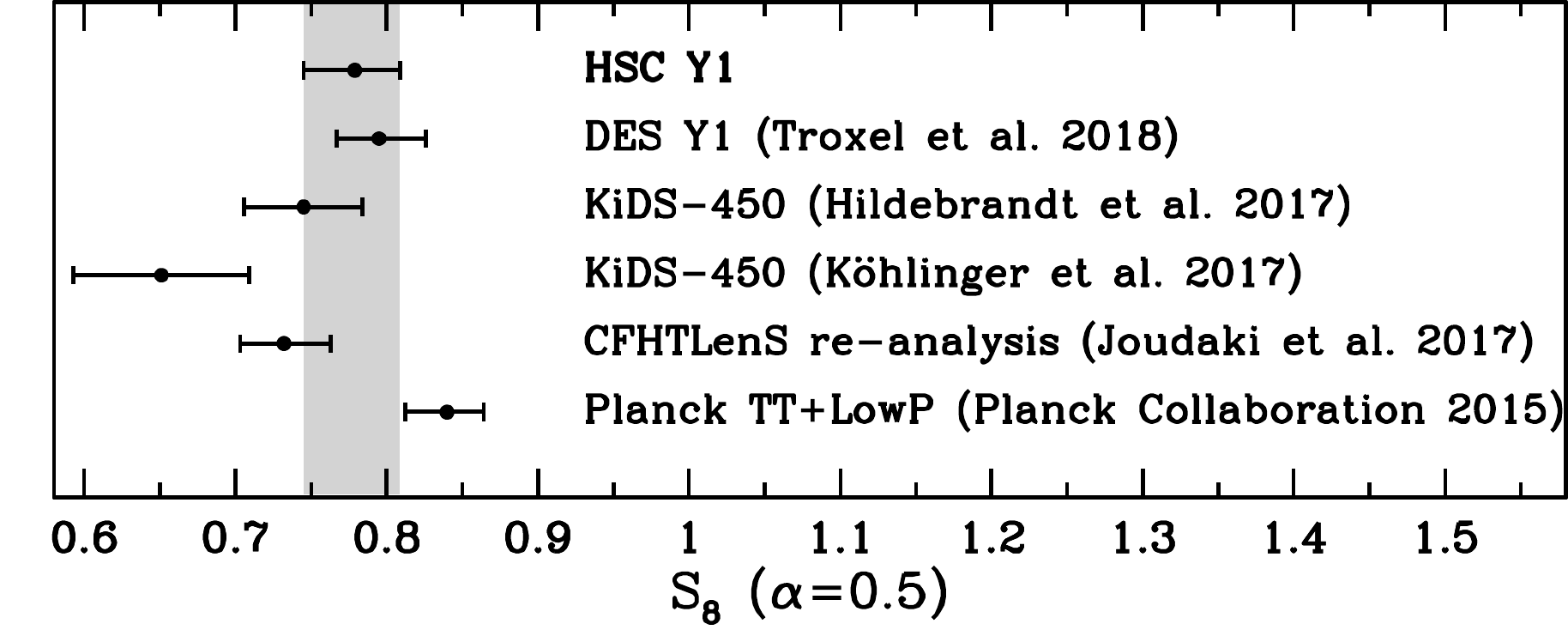}
\end{center}
\caption{The 68\% credible interval on $S_8(\alpha =0.5)$
  from the HSC first-year data in the $\Lambda$CDM model as well as from several
  literature. }
\label{fig:S8_LCDM_ref}
\end{figure*}

\section{Cosmological constraints from tomographic cosmic shear power spectra}
\label{sec:cosmology}
In this section, we present cosmological constraints from tomographic
cosmic shear power spectra measured with the HSC first-year data in
the $\Lambda$CDM model.  We also test the robustness of our
cosmological constraints against various systematics including shape
measurement errors, photo-$z$ uncertainties, intrinsic alignment
modeling, baryon physics, and uncertainty in neutrino mass.  We next
perform internal consistency checks by adopting different ranges of
$\ell$ and $z$ bins. We then extend our fiducial analysis to $w$CDM,
as well as combine our results with the {\it Planck} CMB result to
check the consistency of the results and to constrain
cosmological parameters including neutrino mass, intrinsic alignment
and the baryon feedback parameter.

\subsection{Cosmological constraints in the $\Lambda$CDM model}
\label{sec:cosmo_const}

First, we compare the measured cosmic shear power spectra with our
best-fitting model in Figure~\ref{fig:cl_tomography}.
We compute $\chi^2$ of our best-fitting model as
\begin{eqnarray}
\chi^2&=&\chi^2_{\rm data}+\chi^2_{\rm Gauss}, \\
\chi^2_{\rm data}&=&\sum_{bb'}({\cal C}_b^{\rm obs}-{\cal C}_b^{\rm model})[{\rm Cov}]^{-1}
({\cal C}_{b'}^{\rm obs}-{\cal C}_{b'}^{\rm model}), \\
\chi^2_{\rm Gauss}&=&\sum_j[(p_j^{\rm obs}-\bar{p}_j)/\sigma_j]^2.
\end{eqnarray}
where $\chi^2_{\rm data}$ comes from the tomographic E-mode spectra
and $\chi^2_{\rm Gauss}$ comes from parameters with
Gaussian priors, i.e., $\tilde{\alpha}$, $\tilde{\beta}$, $\Delta m$,
and $\Delta z_i$ ($i=1-4$), and $\bar{p}_j$ and $\sigma_j$ indicate
the mean value and the standard deviation of each prior, respectively
(see Table~\ref{tab:params}). The degree-of-freedom (DOF) is computed
as
\begin{equation}
{\rm DOF}=N_{\rm data} - N_{\rm eff},
\end{equation}
where $N_{\rm data}$ is 60, corresponding to the number of data points
of the tomographic E-mode spectra in the 4 tomographic bins
(Figure~\ref{fig:cl_tomography}), and $N_{\rm eff}$ represents the
effective number of parameters that the data constrain. We compute
$N_{\rm eff}$ as \citep{RaveriHu18}
\begin{equation}
N_{\rm eff}=N_{\rm para}-{\rm tr}[{\cal C}_{\rm prior}^{-1}{\cal C}_{\rm post}],
\end{equation}
where $N_{\rm para}$ is the number of parameters including both
cosmology and nuisance parameters (14 in the fiducial setup),
${\cal C}_{\rm prior}$ is the covariance of prior distributions, and ${\cal
C}_{\rm post}$ is the covariance of posterior distributions. The effective
number of parameters account for parameters that are dominated by the
parameters whose posteriors are driven by data rather than the priors.  We find
that $N_{\rm eff}$ is 3.1, which results in DOF of 56.9. The difference between
$N_{\rm eff}$ and the total number of parameters in our model reflects the fact
that a number of our model parameters are prior-dominated.

We find that our model well reproduces the observed power spectra
quite well.  Our maximum-likelihood case in the fiducial $\Lambda$CDM
model has a minimum $\chi^2$ of 45.4 for 56.9 DOF ($p$-value is 0.86),
which is a very acceptable fit\footnote{Our choice of using $N_{\rm
    eff}$ to compute the degrees of freedom is different from the
  choice of using the total number of parameters made by contemporary
  weak lensing analyses \citep{Troxel18}. Regardless of which
  definition we use, it does not change our conclusion about the
  goodness of fit. For instance, even if we conservatively include all
  parameters without the Gaussian priors to $N_{\rm eff}$, we have 53
  DOF and the resulting $p$-value is 0.76, which is also a very
  acceptable fit.}. Using the covariance assuming {\it Planck}
cosmology, the total signal-to-noise ratio, estimated as
$[\sum_{bb'}{\cal C}_b^{\rm obs}[{\rm Cov}]^{-1}{\cal C}_{b'}^{\rm
    obs}]^{1/2}$, in the four bin tomographic lensing spectra is
$15.6$ in the fiducial range of multipoles. The signal-to-noise ratios
of the cosmic shear auto spectra in individual redshift bins are 4.9,
9.2, 12.3, and 11.5 from the lowest to the highest redshift bins,
respectively. Although the number of source galaxies in the higher
redshift bins is less than in the lower redshift bins, the
signal-to-noise ratios of the measurements are higher due to the
higher amplitudes of the cosmic shear power spectra.

We derive marginalized posterior contours in the
$\Omega_{\rm m}$-$\sigma_8$ plane from our tomographic cosmic shear
power spectrum analysis in the fiducial $\Lambda$CDM
model. Constraints from cosmic shear are known to be degenerate in
the $\Omega_{\rm  m}$-$\sigma_8$ plane. Cosmic shear can tightly
constrain a combination of cosmological parameters $S_8(\alpha)\equiv\sigma_8
(\Omega_{\rm m}/0.3)^\alpha$, which we adopt to quantify cosmological
constraints from the HSC first year data. By carrying out a linear fit of the
logarithm of the posterior samples of $\Omega_{\rm m}$ and $\sigma_8$, we find
that the tightest constraints for $S_8$ are obtained with
$\alpha=0.45$. However, the previous studies by DES \citep{Troxel18} and
KiDS \citep{Hildebrandt17, Kohlinger17} have presented constraints on $S_8$
with $\alpha=0.5$. To present best constraints as well as constraints
that can be directly compared with these previous cosmic shear
results, in this paper we present our results of $S_8$ both for
$\alpha=0.45$ and $\alpha=0.5$.

In Figure~\ref{fig:Omm-sig8_LCDM}, we show our marginalized
constraints in $\Omega_{\rm m}$-$\sigma_8$ and $\Omega_{\rm
  m}$-$S_8(\alpha=0.45)$ planes. As expected, there is no strong
correlation between $\Omega_{\rm m}$ and $S_8$. We find
$S_8(\alpha=0.45)=0.800^{+0.029}_{-0.028}$ and $\Omega_{\rm
  m}=0.162^{+0.086}_{-0.044}$.  Our HSC first-year cosmic shear
analysis places a 3.6\% fractional constraint on $S_8$, which is
comparable to the results of DES \citep{Troxel18} and KiDS
\citep{Hildebrandt17}. For comparison, we find a slightly degraded
constraint on $S_8(\alpha=0.5)=0.780^{+0.030}_{-0.033}$ for
$\alpha=0.5$. We compare our constraints in the $\Omega_{\rm
  m}$-$\sigma_8$ and $\Omega_{\rm m}$-$S_8(\alpha=0.5)$ planes with
cosmic shear results from DES Y1 \citep{Troxel18b} and also from
KiDS-450 with two different methods, correlation functions
\citep[CF;][]{Hildebrandt17} and quadratic estimators
\citep[QE;][]{Kohlinger17}. Note that the plotted results from DES Y1
use the same set of cosmological parameters and priors as adopted in
this paper, and are different from the fiducial constraints in
\citet{Troxel18b}. For the KiDS results, we show the same constraints
as shown in the literature but not corrected for the noise covariance
\citep{Troxel18b}. We also note that there are also some
  differences in the choice of parameters and priors. KiDS adopt wider
  priors on $\Omega_ch^2$ and $n_s$, while $h$ prior is narrower and
  non-flat (instead KiDS adopt a flat prior on the approximated sound
  horizon scale $\theta_{\rm MC}$). The impact of the difference
  between $\theta_{\rm MC}$ and $h$ is found to be small
  \citep{Hildebrandt17,Troxel18b}. The choice of priors and their
  ranges can impact especially on the degeneracy direction between
  $\Omega_m$ and $\sigma_8$ \citep[e.g.,][]{Chang19}.
Figure~\ref{fig:S8_LCDM_ref} compares the values of $S_8 (\alpha=0.5)$
and their 1-$\sigma$ errors among recent cosmic shear studies. We find
that there is no significant difference between the $S_8$ values
obtained by these independent studies. Our result for $S_8$ is smaller
than the DES results by 0.5$\sigma$ and larger than the KiDS results
by 1$\sigma$ or 2$\sigma$ for the CF and QE estimators. The difference
becomes smaller when the shape noise covariance is corrected in the
KiDS results \citep{Troxel18b}.

Our best-fit $\Omega_{\rm m}$ value is smaller
than that obtained by other lensing surveys. More specifically, the
best-fit $\Omega_{\rm m}$ values from DES and KiDS are 1.3$\sigma$
and 1$\sigma$ larger than our best-fit value, respectively. 
We also compare these cosmic shear constraints with {\it Planck} CMB constraints
\citep{Planck16}. We find that {\it Planck} best-fit values of $S_8$
and $\Omega_{\rm m}$ are 2$\sigma$ and 1.8$\sigma$ higher than our
best-fit values, where we take into account of the uncertainties in
both measurements.  We check the consistency between our result and
{\it Planck} CDM constraints more quantitatively in
Section~\ref{sec:consistency}. Our result prefers slightly lower $S_8$
and $\Omega_{\rm m}$ values than the {\it Planck} CMB results, which is
qualitatively consistent with other cosmic shear results from DES and
KiDS, as discussed in the recent {\it Planck} 2018 paper
\citep{Planck18_cosmology}.

As discussed above, our result shows a small best-fit value of
$\Omega_{\rm m}$, as well as a relatively tight constraint on
$\Omega_{\rm m}$.  While tomographic analysis has been shown to
improve measurements of $\Omega_{\rm m}$ and $\sigma_8$
\citep[e.g.,][]{Simon2004}, we check whether the small errorbar is
expected given our sample size and tomographic bin definition.  For
this purpose, we again use the HSC mock shear catalogs, which were
also used to check the validity and accuracy of the pseudo-$C_\ell$
method (see Section~\ref{subsec:pseudocl}).  We perform nested
sampling analysis on the 100 mock catalogs adopting the same set of
cosmological and nuisance parameters as in the analysis of the real
HSC data, and for each mock we derive both the best-fit value and
1$\sigma$ error of $\Omega_{\rm m}$. We find that the best-fit value
and the 1$\sigma$ error of $\Omega_{\rm m}$ are positively
correlated. Our result indicates that 4 out of 100 mock realizations
have both smaller best-fit values and tighter constraints on
$\Omega_{\rm m}$ than the observed values. Although this mock catalog
analysis is based on a specific cosmological model ({\it WMAP9}), it
suggests that the observed small best-fit value and the tight
constraint on $\Omega_{\rm m}$ can be explained by a statistical
fluctuation at the 2$\sigma$ level, and therefore they do not
necessarily indicate the presence of systematic errors (see also
Section~\ref{subsec:robustness}).

\begin{table*}
\caption{Summary of the median and 68\% range of uncertainties of
  $S_8(\alpha)$ with $\alpha=0.45$ and $0.5$ and $\Omega_{\rm m}$, as
  well as their robustness against various systematics, in the
  $\Lambda$CDM model from our cosmic shear measurements.$^*$
  Constraints assuming the $w$CDM cosmology are also shown in the bottom row of the table.}
\begin{center}
  \begin{tabular}{lccc}
    \hline
setup & $S_8 (\alpha=0.45)$ & $S_8 (\alpha=0.5)$ & $\Omega_{\rm m}$ \\
\hline
Fiducial                                 & $0.800_{-0.028}^{+0.029}$ & $0.780_{-0.033}^{+0.030}$ & $0.162_{-0.044}^{+0.086}$  \\
w/o shape error ($\tilde{\alpha}=\tilde{\beta}=\Delta m=0$)                        & $0.801_{-0.028}^{+0.028}$ & $0.783_{-0.030}^{+0.028}$ & $0.177_{-0.055}^{+0.098}$  \\
w/o photo-$z$ error ($\Delta z_i=0$)                     & $0.804_{-0.028}^{+0.027}$ & $0.784_{-0.033}^{+0.029}$ & $0.162_{-0.046}^{+0.085}$  \\
Ephor AB, stacked                        & $0.818_{-0.029}^{+0.029}$ & $0.799_{-0.032}^{+0.029}$ & $0.172_{-0.049}^{+0.083}$  \\
MLZ, stacked                             & $0.813_{-0.028}^{+0.028}$ & $0.801_{-0.028}^{+0.027}$ & $0.221_{-0.073}^{+0.098}$  \\
Mizuki, stacked                          & $0.807_{-0.027}^{+0.027}$ & $0.791_{-0.027}^{+0.027}$ & $0.195_{-0.055}^{+0.085}$  \\
NNPZ, stacked                            & $0.818_{-0.029}^{+0.029}$ & $0.807_{-0.027}^{+0.027}$ & $0.231_{-0.076}^{+0.112}$  \\
Frankenz, stacked                        & $0.809_{-0.030}^{+0.030}$ & $0.789_{-0.037}^{+0.032}$ & $0.164_{-0.048}^{+0.090}$  \\
DEmP, stacked                            & $0.812_{-0.029}^{+0.028}$ & $0.791_{-0.034}^{+0.029}$ & $0.163_{-0.044}^{+0.076}$  \\
w/o IA ($A_{\rm IA}=0$)                                  & $0.787_{-0.028}^{+0.027}$ & $0.767_{-0.031}^{+0.028}$ & $0.172_{-0.051}^{+0.079}$  \\
$\eta_{\rm eff}$ fixed to be $3$         & $0.800_{-0.032}^{+0.031}$ & $0.776_{-0.038}^{+0.034}$ & $0.152_{-0.041}^{+0.062}$  \\
$\sum m_\nu$ fixed to be 0.06~eV         & $0.797_{-0.029}^{+0.029}$ & $0.777_{-0.034}^{+0.031}$ & $0.166_{-0.047}^{+0.088}$  \\
$\sum m_\nu$ varied                      & $0.785_{-0.031}^{+0.029}$ & $0.765_{-0.038}^{+0.031}$ & $0.175_{-0.049}^{+0.090}$  \\
$A_B$ varied                             & $0.797_{-0.041}^{+0.039}$ & $0.775_{-0.048}^{+0.047}$ & $0.157_{-0.048}^{+0.092}$  \\
AGN feedback model ($A_{\rm B}=1$)                       & $0.818_{-0.029}^{+0.029}$ & $0.804_{-0.031}^{+0.030}$ & $0.201_{-0.064}^{+0.104}$  \\
w/o lowest $z$-bin                       & $0.799_{-0.032}^{+0.032}$ & $0.779_{-0.034}^{+0.033}$ & $0.165_{-0.049}^{+0.097}$  \\
w/o mid-low $z$-bin                      & $0.799_{-0.030}^{+0.030}$ & $0.795_{-0.028}^{+0.028}$ & $0.283_{-0.109}^{+0.109}$  \\
w/o mid-high $z$-bin                     & $0.791_{-0.035}^{+0.033}$ & $0.770_{-0.037}^{+0.034}$ & $0.156_{-0.046}^{+0.107}$  \\
w/o highest $z$-bin                      & $0.797_{-0.034}^{+0.034}$ & $0.780_{-0.033}^{+0.033}$ & $0.186_{-0.058}^{+0.099}$  \\
$\ell_{\rm max}$ extended to 3500        & $0.801_{-0.026}^{+0.027}$ & $0.780_{-0.030}^{+0.028}$ & $0.166_{-0.042}^{+0.060}$  \\
Lower half $\ell$-bin                    & $0.799_{-0.038}^{+0.038}$ & $0.783_{-0.039}^{+0.037}$ & $0.182_{-0.065}^{+0.138}$  \\
Higher half $\ell$-bin                   & $0.799_{-0.031}^{+0.033}$ & $0.789_{-0.029}^{+0.030}$ & $0.235_{-0.079}^{+0.107}$  \\
fixed covariance with best-fit cosmology & $0.806_{-0.031}^{+0.030}$ & $0.785_{-0.033}^{+0.032}$ & $0.173_{-0.049}^{+0.088}$  \\
\hline
$w$CDM                                   & $0.773_{-0.038}^{+0.043}$ & $0.754_{-0.049}^{+0.044}$ & $0.163_{-0.047}^{+0.079}$ \\
    \hline
  \end{tabular}
\end{center}
\begin{tabnote}
  $^*$The values of $S_8 (\alpha=0.45)$ are also illustrated in
  Figure~\ref{fig:S8_LCDM}.
\end{tabnote}
  \label{tab:S8_sys}
\end{table*}

\begin{figure*}
\begin{center}
\includegraphics[width=14cm]{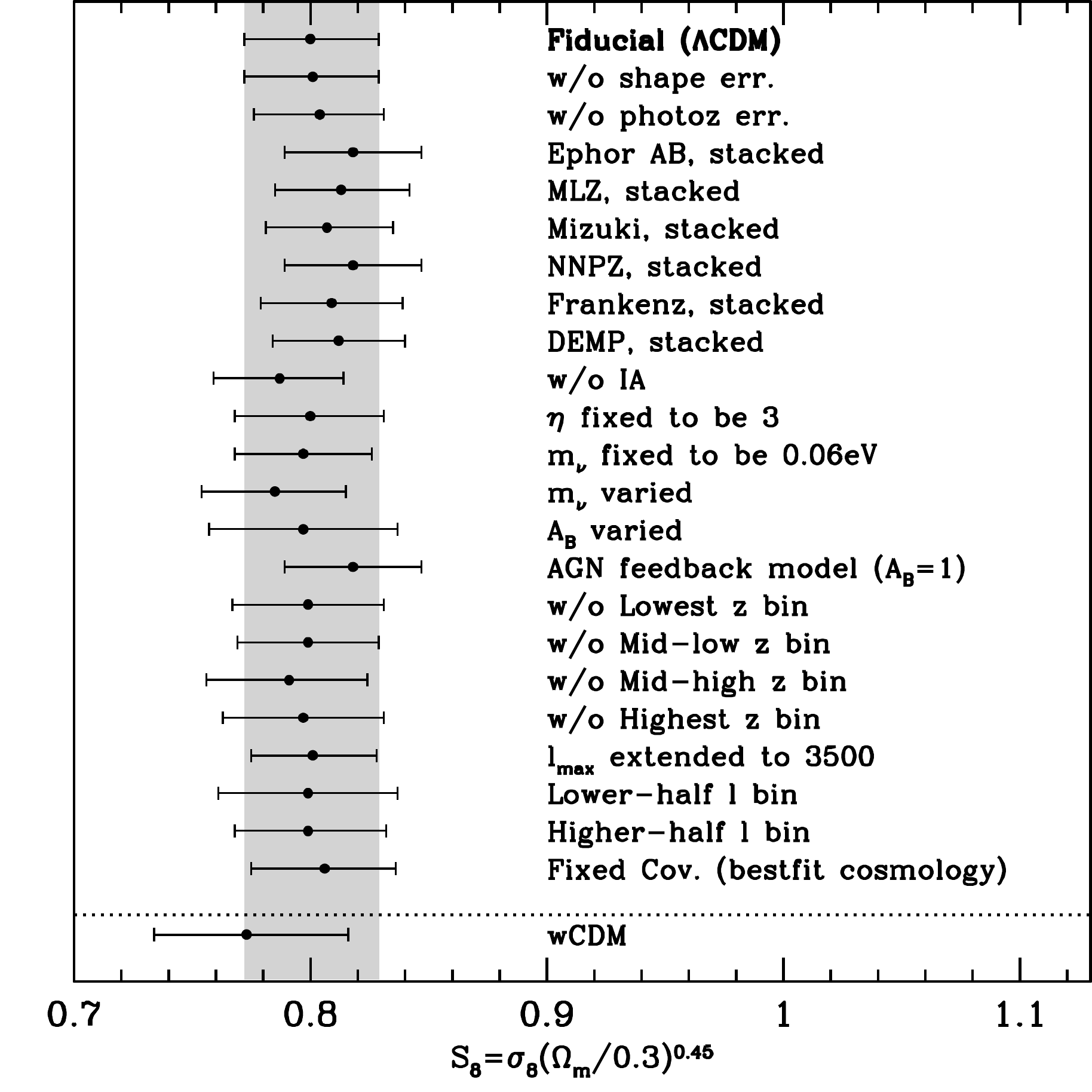}
\end{center}
\caption{Constraints on $S_8(\alpha)\equiv \sigma_8(\Omega_{\rm
    m}/0.3)^{\alpha}$ with $\alpha=0.45$ and their robustness against
  various systematics and modeling choices in the $\Lambda$CDM model.
  The shaded area
  shows the 68\% credible interval for the $S_8$ value for $\alpha=0.45$ in
  our fiducial case. We consider the effects of shape and photo-$z$
  uncertainties, impacts of assumptions on IA modeling, baryonic
  feedback modeling, varying neutrino mass, and different ranges of
  $\ell$ and $z$ bins (see Section~\ref{subsec:robustness} for more
  details).  Constraints based on the $w$CDM model are also shown at the
  bottom.  We find that the systematic differences in the $S_8$ values are
  well within the 1$\sigma$ statistical error, indicating that our
  fiducial constraint on $S_8$ is robust against these systematics.}
\label{fig:S8_LCDM}
\end{figure*}

\begin{figure*}
\begin{center}
\includegraphics[width=7cm]{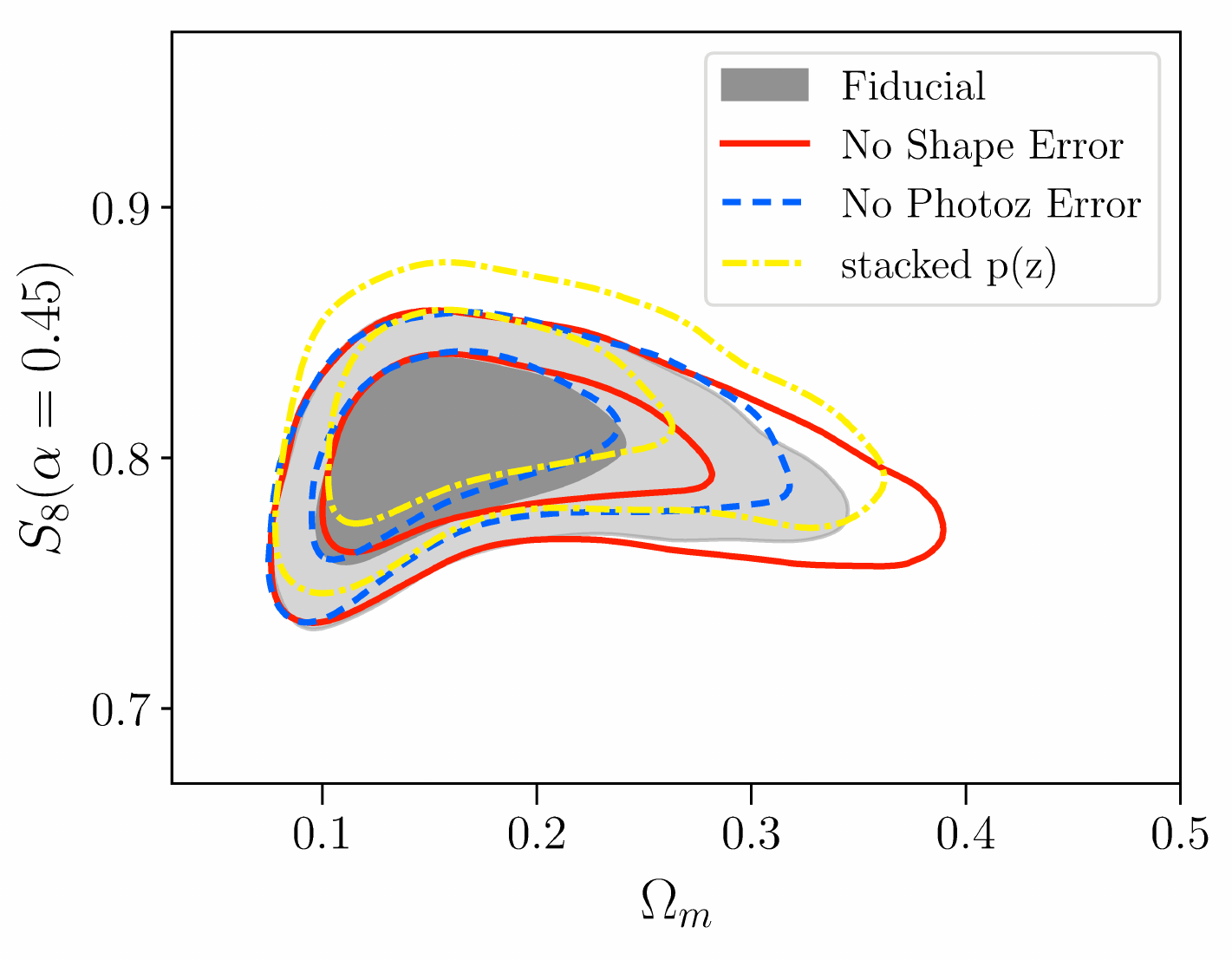}
\includegraphics[width=7cm]{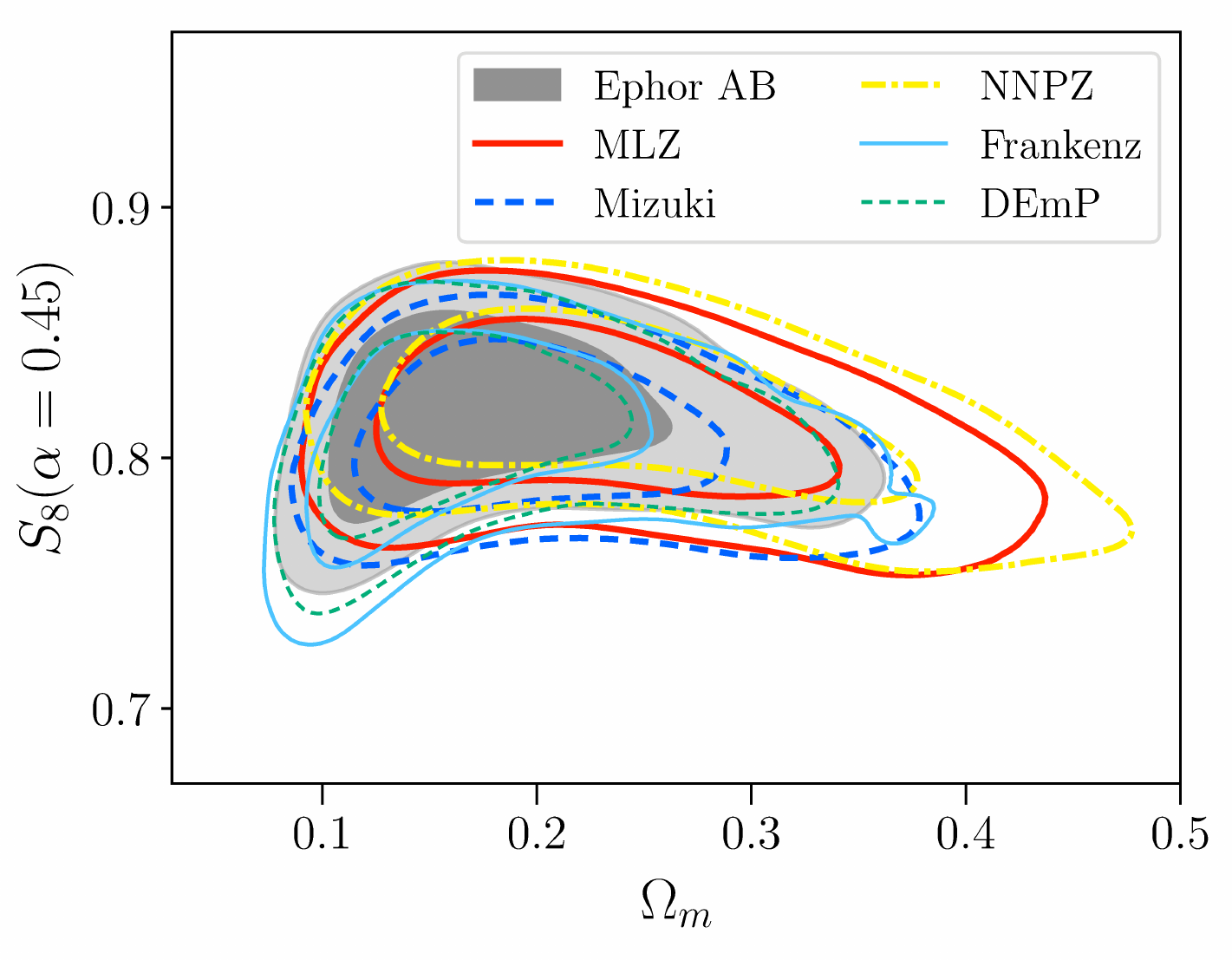}
\includegraphics[width=7cm]{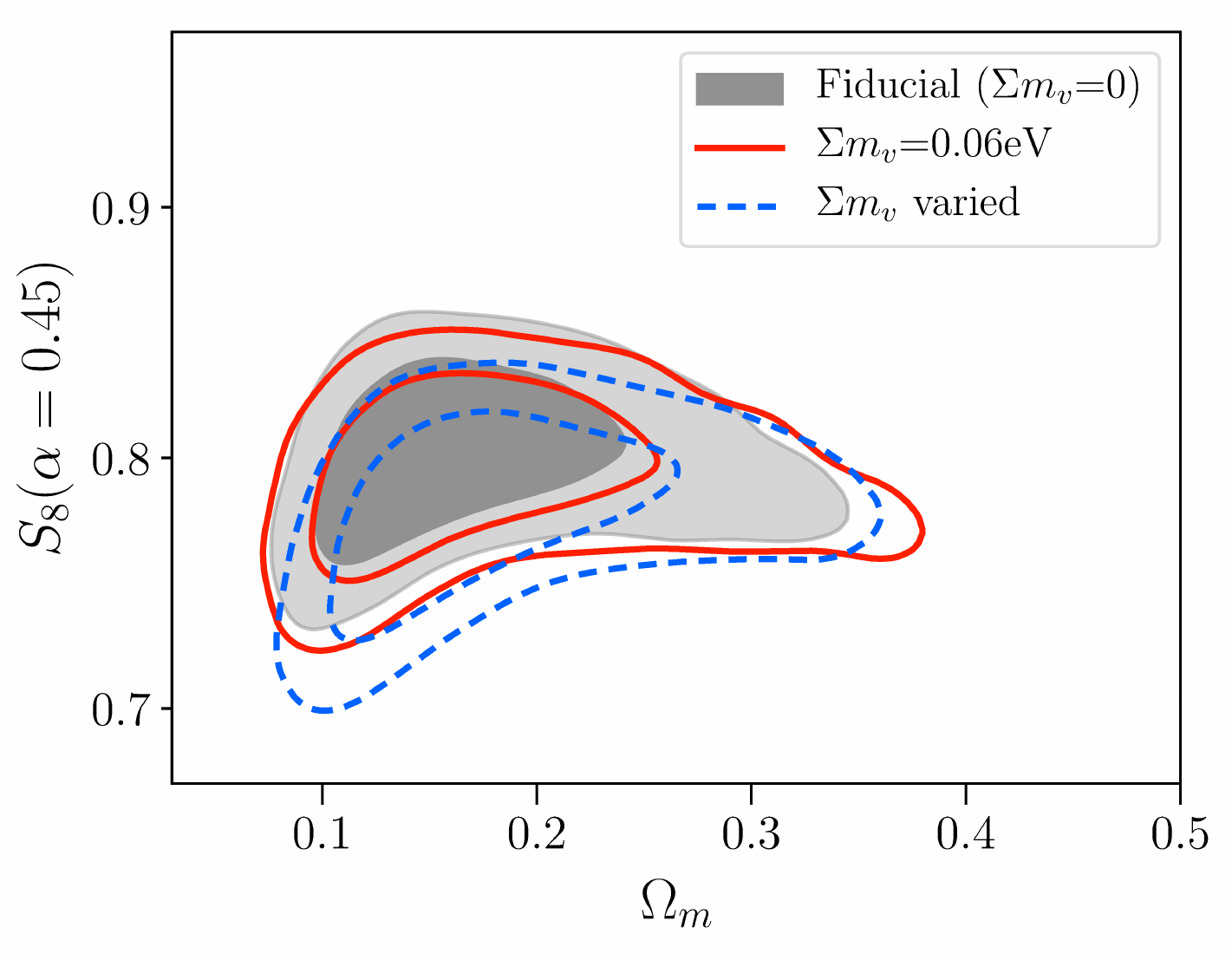}
\includegraphics[width=7cm]{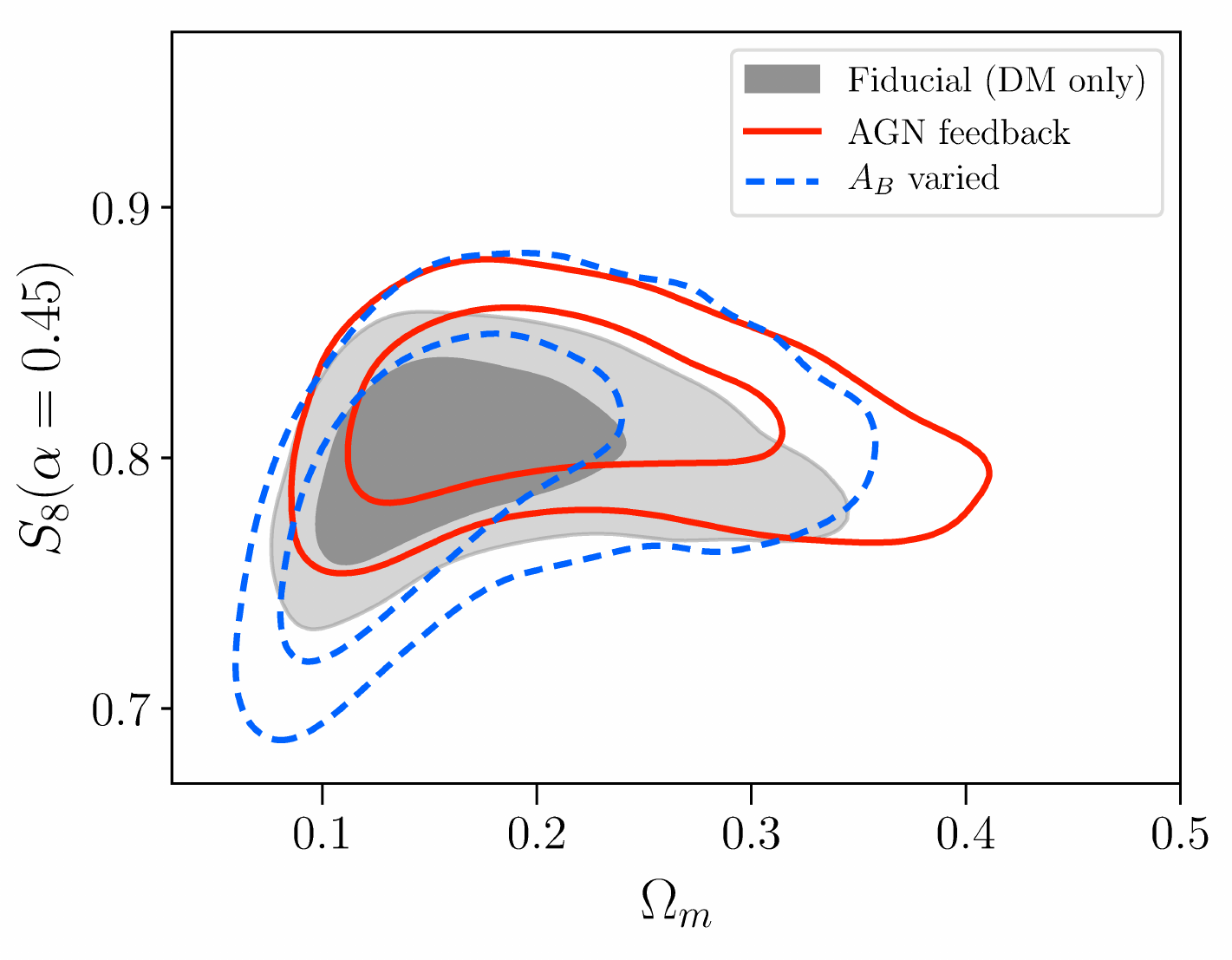}
\includegraphics[width=7cm]{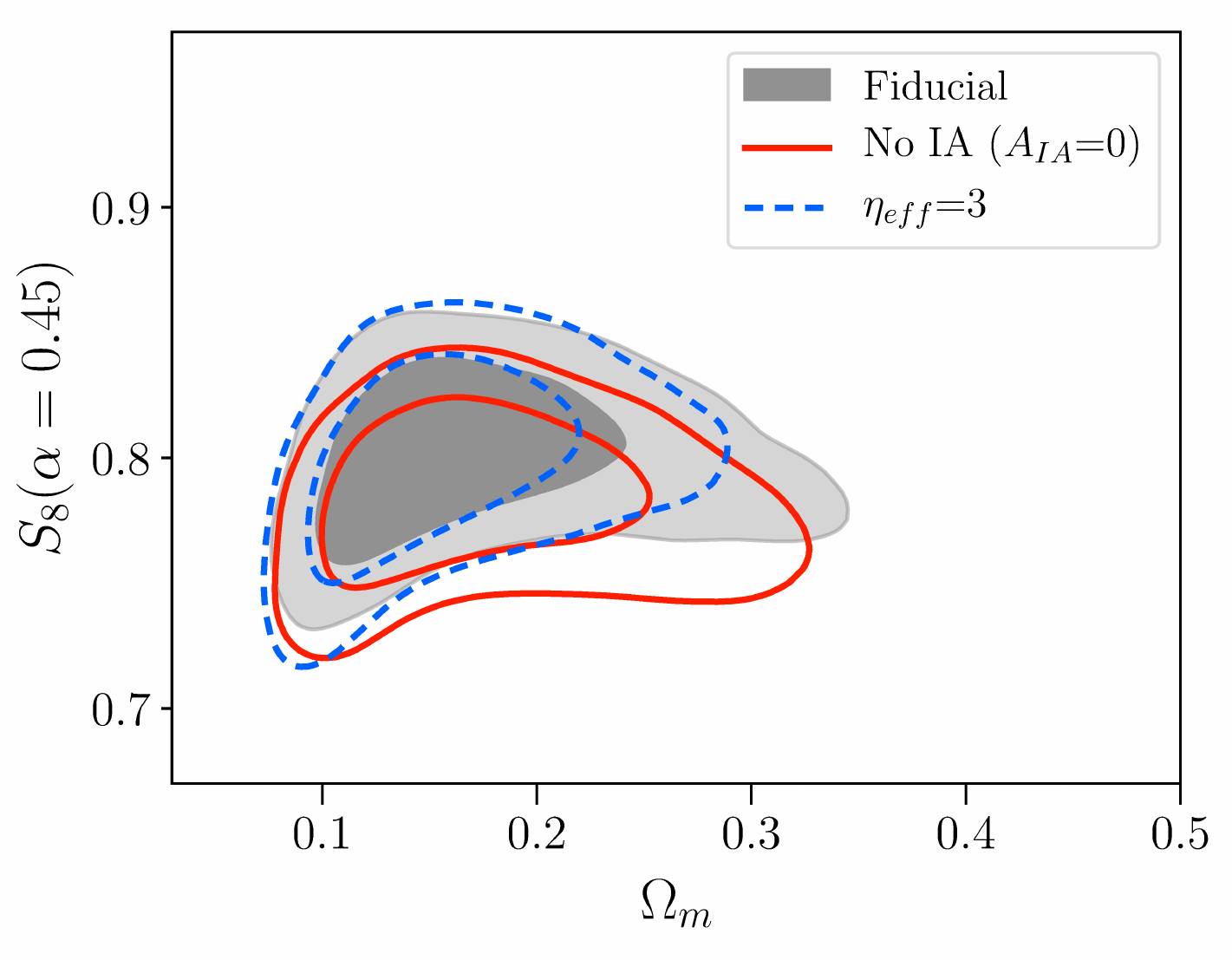}
\includegraphics[width=7cm]{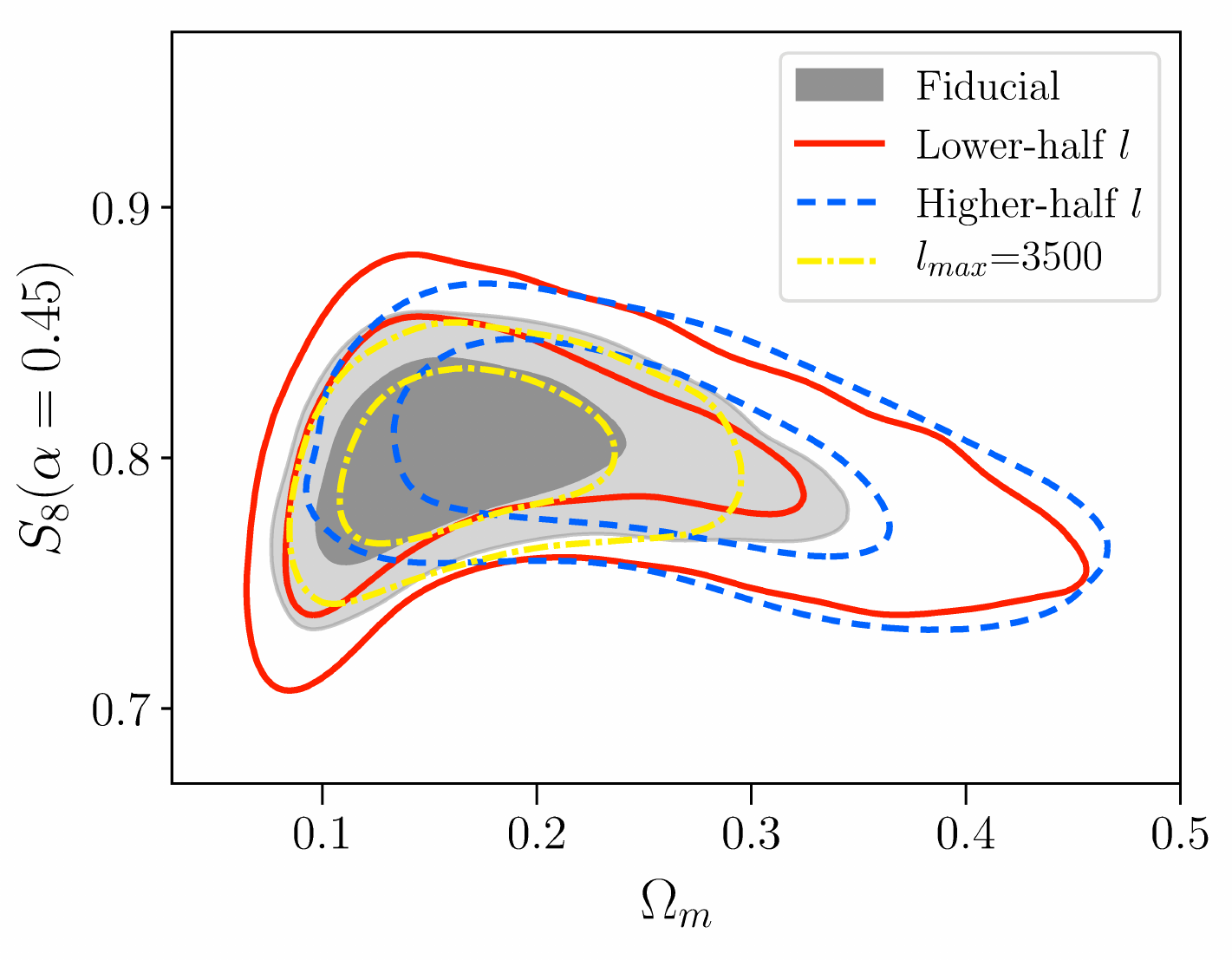}
\includegraphics[width=7cm]{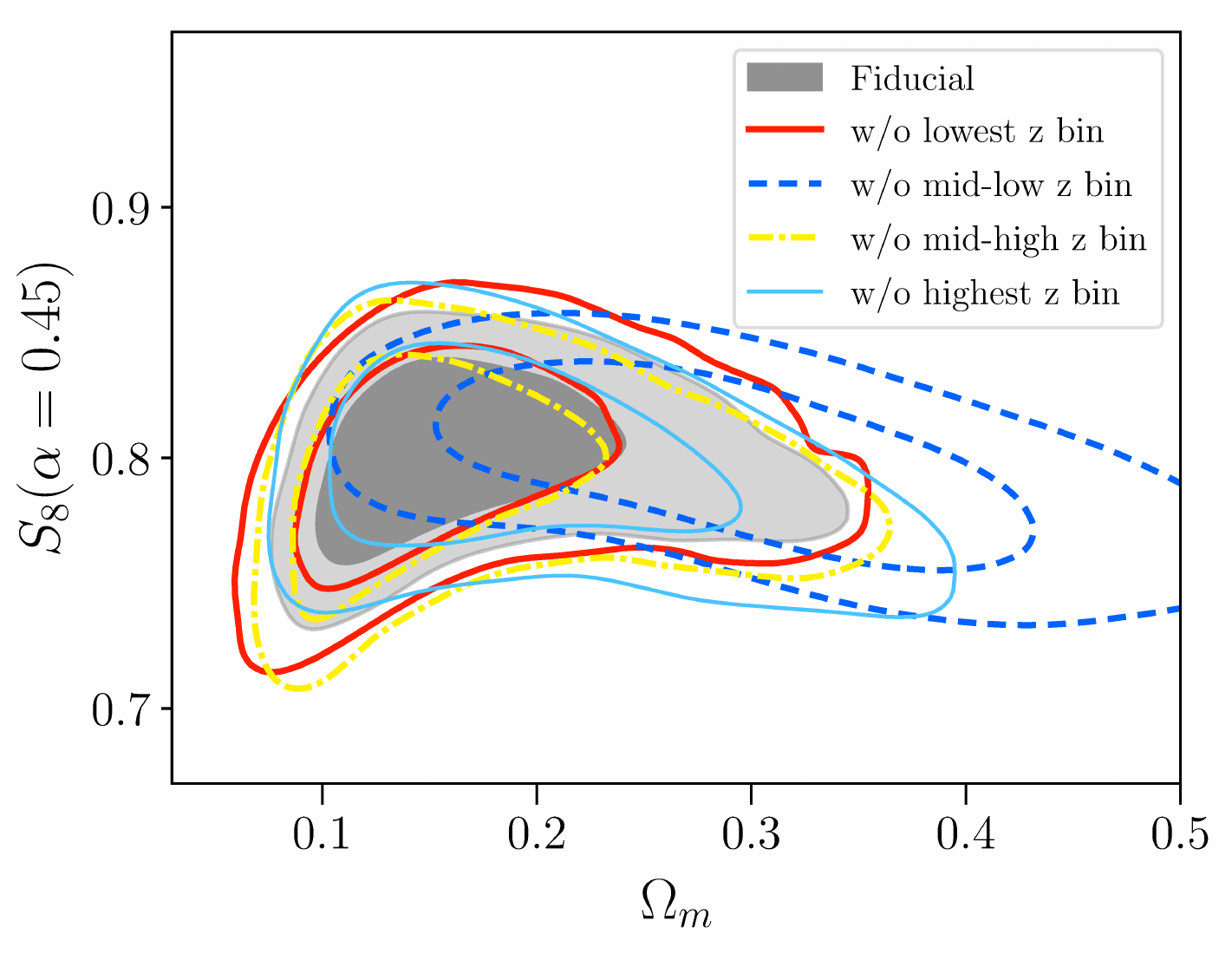}
\includegraphics[width=7cm]{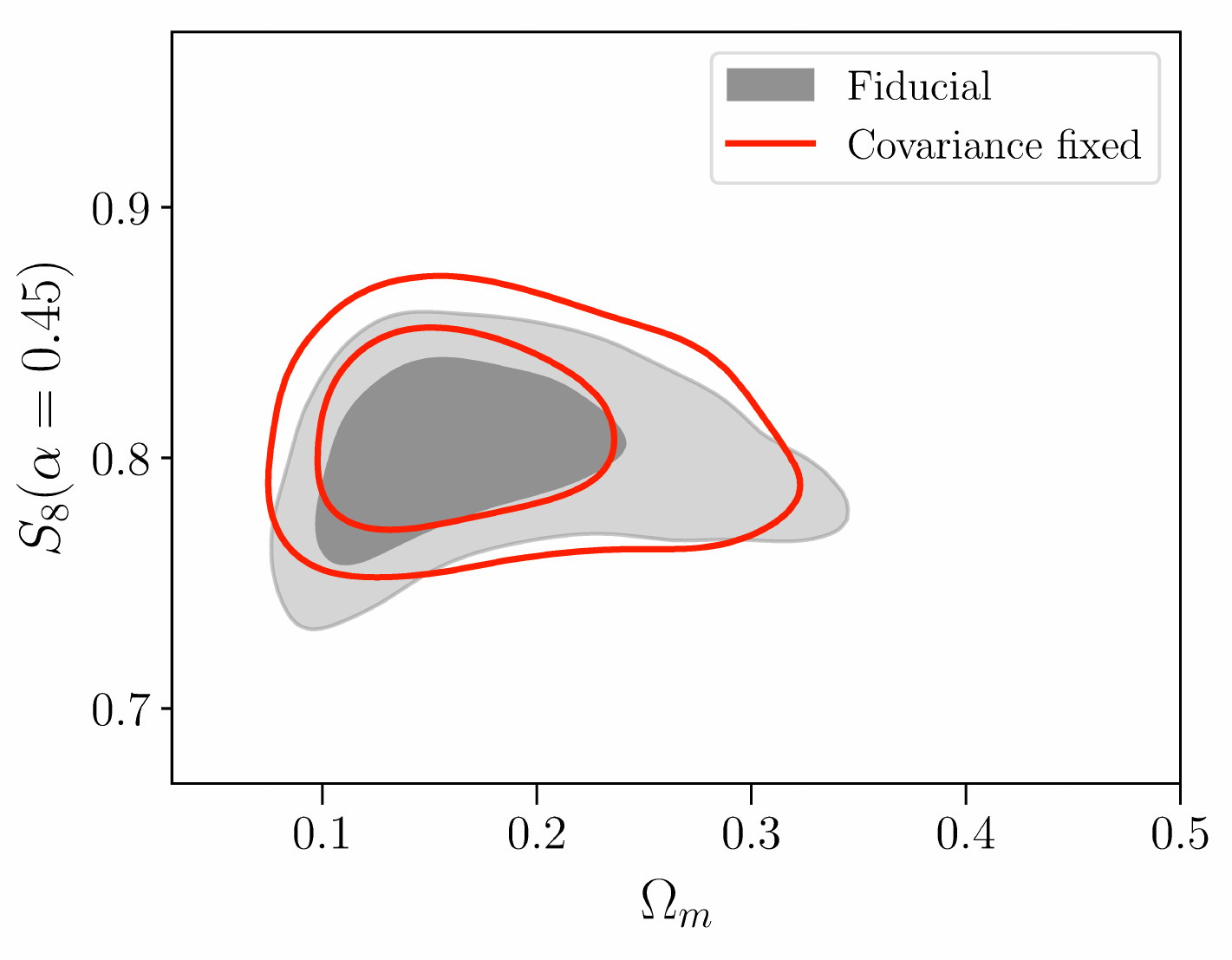}
\end{center}
\caption{The robustness against various systematics in the
  $\Omega_{\rm m}$-$S_8(\alpha=0.45)$ plane in the fiducial
  $\Lambda$CDM model. The contours represent 68\% and 95\%
  credible levels. Different panels show the robustness
  against different systematics and modeling choices (see also Table~\ref{tab:S8_sys}
  and Figure~\ref{fig:S8_LCDM}).}
\label{fig:Omm-S8_sys}
\end{figure*}

\begin{figure*}
\begin{center}
\includegraphics[width=7cm]{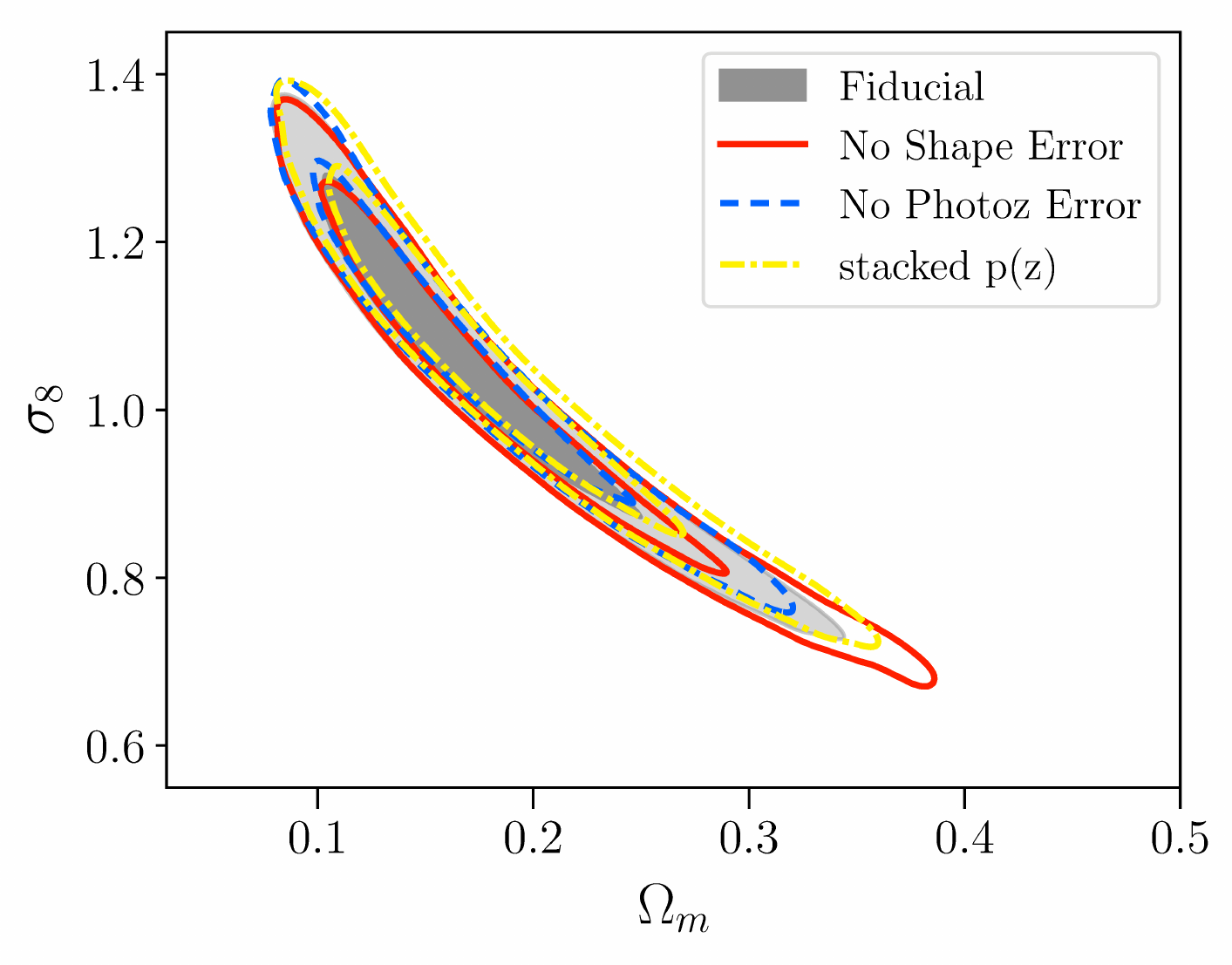}
\includegraphics[width=7cm]{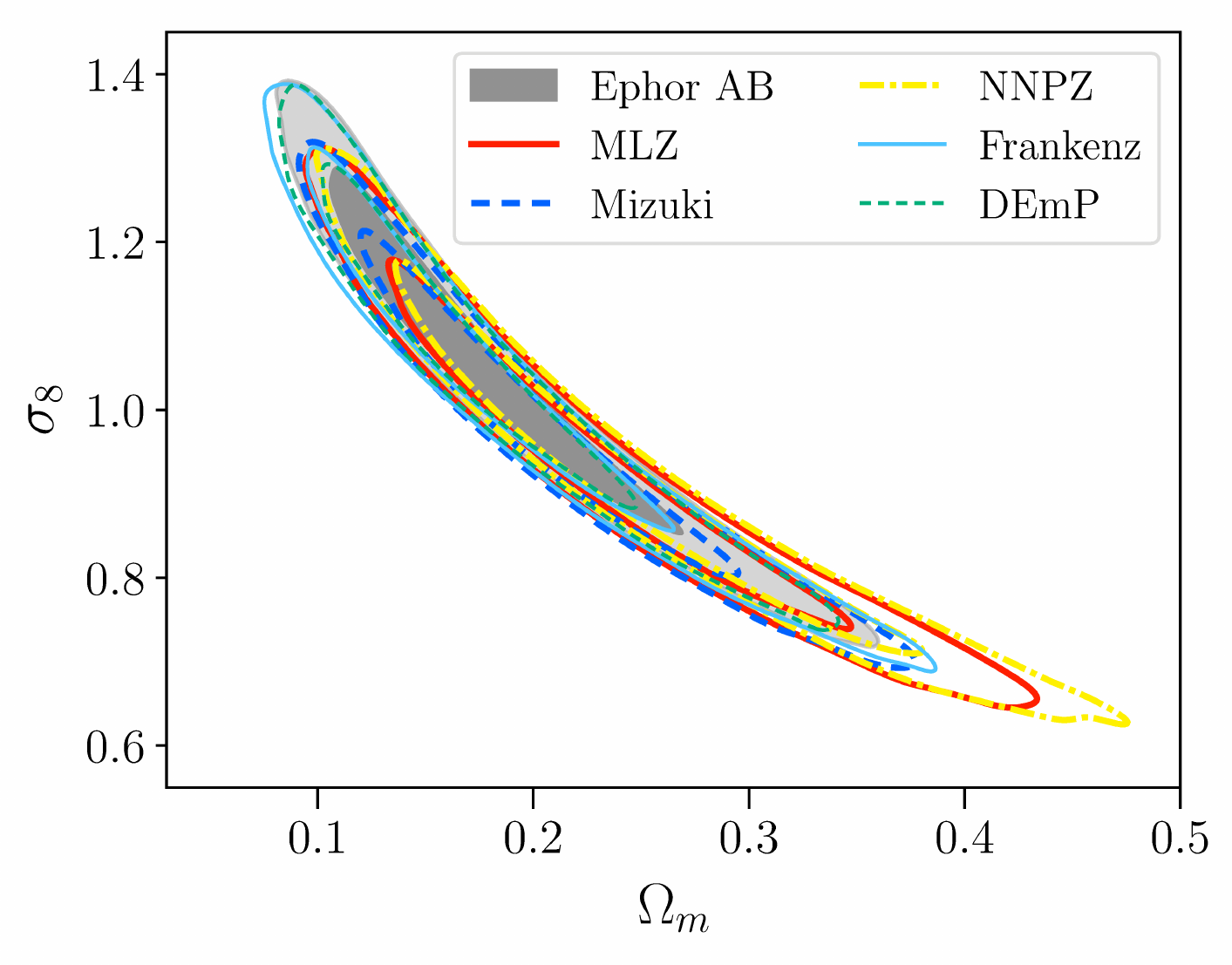}
\includegraphics[width=7cm]{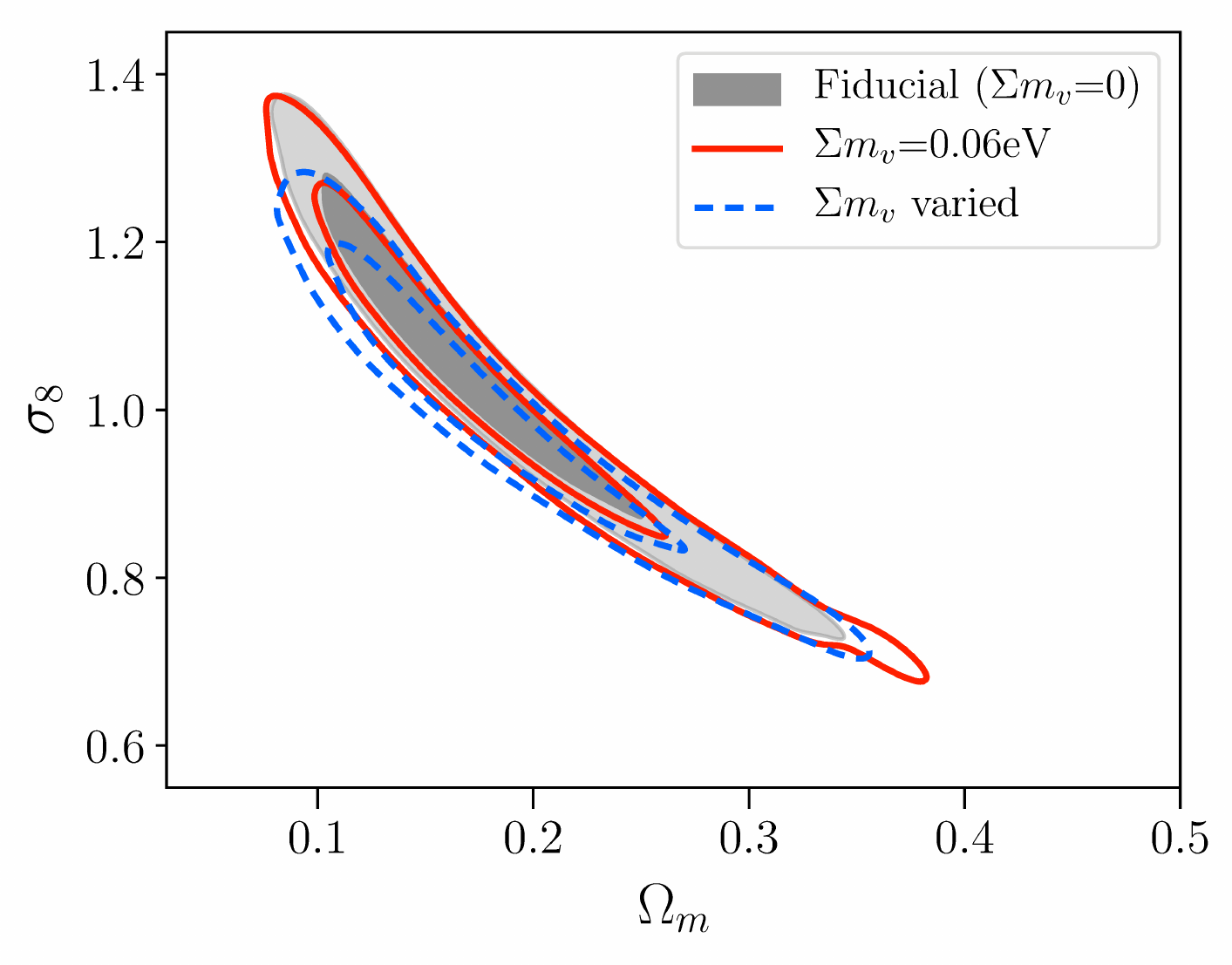}
\includegraphics[width=7cm]{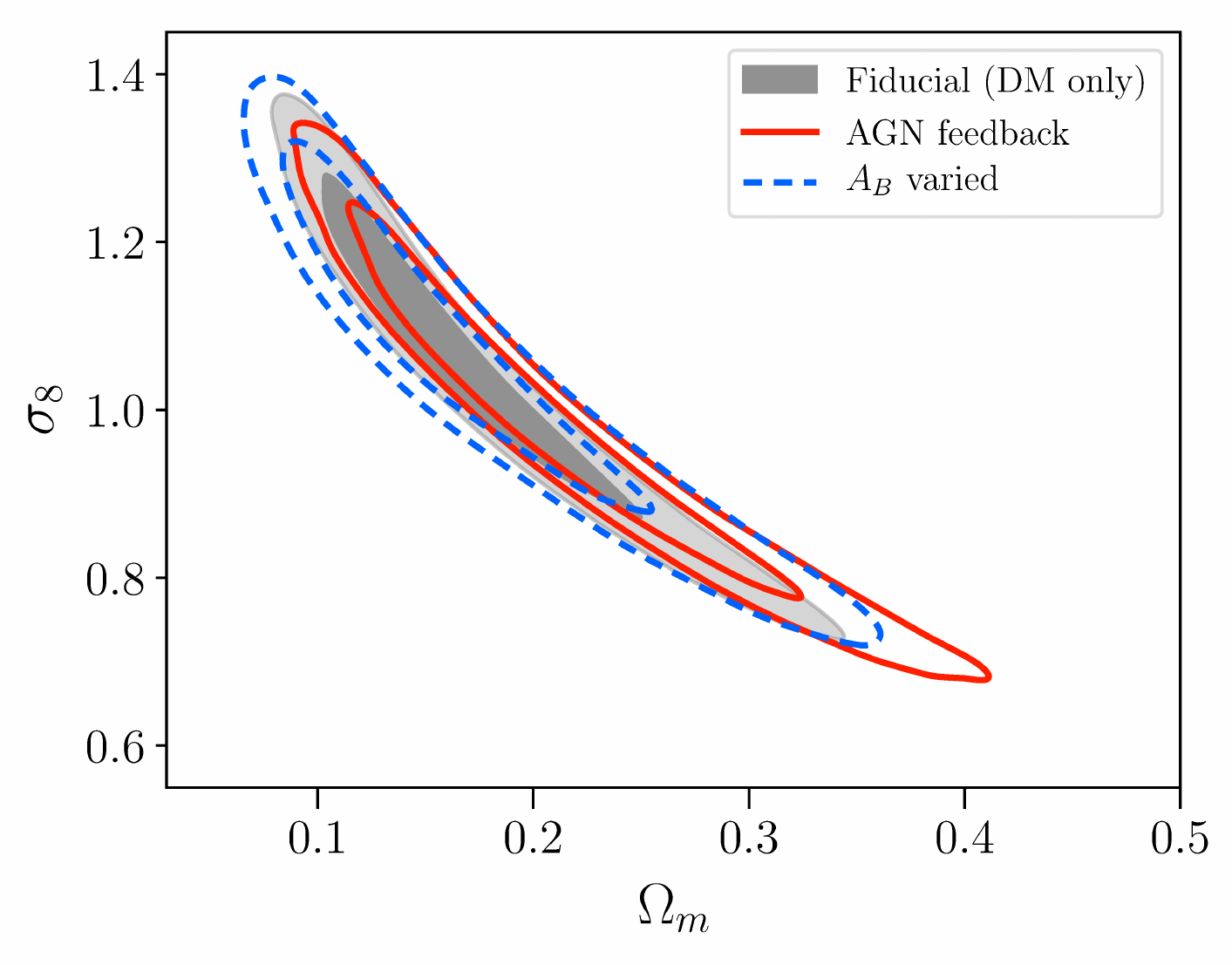}
\includegraphics[width=7cm]{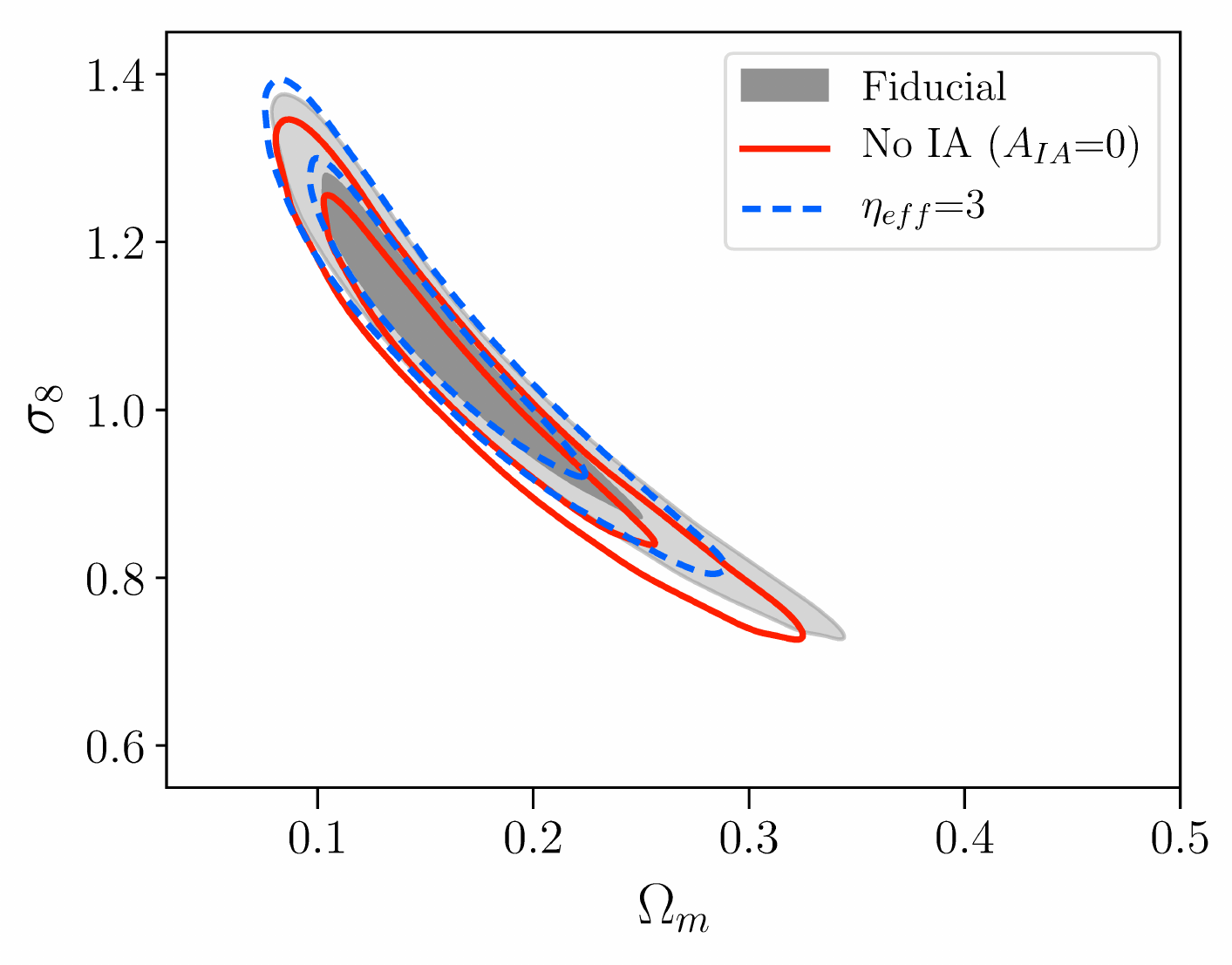}
\includegraphics[width=7cm]{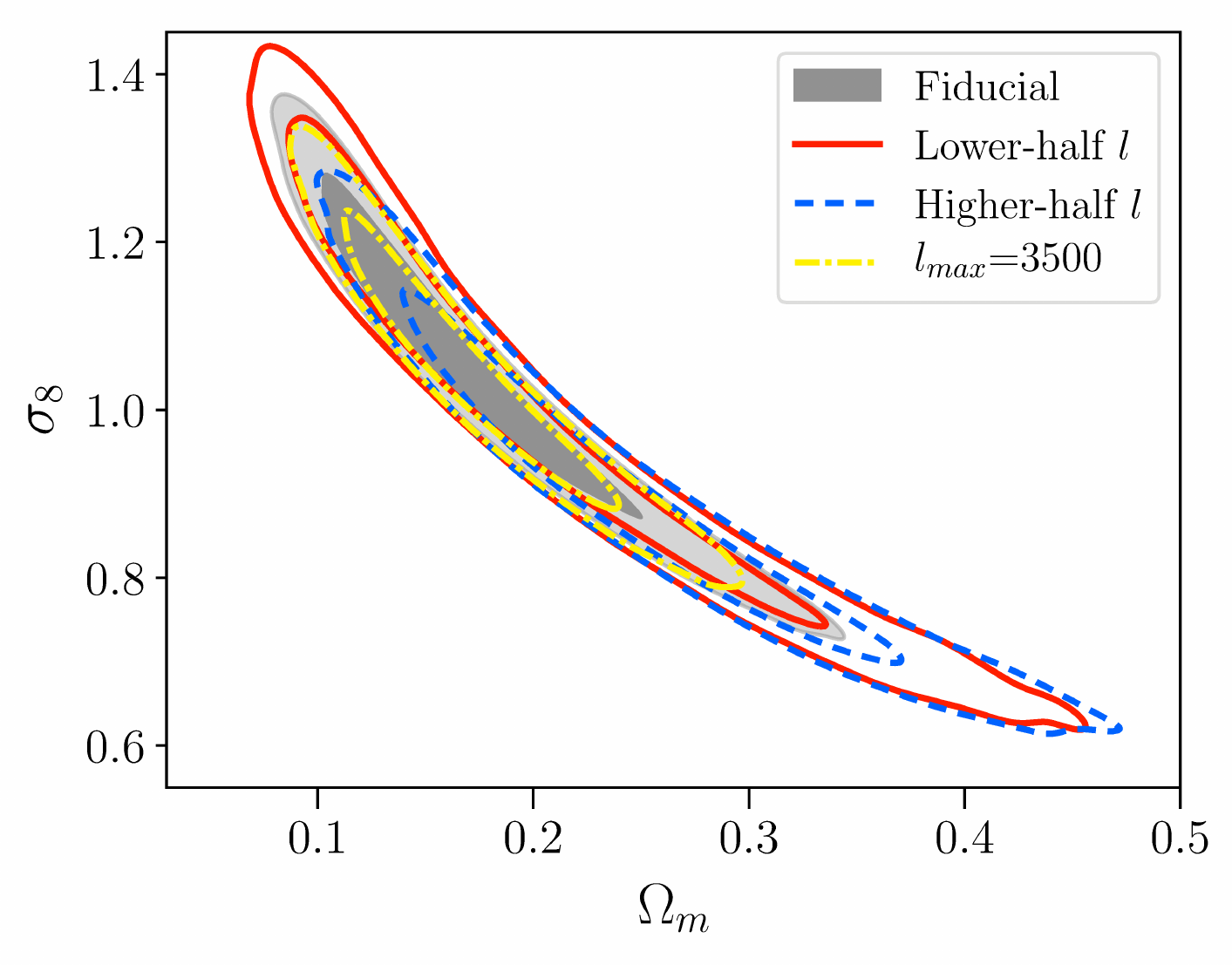}
\includegraphics[width=7cm]{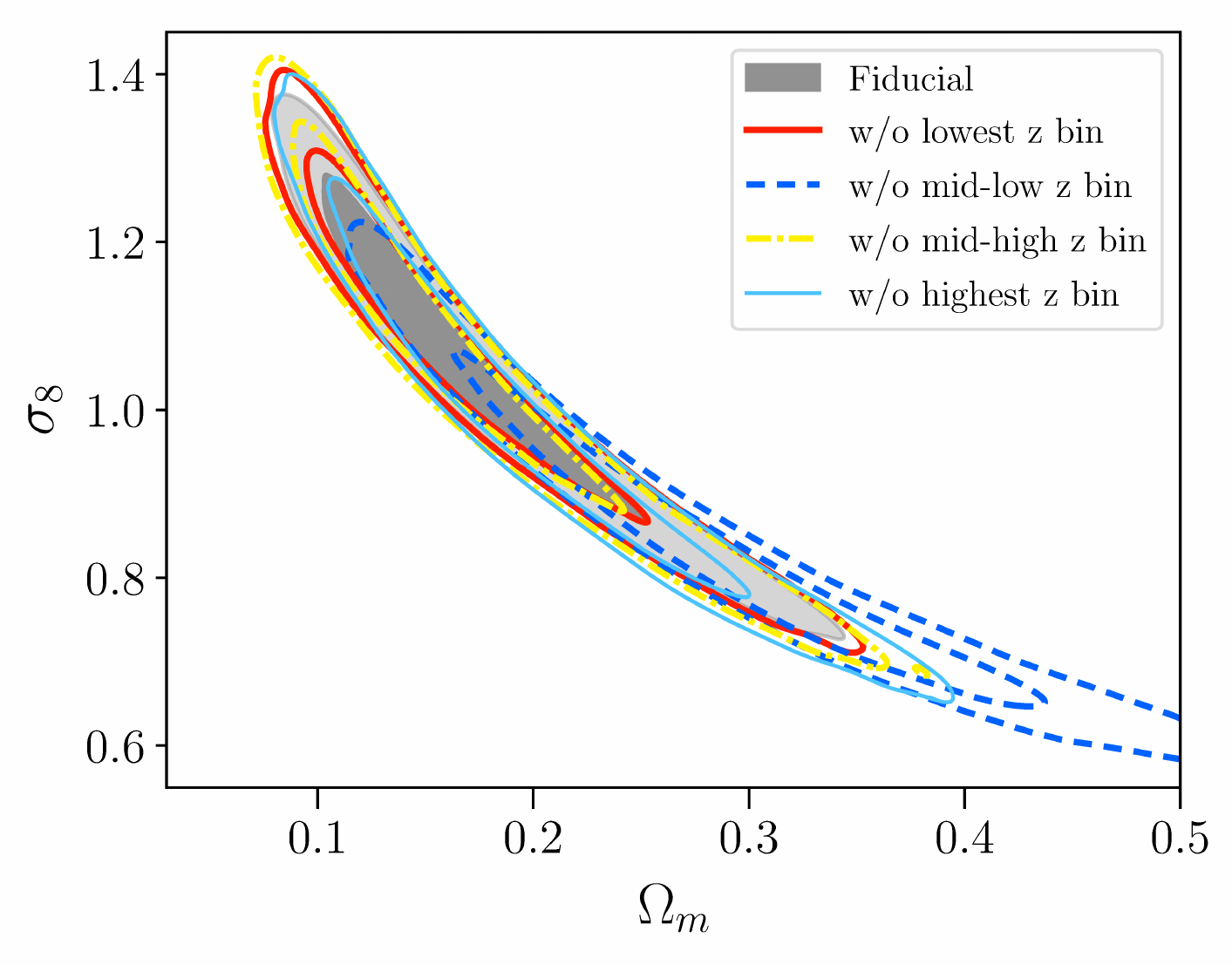}
\includegraphics[width=7cm]{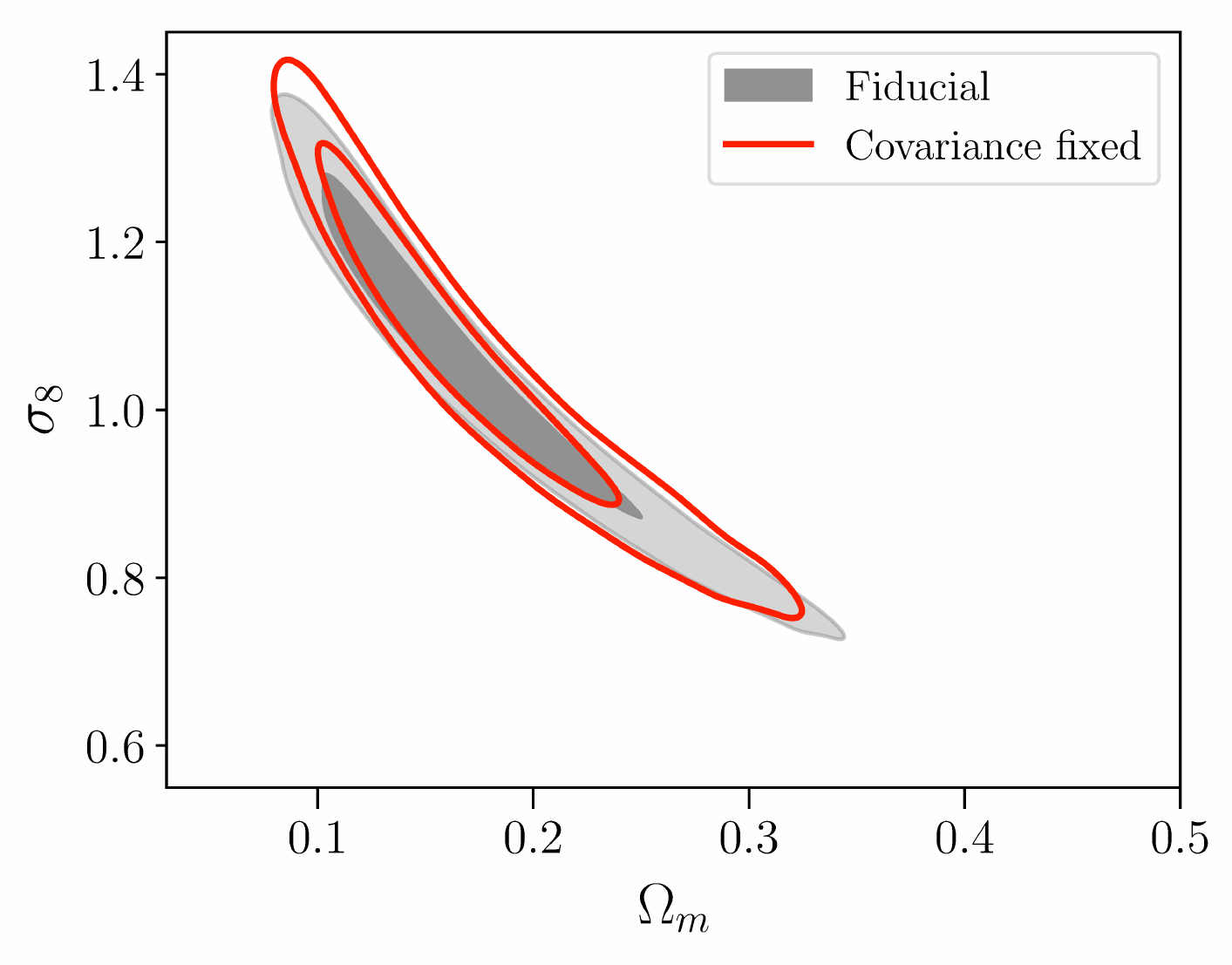}
\end{center}
\caption{Same as Figure~\ref{fig:Omm-S8_sys} but in the $\Omega_{\rm
    m}$-$\sigma_8$ plane.}
\label{fig:Omm-sig8_sys}
\end{figure*}

\subsection{Robustness of our results}
\label{subsec:robustness}
To check the robustness of our results, we change the setup of our
nested sampling analysis in various ways to see the impact on our
cosmological constraints. This includes tests of various systematic
effects such as shear measurement errors, photo-$z$ errors, possible
contamination by IA, a possible modification in modeling of IA,
changes in the matter power spectrum due to baryonic physics and
non-zero neutrino mass, as well as internal consistency checks of our
results by changing the ranges of $\ell$ and $z$ bins over which the
fits are made.

Table~\ref{tab:S8_sys} and Figure~\ref{fig:S8_LCDM} summarize the
results of the robustness check in the $\Lambda$CDM model.  We list
marginalized constraints on $\Omega_{\rm m}$ and $S_8(\alpha)\equiv
\sigma_8(\Omega_{\rm m}/0.3)^{\alpha}$ with $\alpha=0.5$ and 0.45 for
each of the tests. In Figures~\ref{fig:Omm-S8_sys} and
\ref{fig:Omm-sig8_sys}, we show marginalized constraints in the
$\Omega_{\rm m}$-$S_8$ plane and $\Omega_{\rm m}$-$\sigma_8$ for each
setup. We also show one-dimensional and two-dimensional posteriors of
other cosmological parameters and IA parameters in
Appendix~\ref{sec:triangle}. Below we describe each of the different
setups used for the robustness check in detail.

\subsubsection{Shape measurement errors}
In our fiducial analysis, we account for errors due to the PSF leakage
and residual PSF model errors (Section~\ref{subsec:PSF}) and the
uncertainty of the multiplicative bias (Section~\ref{subsec:multip})
by including nuisance parameters to model these errors in the fiducial
nested sampling analysis. To check how uncertainties in these
treatments propagate into our cosmological constraints, we repeat the
nested sampling analysis ignoring these errors (i.e., setting
$\tilde{\alpha}=\tilde{\beta}=\Delta m=0$). We find no significant
shift in the best-fit values of $S_8$ and $\Omega_{\rm m}$ as shown in
Table~\ref{tab:S8_sys} and the top-left panel of
Figure~\ref{fig:Omm-S8_sys} for the $\Lambda$CDM model. The shifts of
both $S_8$ and $\Omega_{\rm m}$ are less than 10\% of the statistical
error.  This indicates that the impact of these shape measurement
errors on the current cosmic shear measurement is negligible.

\subsubsection{Photometric redshift uncertainties}
\label{subsubsec:photoztest}
As discussed in Section~\ref{subsec:photoz}, we include the
photo-$z$ uncertainty in the fiducial nested sampling analysis using the additive
shift parameter $\Delta z_i$. As a simple check, we first ignore
these photo-$z$ errors ($\Delta z_i=0$) and repeat the
nested sampling analysis to find no significant shift or degradation of our
$S_8$ constraints (see e.g., Figure~\ref{fig:Omm-S8_sys}). The result
indicates that the photo-$z$ uncertainty does not have a significant
impact on the $S_8$ constraint compared with the statistical
uncertainty of HSC first year data.

However, as was mentioned in Section~\ref{subsec:photoz}, our approach
to include photo-$z$ uncertainties by single parameters $\Delta z_i$
for individual tomographic bins might be too simplistic.  For
instance, it might be possible that the true redshift distribution has
a small additive shift $\Delta z_i$ as compared with our fiducial
model, but has a significantly larger outlier fraction. In order to
check for possible additional systematics coming from the uncertainty of
the redshift distribution that is not captured by the $\Delta z_i$
parametrization, we replace the redshift distributions of individual
tomographic bins derived from the fiducial COSMOS-reweighted method
with stacked photo-$z$ PDFs from the different photo-$z$ codes mentioned
in Section~\ref{sec:data}, and repeat the nested sampling analysis. In
doing so we set $\Delta z_i=0$.  Since the shapes of the redshift
distributions are slightly different among different estimates of
$\bar{P}(z)$ (see Figure~\ref{fig:fullpz}), we expect that this test
allows us to check the impact of the uncertainty of photo-$z$'s beyond
the $\Delta z_i$ parameter.

Our results shown in Table~\ref{tab:S8_sys} indicate that the changes of the
median $S_8$ values are smaller than the statistical errors, within 0.5$\sigma$
for $\alpha=0.45$.  Constraints in the $\Omega_{\rm m}$-$S_8$ plane shown in
the top-right panel of Figure~\ref{fig:Omm-S8_sys} also indicate that the
effect of adopting different photo-$z$ codes is minor, although we can see some
shifts in the large $\Omega_{\rm m}$ tails of contours along the $S_8$
direction, for some photo-$z$ codes (see Figure~\ref{fig:Omm-sig8_sys}). Larger
effects on the best-fit values of $\Omega_{\rm m}$ can also be seen in e.g.,
Table~\ref{tab:S8_sys}, however they are accompanied with a corresponding
increase in error as well. This is presumably because the minor change of the
setup can shift the best-fit values along the degeneracy direction in the
$\Omega_{\rm m}$-$\sigma_8$ plane, leading to small changes in the $S_8$ value
itself, but larger changes in the tails of the $S_8$ contours along the
degeneracy direction that correspond to larger values of $\Omega_{\rm m}$.

\subsubsection{Non-zero neutrino mass}
\label{subsec:neutrinomass}

In our fiducial setup we assume neutrinos to be massless, i.e., $\sum
m_\nu$ is set to zero. This is mainly because the inclusion of the
minimum value of $\sum m_\nu$ of 0.06~eV is expected to have little
effect on the matter power spectrum, at least as compared with the
statistical error of the HSC first-year cosmic shear analysis.
More generally speaking, the HSC cosmic shear or any other large-scale
structure probe can constrain mainly $\sigma_8$, the present-day
matter fluctuation amplitude. The effect of non-zero neutrino mass on
large-scale structure observables is absorbed by a change in
$\sigma_8$. Only when combining the HSC cosmic
shear result with CMB constraints can we probe neutrino mass, because
a non-zero  neutrino mass leads to a suppression in the matter power
spectrum amplitude at small scales over the range of low redshifts
that the HSC cosmic shear probes, compared to the CMB-inferred
$\sigma_8$. To
check this point explicitly, we include a non-zero neutrino mass of
0.06~eV and repeat the nested sampling analysis, and find that the
median value of $S_8$ is almost unchanged (see e.g.,
Figure~\ref{fig:Omm-S8_sys}).

We also consider the case in which  neutrino mass is allowed to freely vary. While there is a
10\% degradation in the 1$\sigma$ error of the $S_8$ constraint, the
best-fit value is shifted lower by 0.5$\sigma$ (see e.g.,
second left panel of Figure~\ref{fig:Omm-S8_sys}).
This can be explained as follows. Non-zero neutrino mass leads to a suppression in the matter power spectrum
at scales smaller than the neutrino free-streaming scale relative to
large scales within the CDM framework \citep{Takada06}.
Hence, for a fixed $\sigma_8$ that we use to normalize the linear
matter power spectrum at the present, a model with non-zero neutrino
mass leads  to greater amplitudes in the matter power spectrum at
large scales (low $k$) as well as at higher redshifts, compared to
those of the massless neutrino model, i.e. our fiducial model. Hence
varying neutrino mass in the parameter inference prefers a slightly
smaller value in the best-fit $S_8$ in order to reconcile the model
with the measured amplitudes of HSC cosmic shear power spectra  that
are sensitive to the matter power spectrum amplitudes at higher
redshifts such as $z\simgt 1$. For the same reason,
the best-fit value of
$\Omega_{\rm m}$ slightly increases at $\sim 0.2\sigma$ level.
However, we note that varying neutrino mass is a physical extension to the
fiducial model rather than a systematic effect.  When taking into
account the recent upper limits on the neutrino mass $<0.12$~eV (95\%
confidence limit) from {\it Planck} 2018 + BAO
\citep{Planck18_cosmology}, the actual shift of the best-fit due to the
neutrino mass should be smaller than that estimated above.  We will discuss
constraints on the neutrino mass from the HSC cosmic shear, {\it
  Planck}, and their combination in Section~\ref{subsec:jointconst}.

\subsubsection{Baryonic feedback effect}
\label{subsec:baryon_eff}
As described in Section~\ref{subsec:baryon}, we do not include the
modification of the matter power spectrum due to baryonic physics in our
fiducial setup, but we explore the impact of baryonic physics in the
robustness check. As a first check, we fix the matter power spectrum
to that in the AGN feedback model ($A_B=1$), and
find that the $S_8$ value is indeed shifted to a larger value to
compensate for the small-scale suppression of the lensing power. The shift
of the best-fit $S_8$ value is at most $\sim 0.6\sigma$
even in this extreme AGN feedback model and is smaller in other moderate
feedback models as adopted in more recent cosmological hydrodynamical
simulations.  When the baryon parameter $A_B$ is varied using
equation~(\ref{eq:baryon_harnois}), the impact on $S_8$ and
$\Omega_{\rm m}$ is less significant, the shifts of $S_8$ and
$\Omega_{\rm m}$ are at the level of $0.2\sigma$ and
the degradation in the $S_8$ constraints are at the level of
0.1$\sigma$. Constraints in the $\Omega_{\rm m}$-$S_8$ plane shown in
the middle-right panel of Figure~\ref{fig:Omm-S8_sys} also show no
significant shift. We obtain a constraint on the value of $A_B=-0.3\pm
1.6$, which  cannot distinguish between our fiducial choice of the
DM-only model ($A_B=0$) and the AGN feedback model ($A_B=1$).

As a further check, we evaluate the baryon feedback effect using a
different fitting formula for the matter power spectra including baryonic
effects, {\tt HMcode} \citep{Mead15}, which is implemented in {\tt
  Monte Python} (Brieden, Archidiacono, Lesgourgues {\em in prep.}). This fitting formula is based on the same set of OWLS
simulations, but includes the cosmological dependence of the baryonic
feedback effect. In {\tt HMcode}, the baryon feedback is characterized
by the minimum concentration parameter $c_{\rm min}$ and the so-called
halo bloating parameter $\eta_0$. \citet{Mead15} find that these two
parameters are degenerate and related to each other as
$\eta_0=1.03-0.11 c_{\rm min}$ among different feedback models.
Following \citet{Hildebrandt17}, we vary just one parameter $c_{\rm
  min}$ by fixing the other parameter $\eta_0$ to $1.03-0.11
c_{\rm min}$. The AGN feedback model in {\tt HMcode} corresponds to $c_{\rm
min}=2.32$, and yields a value for $S_8(\alpha=0.45)=0.824\pm 0.029$, which is
$\sim 0.8\sigma$ larger than our fiducial value. Although the shift of $S_8$
becomes slightly larger than that using equation~(\ref{eq:baryon_harnois}), the
effect of baryons is expected to be smaller than that observed in this extreme
model of the AGN feedback. Indeed, when we vary $c_{\rm min}$ between the
DM-only case ($c_{\rm min}=3.13$) and the AGN feedback case ($c_{\rm
min}=2.32$), we obtain $S_8 (\alpha=0.45)=0.805\pm 0.031$, corresponding to a
0.2$\sigma$ shift of the best-fit value and 0.1$\sigma$ degradation of the
error of $S_8$ compared to our fiducial constraint.

\begin{figure}
\begin{center}
\includegraphics[width=7.5cm]{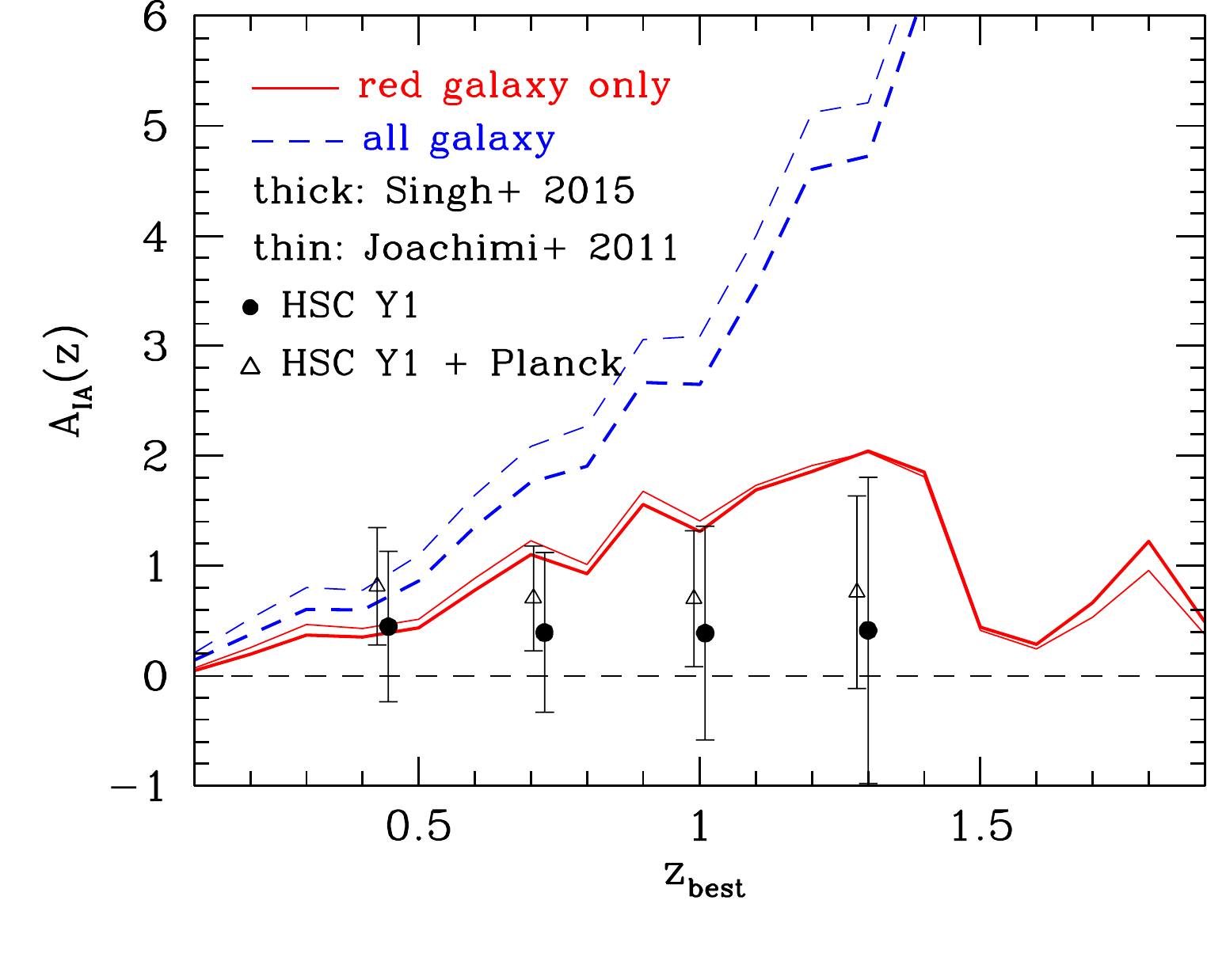}
\end{center}
\caption{The amplitude of the intrinsic alignment at each tomographic
  bin derived from our cosmic shear power spectrum analysis ({\it
    filled circles}) and from the combined analysis with {\it Planck} ({\it
    open triangles}), in which the NLA model is assumed [see
  equation~(\ref{eq:A_IA_z})]. For comparison, we plot IA amplitudes
  extrapolated from the luminosity-dependent IA signals of bright red
  galaxies by \citet{Joachimi11} ({\it thick}) and \citet{Singh15}
  ({\it thin}) assuming that only red galaxies have IA signals ({\it
    solid}) or all galaxies have comparable IA signals ({\it
    dashed}). }
\label{fig:IA}
\end{figure}

\subsubsection{Intrinsic alignment modeling}
In our fiducial analysis, we adopt the NLA model to model intrinsic
alignment contributions to cosmic shear power spectra (see
Section~\ref{subsec:IA}). We include two parameters in the nested
sampling analysis, the overall amplitude $A_{\rm IA}$ and the
power-law index of the redshift dependence of $A_{\rm IA}$, $\eta_{\rm
  eff}$, which are marginalized over when deriving cosmological
constraints.

We test the robustness of our cosmological constrains by adopting IA
models that are different from our fiducial model described above.
First, when the IA contribution is completely ignored, i.e.,
$A_{\rm IA}$ is fixed to $0$ in the nested sampling analysis, the
$S_8$ value decreases by $\sim 0.5\sigma$ (see e.g.,
Table~\ref{tab:S8_sys}). This is because the negative contribution of
the GI term to the measured cosmic shear power spectrum is ignored
in this case. Next, when the IA amplitude $A_{\rm IA}$ is free but the
power-law index of the redshift evolution,  $\eta_{\rm eff}$, is fixed
to $3$, which is a plausible value of $\eta_{\rm eff}$ assuming that
only red galaxies have IA signals
(see Section~\ref{subsec:IA}), we find no significant change of the
best-fit $S_8$ value or $\Omega_{\rm m}$ value.

We now discuss the validity of the IA model parameters derived in the
fiducial nested sampling analysis. We find that the IA amplitude is
consistent with zero, $A_{\rm IA}=0.38\pm 0.70$, for the pivot
redshift $z_0=0.62$. As a further check, when we adopt stacked
photo-$z$ PDFs from several different photo-$z$ codes (see
Section~\ref{subsubsec:photoztest}), we find slightly positive
values of IA signals for some codes, $1.00\pm 0.61$ for {\tt Ephor AB},
$0.50\pm 0.69$ for {\tt MLZ}, $0.62\pm 0.62$ for {\tt Mizuki},
$1.25\pm 0.54$ for {\tt NNPZ}, $0.55\pm 0.67$ for {\tt Frankenz},
and $0.78\pm 0.61$ for {\tt DEmP}, although the significance of
non-zero $A_{\rm IA}$ is not high in any case. This may
indicate that constraint on the IA amplitude is more sensitive to
the photo-$z$ uncertainty than that on $S_8$.

Together with derived constraints on the redshift dependence,
$\eta_{\rm eff}$, we derive $A_{\rm IA}$ amplitudes and their
1$\sigma$ errors for individual tomographic bins as
\begin{equation}
\label{eq:A_IA_z}
A_{\rm IA}(z)=A_{\rm IA} ((1+\langle z\rangle)/(1+z_0))^{\eta_{\rm eff}}
\end{equation}
where $\langle z\rangle$ is the weighted average of source redshifts
in each bin. Figure~\ref{fig:IA} shows the results of $A_{\rm IA}(z)$
derived from our nested sampling. We do not find
significant redshift evolution of $A_{\rm IA}$. We compare this
redshift evolution with the extrapolation of IA amplitudes from the
luminosity-dependent IA signals of bright red galaxies
\citep{Joachimi11,Singh15}. We consider two cases, one in which only
red galaxies have IA signals and the other in which all galaxies
including blue galaxies have comparable IA signals. The details of the
model are given in Section~\ref{subsec:IA}. Figure~\ref{fig:IA}
shows that our results are more consistent with the former model.
This result is consistent with the idea that blue galaxies do not make
a significant contribution to the overall amplitude of the observed IA
signal. While tidal torquing aligns blue galaxies with large-scale
structure, the IA of blue galaxies has not yet been detected in
observations \citep{Joachimi15}. In either case, more accurate
measurements of cosmic shear signals are necessary for further
analysis of the IA. Our conclusion here is that the IA signals from
our fiducial analysis appear to be reasonable.  We will discuss the IA
result when combining HSC first year cosmic shear with {\it Planck} in
Section~\ref{subsec:jointconst}.

\begin{figure}
\begin{center}
\includegraphics[width=7cm]{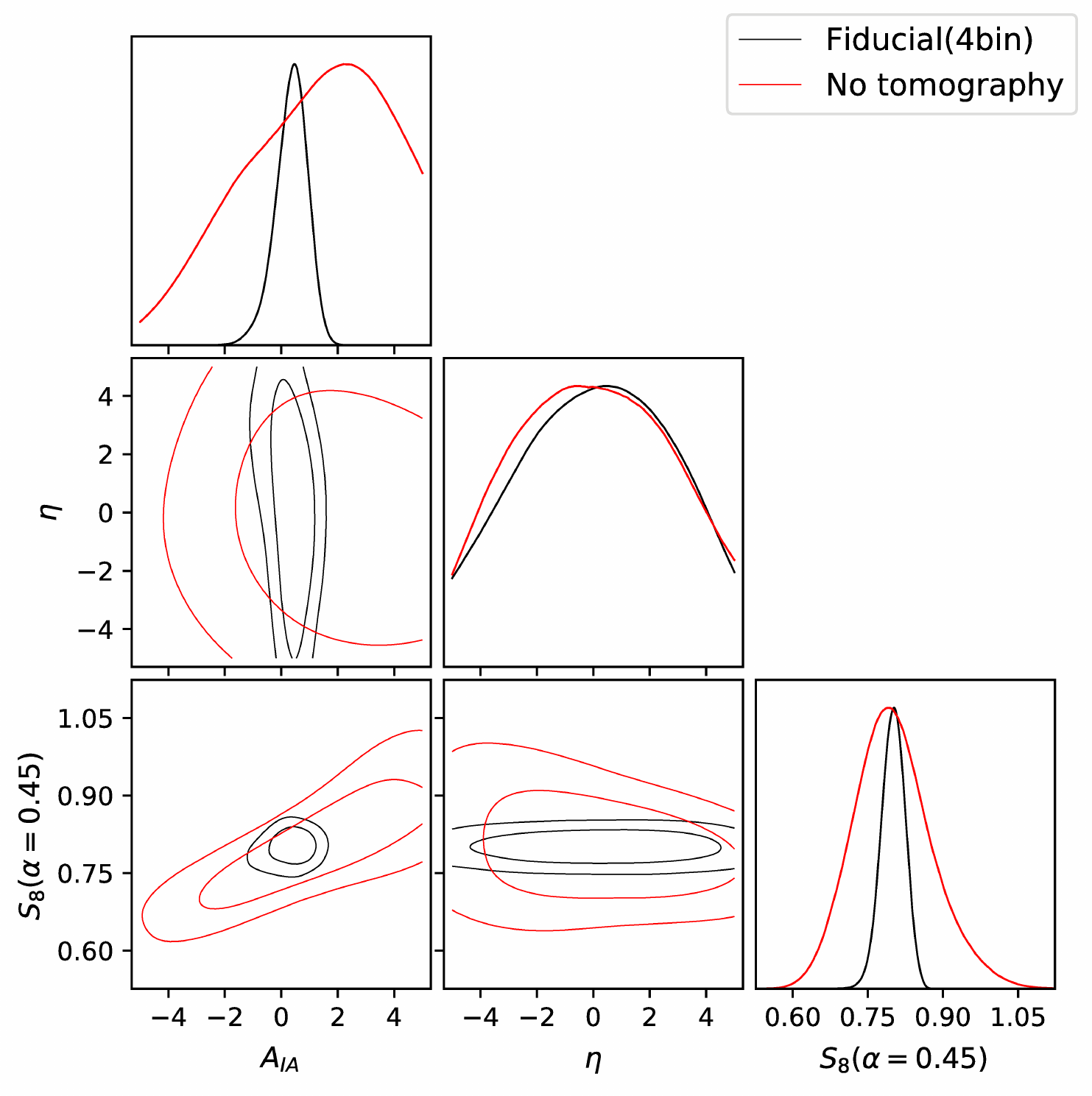}
\end{center}
\caption{Joint constraints and marginalized one-dimensional posteriors
  for the amplitude of the intrinsic alignment $A_{\rm IA}$, the
  power-law index of the redshift evolution, $\eta_{\rm eff}$, and
  $S_8$ in our fiducial four tomographic redshift bin analysis are
  compared with those from non-tomographic cosmic shear analysis. The
  contours represent 68\% and 95\% credible levels. It clearly
  demonstrates that the lensing tomography helps break the degeneracy
  between $A_{\rm IA}$ and $S_8$. }
\label{fig:IA_S8_LCDM}
\end{figure}

Finally, we show that the tomographic analysis helps break the
degeneracy between $S_8$ and the IA
parameters. Figure~\ref{fig:IA_S8_LCDM} shows the joint constraints on
the amplitude parameter of intrinsic alignment $A_{\rm IA}$, the
power-law index of the redshift evolution $\eta_{\rm eff}$, and $S_8$
with and without tomography. It clearly demonstrates the power of lensing tomography to break the
degeneracy between $A_{\rm IA}$ and $S_8$.

\subsubsection{Internal consistency among different redshift and multipole bins}
In our fiducial analysis, we adopt four tomographic bins to constrain
the redshift dependence of cosmic shear power spectra. The comparison
of our cosmological results among these tomographic bins serves as an
important internal consistency check\footnote{For these exercises, we
  will merely examine the amount of shift of the best fit values in
  terms of our statistical errors.  When excluding certain parts of
  the data set, the resultant measurement is quite correlated with the
  entire measurement, so the shifts should not be interpreted as
  measures of statistical significance.}
\citep[e.g.,][]{Efstathiou18,Kohlinger18}.  For this purpose, we
exclude one photo-$z$ bin at a time and see whether the results are
consistent with the fiducial one as shown in the bottom-left panels of
Figures~\ref{fig:Omm-S8_sys} and \ref{fig:Omm-sig8_sys}. We find that
the results of $S_8$ are consistent within 0.3$\sigma$ when any one of
the four tomographic bins is excluded (see e.g.,
Table~\ref{tab:S8_sys}). This indicates that our $S_8$ results do not
significantly rely on the cosmic shear power spectrum at any specific
redshift bin. In contrast, we find that the shift of the best-fit
$\Omega_{\rm m}$ value is relatively larger than $S_8$. In particular,
when the measurement at mid-lower redshift bin is excluded, the
best-fit $\Omega_{\rm m}$ value shifts to a value $\sim 1\sigma$
higher, although the error on $\Omega_{\rm m}$ also increases by 60\%.
This suggests that the constraint on $\Omega_{\rm m}$ is driven by the
relative amplitudes of cosmic shear power spectra between low and high
redshift bins.

As done in Section~\ref{sec:cosmo_const}, we can use the HSC mock
shear catalogs to see whether the large shift in the best-fit
$\Omega_{\rm m}$ value when excluding a single redshift bin is simply
explained by a statistical fluctuation. We find that 9 out of 100 mock
realizations show more than a $1\sigma$ shift of the best-fit
$\Omega_{\rm m}$ value by excluding the measurement at the mid-lower
redshift bin. Thus we conclude that the observed large shift of
$\Omega_{\rm m}$ value can be explained by a $<2\sigma$ statistical
fluctuation.

We also check the internal consistency among different multipole bins.
We first split the fiducial multipole range ($300<\ell<1900$) into
half, i.e., $300<\ell<800$ (a lower-half bin) and $800<\ell <1900$ (a
higher-half bin) as shown in the middle-right panels of
Figures~\ref{fig:Omm-S8_sys} and \ref{fig:Omm-sig8_sys}. We find no
significant shift of either $S_8$ or $\Omega_{\rm m}$. Although the
measurements at the higher-half $\ell$ prefer higher values of
$\Omega_{\rm m}$, once the larger statistical error is taken into
account the significance is less than 1$\sigma$. Next we extend the
upper limit of the $\ell$ range from $1900$ to $3500$ and repeat the
nested sampling analysis.  We find that neither best-fit values of
$S_8$ and $\Omega_{\rm m}$ change significantly, although this
modification leads to a 10\% smaller statistical error for $S_8$
and 20\% smaller error for $\Omega_{\rm m}$ than our fiducial case. This
indicates that our result is insensitive to the choice of the
multipole range and the small-scale physics such as one-halo term of
the intrinsic alignment and baryon physics. This test suggests that
our fiducial choice of the multipole range of $300<\ell<1900$ is
conservative.

\subsubsection{Covariance}
In our analysis we use an analytic model of the covariance matrix
(Section~\ref{subsec:cov}) that includes the dependence of
the covariance matrix on cosmological parameters in the likelihood
function [equation~(\ref{eq:likelihood})]. Here we check how the
results differ when the covariance is fixed to that in the best-fit
cosmology, as has been assumed in most previous cosmic shear
analyses. We find that both $S_8$ and $\Omega_{\rm m}$ agree within
20\% of the statistical uncertainty, and sizes of the errors also
agree within 10\%. This indicates that the effect of the cosmology
dependence of the covariance is subdominant in our analysis.

\subsubsection{Summary of the robustness checks}
The robustness checks presented above indicate that our results are
indeed robust against various systematics and modeling choices?. Among others, the most
significant sources of possible systematics are
photo-$z$ uncertainties, intrinsic alignment modeling, and the effect
of baryonic physics on the matter power spectrum, which can shift the
best-fit values of $S_8(\alpha=0.45)$ by up to $\sim 0.6\sigma$
of the statistical errors in most extreme cases examined in this
paper. We find that constraints on $\Omega_{\rm m}$ are more sensitive to
various systematics in that they shift the best-fit values up to
$\sim 1\sigma$ of the statistical errors. This is because
these systematics tend to move the best-fit values along the
degeneracy direction in the  $\Omega_{\rm m}$-$\sigma_8$ plane. In
addition, we conduct internal consistency checks among different
redshift and multipole bins, and find no sign of systematics.

\subsection{Consistency between HSC cosmic shear and CMB data}
\label{sec:consistency}
We evaluate the consistency between our HSC cosmic shear measurements
and CMB datasets from both {\it Planck} and {\it WMAP9}. The
consistency between different datasets is often judged with the
following Bayesian evidence ratio \citep{Marshall06}:
\begin{equation}
R=\frac{P(\mathbf{D_1,D_2}|{\rm M})}{P(\mathbf{D_1}|{\rm M})P(\mathbf{D_2}|{\rm M})},
\end{equation}
where $D_1$ and $D_2$ denote the two datasets and $M$ indicates the
cosmological model (either $\Lambda$CDM or $w$CDM). The numerator is
the evidence that the two datasets $D_1$ and $D_2$ share the same
cosmological parameters in a given model $M$, while the denominator is
the evidence that the two datasets have different cosmological
parameters in the model $M$. A positive (negative) value of $\log R$
would indicate that the two datasets are (in)consistent.

\begin{table}
  \caption{Bayesian evidence ratios $R$ and the differences of
    log-likelihoods at the maximum a posteriori point $Q_{\rm DMAP}$ in
    different combinations of datasets based on the $\Lambda$CDM or
    $w$CDM cosmology. The values listed here indicate that all dataset
    combinations are consistent.}
\begin{center}
   \begin{tabular}{cccc}
    \hline
    model & datasets & $\log R$ & $Q_{\rm DMAP}$ \\
    \hline
    $\Lambda$CDM & HSC + {\it Planck} & 3.7 & 2.4 \\
    $\Lambda$CDM & HSC + {\it WMAP9} & 4.1 & 1.5 \\
    $w$CDM & HSC + {\it Planck} & 5.3 & 0.5 \\
    \hline
  \end{tabular}
\end{center}
  \begin{tabnote}
  \end{tabnote}
 \label{tab:concordance}
\end{table}

Table~\ref{tab:concordance} lists values of $\log R$ in different
combinations of datasets and cosmological models. We obtain a positive
value of $\log R$ for HSC and {\it Planck} in the
$\Lambda$CDM model, which indicates that the tension in the two datasets, if
any, does not rise to a significant level. We also find that the value of $R$
is even larger for HSC and {\it WMAP} combinations and also for the $w$CDM
model. In all the cases examined here, the Bayesian evidence ratio does not
signal inconsistency between HSC cosmic shear measurements and {\it Planck} and
{\it WMAP9} CMB datasets.

As a further test, we employ another criterion of consistency using
differences of log-likelihoods at the maximum a posteriori (MAP) point
$\theta_p$ in parameter space \citep{RaveriHu18}
\begin{equation}
Q_{\rm DMAP} = -2\ln {\cal L}_{12}(\theta_p^{12})
+ 2\ln {\cal L}_{1}(\theta_p^{1}) + 2\ln {\cal L}_{2}(\theta_p^{2}).
\end{equation}
This can be interpreted as the difference of $\chi^2$ values $\Delta
\chi_{\rm eff}^2 = \chi^2_{\rm eff,12} - \chi^2_{\rm eff,1} -
\chi^2_{\rm eff,2}$ where $\chi^2_{\rm eff,i}$ is defined as
$-2\ln{\cal L}_i(\theta_p^i)$. This criterion was also used as a
consistency check by the recent {\it Planck} 2018 paper
\citep{Planck18_cosmology}. We show the values of $Q_{\rm DMAP}$ in
Table~\ref{tab:concordance}.  In this calculation, we use the
covariance assuming {\it Planck} cosmology. The statistic $Q_{\rm DMAP}$
is expected to follow a $\chi^2$ distribution with
$N_{\rm eff}^{1}+N_{\rm eff}^{2}-N_{\rm eff}^{12}$,
which becomes about 2 degrees of freedom \citep{RaveriHu18}.
We find $Q_{\rm DMAP}=2.4$ for HSC and {\it
  Planck}, which corresponds to a p-value of $0.30$.  In conclusion,
we do not find any signs of significant inconsistency between HSC and
{\it Planck} and {\it WMAP9} CMB datasets for both the $\Lambda$CDM
and $w$CDM models.

\begin{figure}
\begin{center}
\includegraphics[width=8cm]{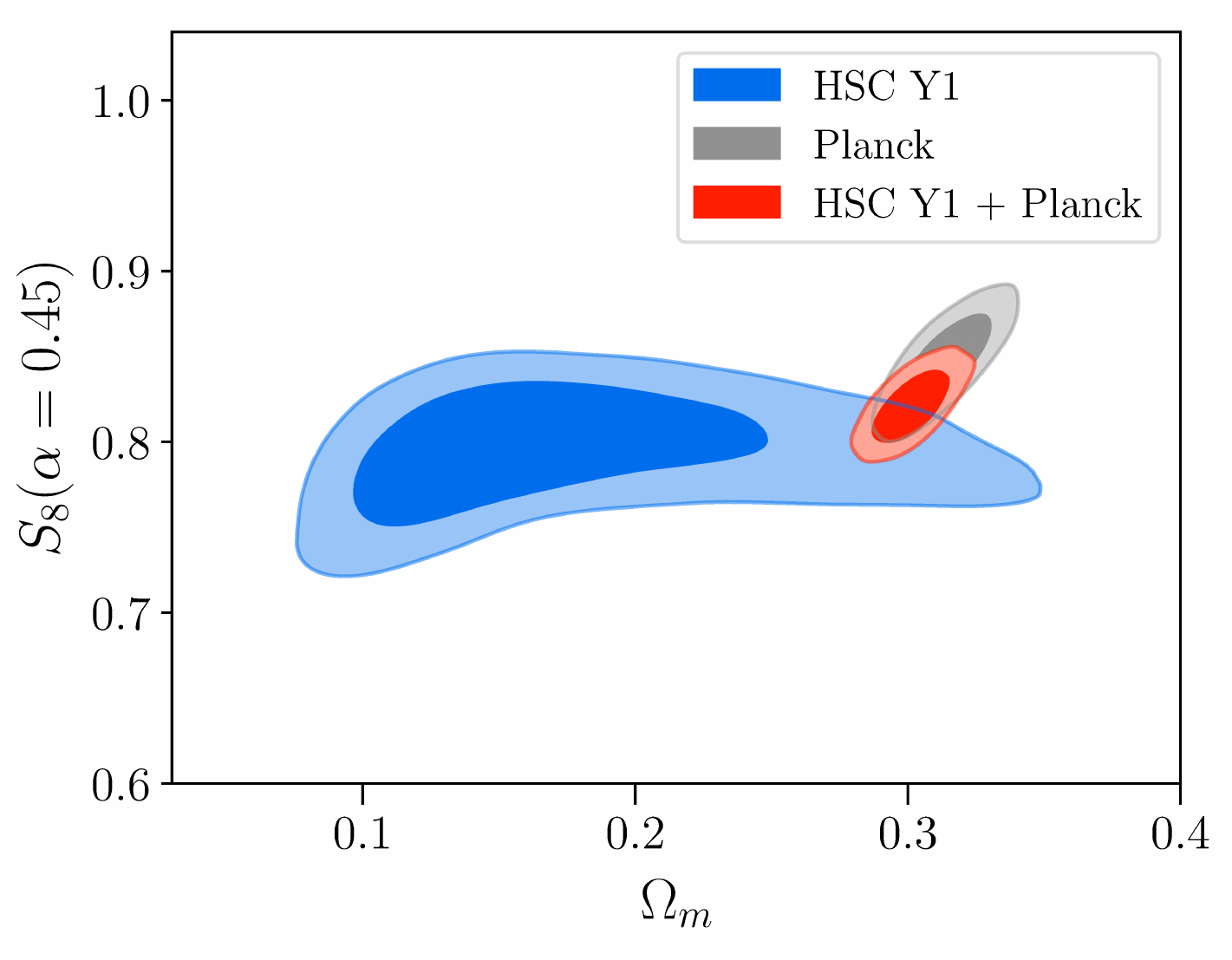}
\end{center}
\caption{Marginalized constraints in the $\Omega_{\rm
    m}$-$S_8(\alpha=0.45)$ plane for HSC, {\it Planck}, and their
  joint analysis. The contours represent 68\% and 95\% credible
  levels. The contours for the HSC alone and {\it Planck} alone are
  same as those plotted in Figure~\ref{fig:Omm-sig8_LCDM}.}
\label{fig:joint_Omm-S8}
\end{figure}

\begin{figure}
\begin{center}
\includegraphics[width=8cm]{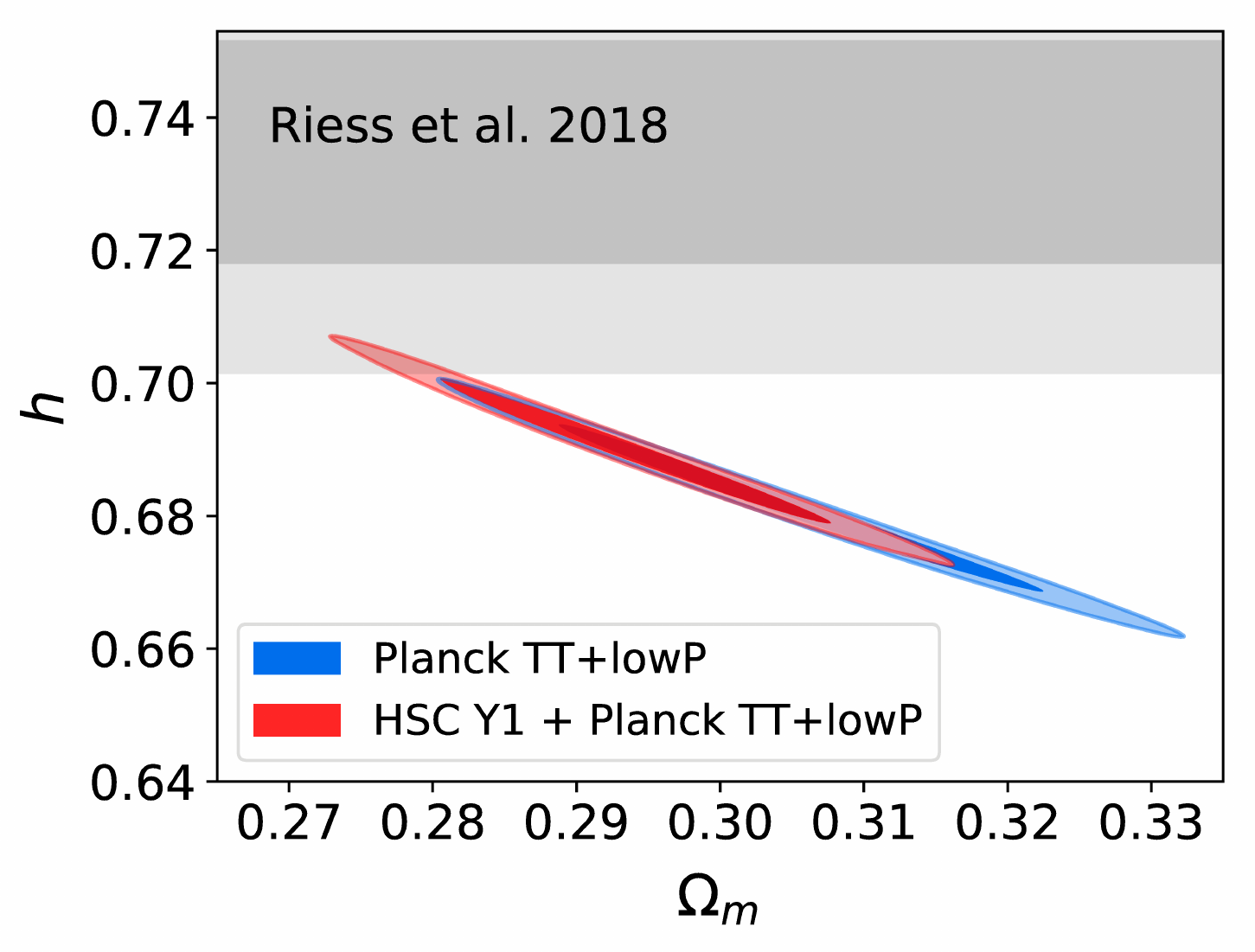}
\end{center}
\caption{Marginalized constraints in the $\Omega_{\rm m}$-$h$ plane
in the $\Lambda$CDM model for {\it Planck} alone and {\it Planck} +
HSC Y1. The contours represent 68\% and 95\% credible levels. We
also show the local Hubble constraint measurement by \citet{Riess18}
for comparison.}
\label{fig:joint_Omm-h}
\end{figure}

\begin{figure}
\begin{center}
\includegraphics[width=8cm]{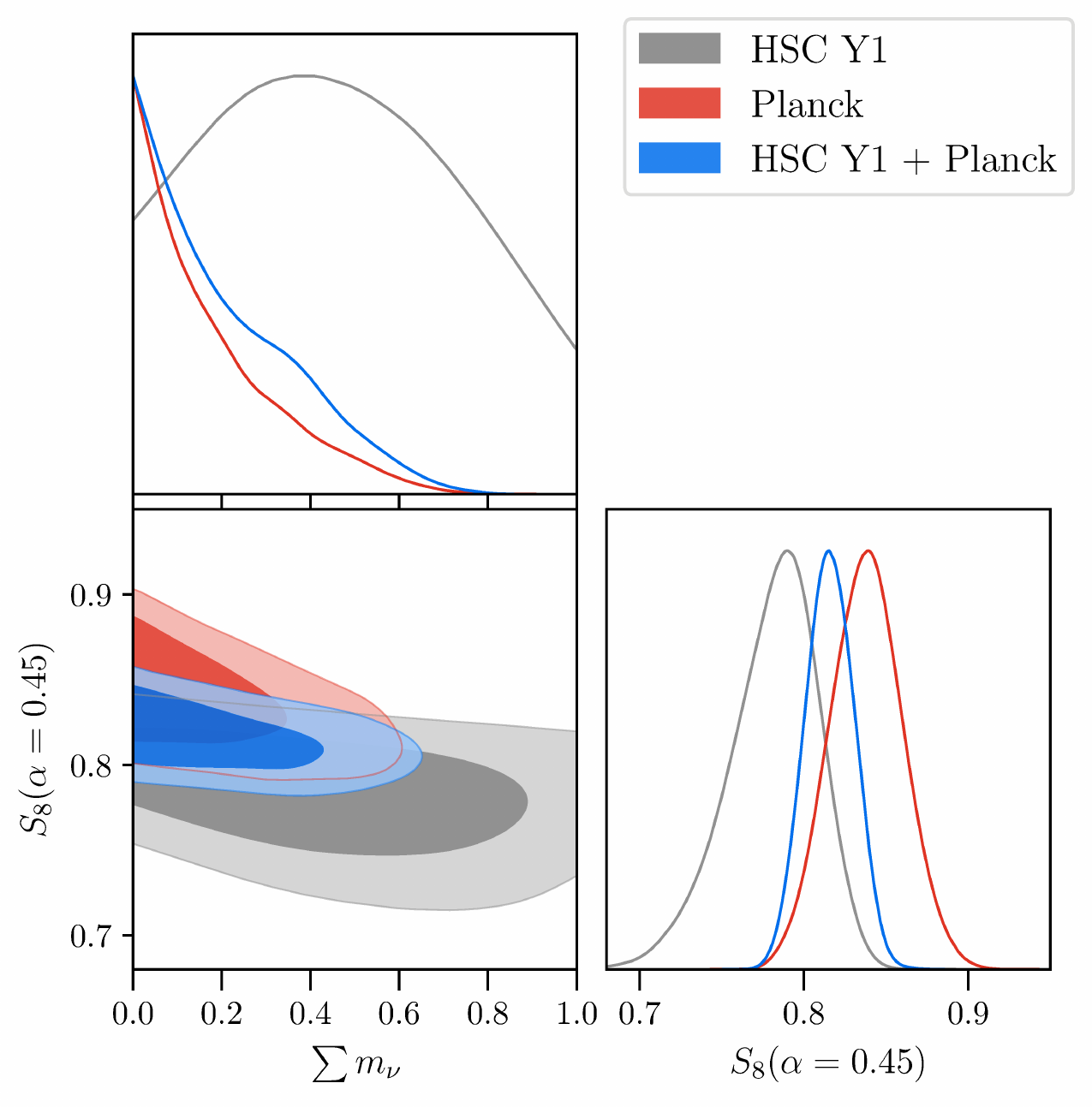}
\end{center}
\caption{Marginalized constraints and one-dimensional posteriors of
  the sum of neutrino masses $\sum m_\nu$ and $S_8(\alpha=0.45)$ from HSC,
  {\it Planck}, and their joint analysis. The contours
represent 68\% and 95\% credible levels. Increasing $\sum m_\nu$
  decreases $S_8$.}
\label{fig:joint_mnu-S8}
\end{figure}

\subsection{Joint constraints combining HSC with external datasets}
\label{subsec:jointconst}
Since we do not see any hints for inconsistencies between HSC and {\it
  Planck}, we now conduct a joint cosmology analysis by combining the
HSC cosmic shear and {\it Planck} 2015 results.  In addition, we
present results combining the distance measurements using a set of
baryonic acoustic oscillations (BAOs) and 740 Type Ia supernovae (SNe)
via the Joint Lightcurve Analysis (JLA) data \citep{JLA}. The BAO
dataset includes the measurements of angular diameter distances from
the 6dF Galaxy Survey \citep{Beutler11}, the SDSS DR7 Main Galaxy
Sample \citep{Ross15}, and BOSS LOWZ and CMASS DR12 sample
\citep{Alam17}. In this subsection, we take into account the lower
bound of the absolute sum of neutrino mass, $\sim 0.06$~eV to match
the fiducial setup in \cite{Planck16}.

Figure~\ref{fig:joint_Omm-S8} shows marginalized constraints on
$\Omega_{\rm m}$ and $S_8$ from HSC and {\it Planck} in the flat
$\Lambda$CDM model. When combining
HSC and {\it Planck}, constraints on $S_8$ and $\Omega_{\rm m}$
both improve, to $0.821\pm 0.017$ and $0.301\pm 0.010$, respectively,
compared to {\it Planck} alone \citep{Planck16} where $S_8=0.848\pm 0.024$
and $\Omega_{\rm m}=0.315\pm 0.013$. Since the well-determined CMB peak
locations are highly sensitive to the combination of $\Omega_{\rm m}h^3$
\citep{Percival02}, lower $\Omega_{\rm m}$
slightly increases the Hubble parameter $H_0$ to $68.36\pm 0.81$~km/s
from the joint analysis of HSC and {\it Planck}, as shown in
Figure~\ref{fig:joint_Omm-h}. The reported 3.7$\sigma$ tension of
$H_0$ between {\it Planck} and the local measurement \citep{Riess18}
reduces to 3.1$\sigma$ by adding our cosmic shear measurement. The
difference is still present, albeit at a slightly smaller significance.

Figure~\ref{fig:joint_mnu-S8} shows marginalized constraints from HSC,
{\it Planck}, and their combination when the sum of neutrino masses
$\sum m_\nu$ is allowed to vary. The constraint on $\sum m_\nu$ from
the HSC cosmic shear measurement alone is very weak with the peak of the
posterior around 0.45~eV, for the reasons described in
Section~\ref{subsec:neutrinomass}. Increasing the sum of the neutrino mass
damps the amplitude of linear matter fluctuations and thus decreases
the constraints on $S_8$ for both {\it Planck} and HSC.
Since the HSC cosmic shear constraints favor a higher $\sigma_8$ value
than that from the {\it Planck} data as shown in
Figure~\ref{fig:Omm-sig8_LCDM}, we cannot expect that combining the
HSC and {\it Planck} constraints improves the neutrino mass
constraint. For this reason, the
posterior distribution of $\sum m_\nu$ does not significantly change
between {\it Planck} and the joint constraint from HSC and {\it
  Planck}, although a slight increase of the probability at larger
$\sum m_\nu$ is seen when the HSC constraint is added.

We obtain the intrinsic alignment amplitude of $A_{\rm IA}=0.73\pm 0.46$
when combining HSC cosmic shear with {\it Planck}. The amplitude is
slightly higher than the value from the HSC cosmic shear analysis
alone, $A_{\rm IA}=0.38\pm 0.70$. The intrinsic alignment amplitudes
of individual redshift bins shown in Figure~\ref{fig:IA} indicate that
our result is consistent with the hypothesis that only red galaxies
have significant IA signals (see Section~\ref{subsec:IA} for more details).

When varying the baryonic feedback parameter [see
equation~(\ref{eq:baryon_harnois})], we obtain $A_B=1.1\pm 0.8$ from the
joint analysis of the HSC cosmic shear and {\it Planck}, which
appears to slightly prefer the presence of baryon feedback effects
on the matter power spectrum.  This is because baryon feedback
increases the HSC-inferred values of $S_8$ and $\Omega_{\rm m}$,
making them more consistent with the {\it Planck} best-fit values, as
discussed in Section~\ref{subsec:baryon_eff}. We check whether the
preference of the model including the baryon feedback parameter is
significant or not using the following Bayesian evidence ratio
\begin{equation}
\label{eq:Bayesfac}
K_B(\mathbf{D})=\frac{P(\mathbf{D}|\Lambda {\rm CDM}
  + A_B)}{P(\mathbf{D}|\Lambda {\rm CDM})}.
\end{equation}
The Bayesian evidence $P(\mathbf{D}|{\rm M})$ indicates the
probability of obtaining dataset $\mathbf{D}$ in the model ${\rm M}$
\begin{equation}
P(\mathbf{D}|{\rm M})=\int \mathbf{dp}
P(\mathbf{D}|\mathbf{p},{\rm M})P(\mathbf{p}|{\rm M}),
\end{equation}
where $P(\mathbf{D}|\mathbf{p},{\rm M})$ is the likelihood and
$P(\mathbf{p}|{\rm M})$ is the prior of the parameter set of
$\mathbf{p}$. A value of $K_B$ larger than unity would indicate that the
model with the baryon feedback parameter is preferred. The substantial
evidence of the preference of the model with the baryon feedback
parameter is found when $K_B>\sqrt{10}$ and strong evidence when
$K_B>10$, based on Jeffreys' scale, and vice versa.
However, we find $K_B=0.46$, which indicates that there is no
preference for the model including the baryon feedback parameter, and
the improvement in the fit to the data is not significant enough to
justify addition of another parameter.

\begin{figure*}
\begin{center}
\includegraphics[width=8cm]{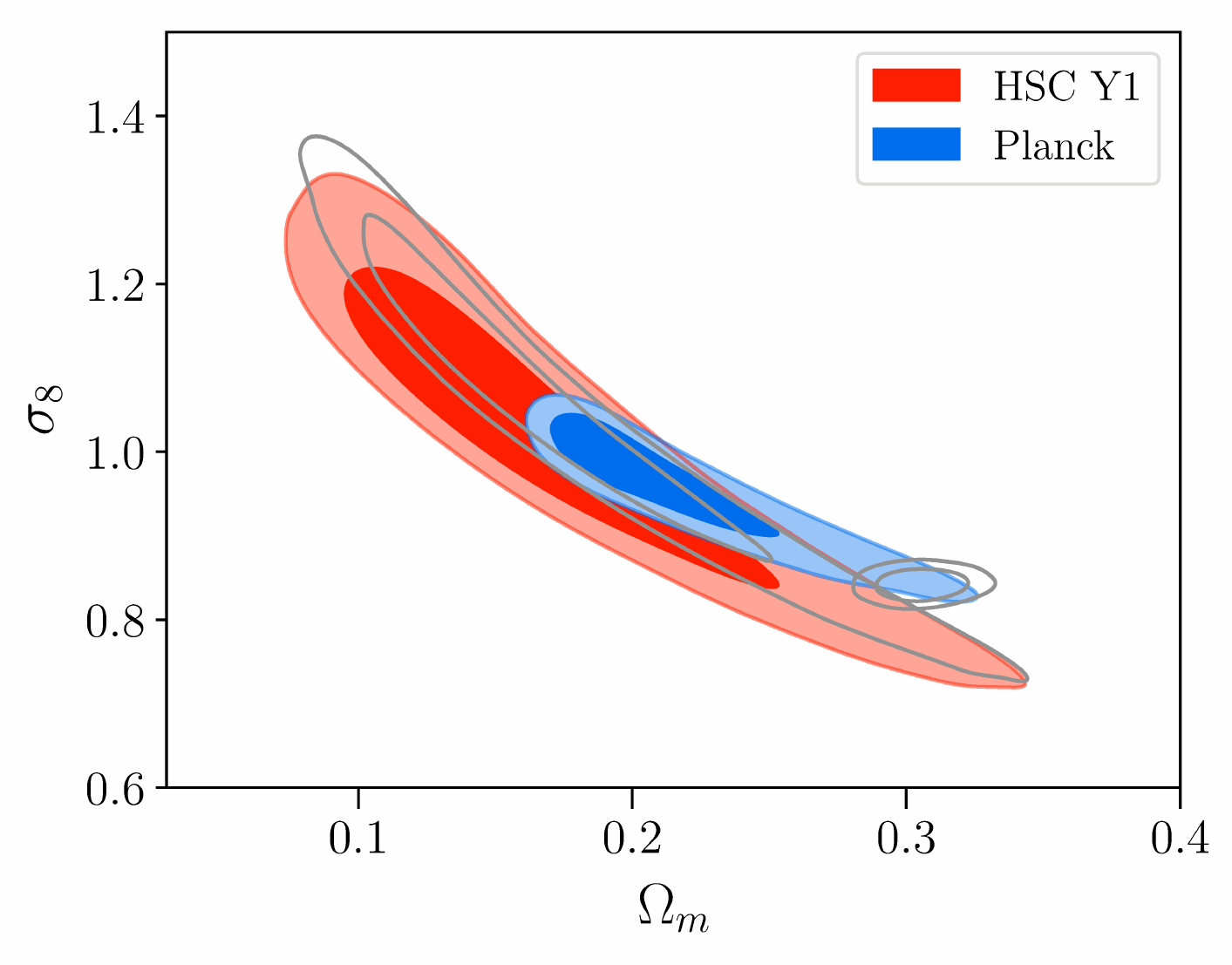}
\includegraphics[width=8cm]{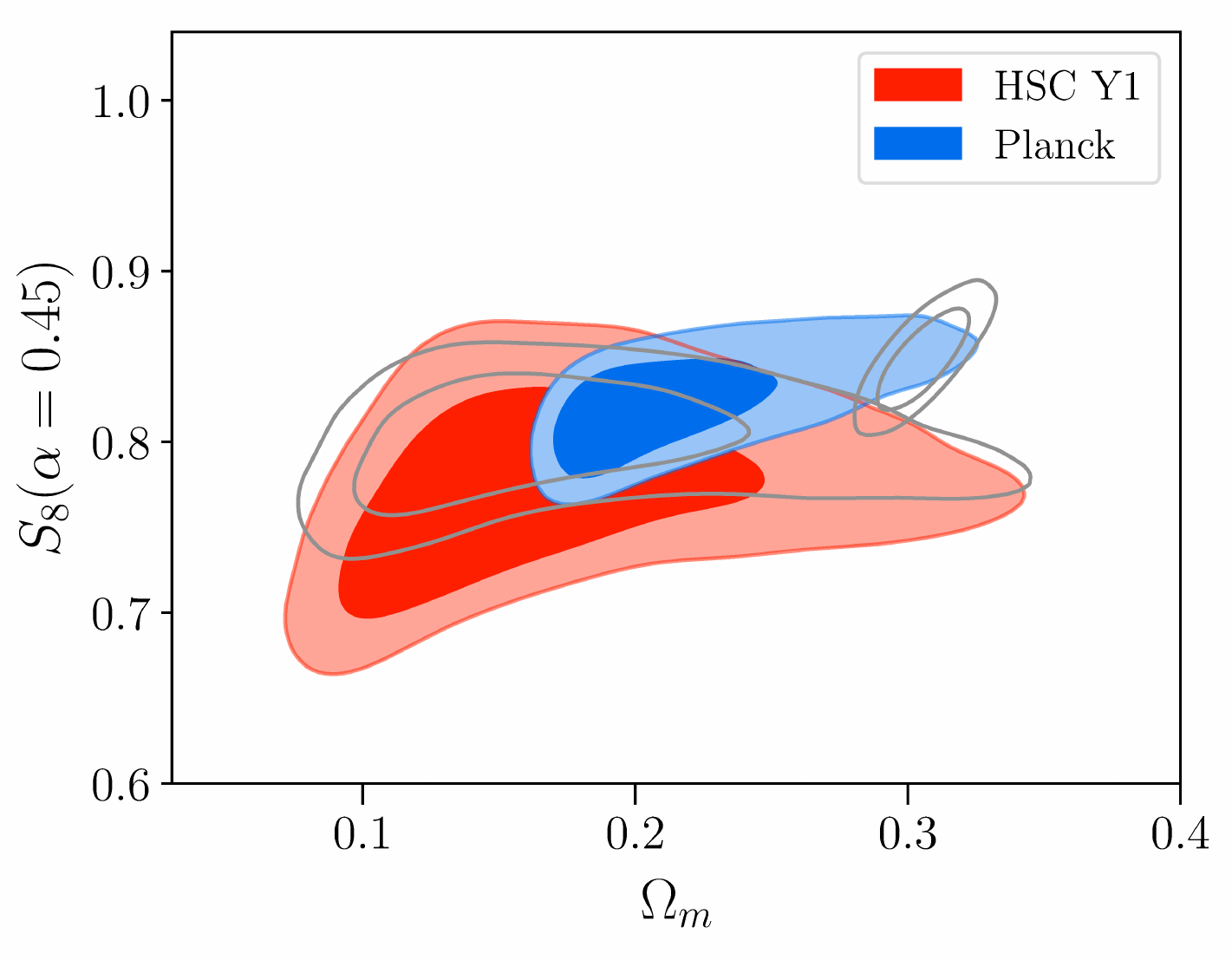}
\end{center}
\caption{Same as Figure~\ref{fig:Omm-sig8_LCDM}, but for the constraints
  from HSC and {\it Planck} in the $w$CDM model. For reference,
  the constraints in the fiducial $\Lambda$CDM model are shown by gray
  contours. }
\label{fig:Omm-sig8_wCDM}
\end{figure*}

\begin{figure}
\begin{center}
\includegraphics[width=8cm]{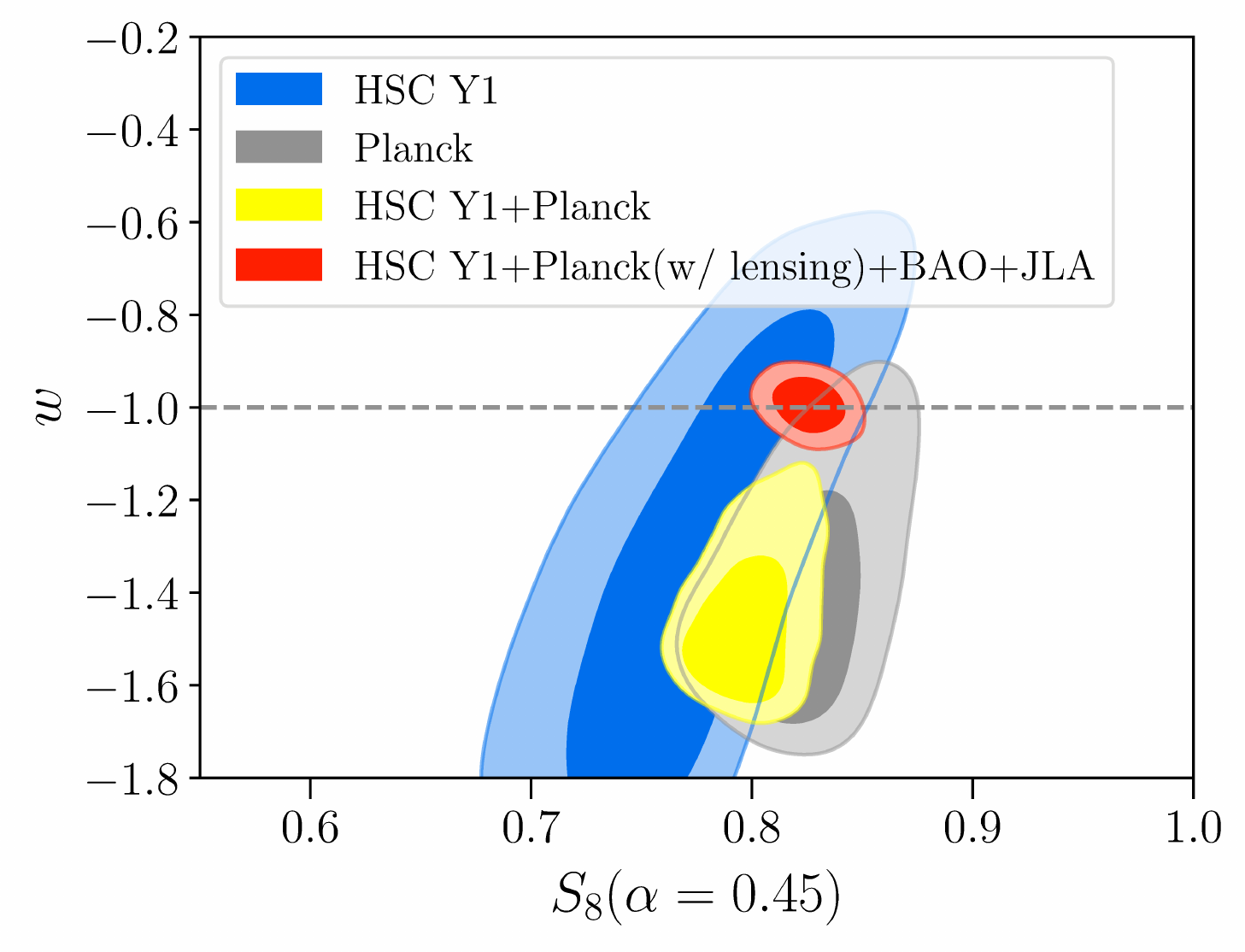}
\end{center}
\caption{Marginalized constraints in the $S_8(\alpha=0.45)$-$w$ plane
  for the $w$CDM model from HSC, {\it Planck}, and their
  joint analysis. The contours represent 68\% and 95\% credible
  levels.}
\label{fig:S8-w}
\end{figure}

\subsection{Cosmological constraints in the $w$CDM model}
Next we allow the dark energy equation of state constant $w$ to be a free
parameter in our model, instead of our fiducial choice of fixing its
value to $-1$.  We find
that $S_8 (\alpha=0.45)=0.773^{+0.043}_{-0.038}$ and
$\Omega_{\rm m}=0.163^{+0.079}_{-0.047}$.  Figure~\ref{fig:Omm-sig8_wCDM}
shows marginalized constraints in the $\Omega_{\rm m}$-$\sigma_8$
and $\Omega_{\rm m}$-$S_8(\alpha=0.45)$ planes.
Adding $w$ as a model parameter significantly degrades constraints on
$S_8$ compared with the $\Lambda$CDM case. This indicates that $w$ is
not well- constrained from the cosmic shear analysis alone.

The degeneracy between $S_8$ and $w$ is clearly seen in
Figure~\ref{fig:S8-w}, which shows the constraints in the $S_8$-$w$
plane.  The constraint on $w$ is $w=-1.37^{+0.43}_{-0.37}$ from HSC
cosmic shear measurements alone. We find that both HSC and {\it
  Planck} prefer $w< -1$, leading to their joint constraint of
$w=-1.45^{+0.16}_{-0.10}$, i.e., 3$\sigma$ deviation from $w=-1$.  The
similar deviation was found also in the previous lensing surveys by
\citet{Joudaki17b,DES17}. As discussed in \citet{Planck16}, this comes
from the strong degeneracy between $w$ and $h$. When we add the BAOs
and SNe data, we obtain $w=-1.006\pm 0.044$.  This constraint on $w$
is similar to that from {\it Planck}+BAOs+SNe without the HSC result,
$w=-1.012\pm 0.046$ \citep{Planck16,Alam17}.

We evaluate the possible preference for the $w$CDM model over the
$\Lambda$CDM model based on the
Bayes factor similar to equation~(\ref{eq:Bayesfac})
\begin{equation}
  K_{w{\rm CDM}}(\mathbf{D})
  =\frac{P(\mathbf{D}|w{\rm CDM})}{P(\mathbf{D}|\Lambda {\rm CDM})}.
\end{equation}
We obtain $K_{w{\rm CDM}}=0.73\pm 0.05$ from our HSC cosmic shear
analysis, which indicates that there is no preference for $w\ne -1$
from the cosmic shear alone.  When combining HSC with {\it Planck}, the
value of $K_{w{\rm CDM}}$ increases to 2.7, which indicates that the $w$CDM
model is favored, although the significance is small.  When combining
HSC with {\it Planck} including lensing and with BAOs and SNe,
however, $K_{w{\rm CDM}}$ decreases to 0.091, indicating that the
$w$CDM model is not preferred. This is not surprising given that our
best fit $w$CDM model when including all the probes was not
significantly different from our fiducial choice of dark energy being
a cosmological constant, i.e. $w=-1$.

\section{Summary and conclusions}
\label{sec:summary}
We have presented results of the cosmic shear power spectrum analysis
using the HSC first-year data over 137~deg$^2$ of sky. The exquisite
depth and image quality of the HSC survey images allow us to construct
a source galaxy sample with an effective number density of
$16.5$~arcmin$^{-2}$ even after conservative cuts for constructing the
accurate shear catalog \citep{Mandelbaum18a} as well as the redshift
cut of $0.3\leq z\leq 1.5$ for the tomographic analysis.

We have measured cosmic shear power spectra from the HSC first-year
shear catalog using the pseudo-$C_\ell$ method that corrects for the
non-uniform mask effect due to the survey geometry, the bright star
masks, and the non-uniformity of the number density of sources.
Using realistic HSC mock shear catalogs, we have demonstrated that
this method separates E- and B-modes and recovers input spectra in an
unbiased way.

The measured tomographic cosmic shear power spectra have a
total signal-to-noise ratio of 16, whereas the B-mode power
spectra are consistent with zero within the multipole range
$300 \leq \ell \leq 1900$. We fit the spectra using a model
that includes contributions from intrinsic alignments of galaxies, as
well as seven additional nuisance parameters to account for shape
measurement errors and photo-$z$ errors. We use an analytic model for
the covariance matrix,  whose accuracy has also been tested against
the HSC mock shear catalogs. We have found that our model fits the
measured cosmic shear power spectra quite well, with a minimum
$\chi^2$ of 45.4 for 56.9 degrees of freedom for our fiducial
$\Lambda$CDM model.

Our cosmological constraints are well encapsulated by constraints on
the parameter $S_8=\sigma_8(\Omega_{\rm m}/0.3)^\alpha$, with
$\alpha=0.45$ being the best choice for our case. Assuming a flat
$\Lambda$CDM model, we have found
$S_8(\alpha=0.45)=0.800^{+0.029}_{-0.028}$ and
$S_8(\alpha=0.5)=0.780^{+0.030}_{-0.033}$. When the dark energy
equation of state $w$ is allowed to deviate from $-1$, the constraint
on $S_8$ is degraded to $S_8 (\alpha=0.45)=0.773^{+0.043}_{-0.038}$.
The dark energy equation of state is not very well constrained from
the HSC cosmic shear analysis alone, $w=-1.37^{+0.43}_{-0.36}$. Our
constraints on $S_8$ agree with other recent cosmic shear analysis
results \citep{Hildebrandt17,Kohlinger17,Troxel18b} as well as
results of extended analyses including galaxy-galaxy lensing
\citep{Joudaki18,KiDS18,DES17,DLS18}. The 3.6\% fractional constraint on $S_8$
represents one of the tightest constraints on $S_8$ from cosmic shear
analyses conducted to date.

We have carefully checked the consistency between our HSC cosmic shear
results and CMB datasets both for {\it Planck} 2015 and {\it WMAP9}.
Both the results of the Bayesian evidence ratio and the difference of
log-likelihoods at the maximum a posteriori point indicate that the
HSC cosmic shear results and these CMB datasets are consistent with
each other. The best-fit value of $S_8$ from the HSC cosmic shear
analysis is also consistent within 2$\sigma$ with those from  {\it
  Planck} 2015 and {\it WMAP9}.

To check the robustness of our results, we have conducted a number of
systematics tests and tests for our sensitivity to modeling
assumptions such as additional photo-$z$ uncertainties, the
uncertainty of IA modeling, and the effect of baryon physics on the
matter power spectrum. We find that our results are robust against these
systematics in that they can shift the best-fit values of $S_8$ by no
more than $\sim 0.6\sigma$ of the statistical errors. We have
also found our cosmological results are consistent among different
redshift and multipole bins.

In our cosmological analysis we simultaneously fit the IA amplitude as
well as its redshift evolution. The best-fit IA amplitude at the pivot
redshift of $z_0=0.62$ and its 1$\sigma$ error is found to be
$A_{\rm IA}=0.38\pm 0.70$, which indicates that the IA
amplitude is consistent with zero. We have found that the redshift
evolution of $A_{\rm IA}$ is weak. These results are consistent with a
model that the IA amplitude is dominated by that of red galaxies,
although this conclusion is based on the extrapolation of IA amplitude
scalings that are calibrated using luminous red galaxies. We have also
confirmed that our constraints on $S_8$ are insensitive to the priors
we use on the parameters of our IA model.

While our results are statistically consistent with the {\it Planck}
2015 results, both the best-fit values of
$S_8(\alpha=0.45)=0.800^{+0.029}_{-0.028}$ and $\Omega_{\rm
  m}=0.162^{+0.086}_{-0.044}$ are lower than the {\it Planck}
result. Such lower best-fit values of $S_8$ and $\Omega_{\rm m}$ are
also found in other recent weak lensing analyses
\citep[e.g.,][]{DES17}. Although our consistency and mock analysis
indicates that these lower values could just be a statistical
fluctuation, there is a possibility that these lower values originate
from systematic effects that are unaccounted for in our current cosmic
shear analysis, or more interestingly, from a possible failure of our
fiducial neutrino mass-free $\Lambda$CDM model.  For instance, our
robustness tests have shown that systematic errors and modeling
choices tend to have a larger impact on $\Omega_{\rm m}$ than on
$S_8$, which could imply that the low $\Omega_{\rm m}$ of our result
might partly be due to additional systematic errors. However, it is
also possible that the lower values of $S_8$ and $\Omega_{\rm m}$ have
a physical origin.  The lower best-fit value of $\Omega_{\rm m}$ seen
in our measurements of the HSC cosmic shear power spectra could be due
to a slower redshift evolution of density fluctuations than predicted
by the $\Lambda$CDM model with {\it Planck} best-fit cosmological
parameters. Hence it is of great importance to improve constraints
from weak lensing in order to discriminate between these
possibilities.

This paper presents cosmological results from the HSC first-year data.
When the HSC survey is completed in $\sim 2019$, we will have roughly
seven times more area for the cosmic shear analysis. In addition to
increased area, there is room for improvement in several ways.
For instance, we have constructed the HSC first-year shear catalog
adopting a number of conservative cuts on galaxy magnitudes, PSF
sizes, and signal-to-noise ratios. By improving shear measurement
techniques, we can increase the fraction of galaxies that we can use
for the cosmic shear analysis. While we conservatively limit the
multipole range to $\ell<1900$, we can extend the analysis
to higher multipoles once the intrinsic alignment at small scales and
the modification of the matter power spectrum due to baryon physics
are understood better. We plan to continue to improve photo-$z$
measurements in the HSC survey, both by increasing a sample of
galaxies with spectroscopic redshifts for calibrations and by
implementing new techniques such as clustering redshifts. Our
first-year cosmic shear analysis suggests that with the increased area
and continued improvements in analysis methodology, the final HSC weak
lensing analysis has the potential to provide a stringent test of the
concordance cosmological model.

\vspace{1cm}
\begin{ack}
We thank the referee, Catherine Heymans, for very useful comments and
suggestions.  We thank Michael Troxel for kindly providing the outputs
of nested sampling with DES Y1 likelihoods. We also thank Ryuichi
Takahashi, Chihway Chang, David Alonso for their feedback that
improved the quality of the paper. Our likelihood code is partly based
on the likelihood code from \cite{Kohlinger17}, which is publicly
available at
\url{https://bitbucket.org/fkoehlin/kids450_qe_likelihood_public}.

This work was supported in part by World Premier International
Research Center Initiative (WPI Initiative), MEXT, Japan, JSPS KAKENHI
Grant Number JP15H03654, JP16K17684, JP16H01089, JP17H06599, JP18H04348,
JP18K03693, JP18H04350, MEXT Grant-in-Aid for Scientific Research on
Innovative Areas (JP15H05887, JP15H05892, JP15H05893, JP15K21733),
and JST CREST Grant Number JPMJCR1414.
HMi and MSi are supported by the Jet Propulsion Laboratory, California
Institute of Technology, under a contract with the National
Aeronautics and Space Administration.
RMa is supported by the Department of Energy Cosmic Frontier program,
grant DE-SC0010118.

Data analysis were in part carried out on PC cluster at Center for
Computational Astrophysics, National Astronomical Observatory of
Japan. Numerical computations were in part carried out on Cray XC30 at
Center for Computational Astrophysics, National Astronomical
Observatory of Japan.

The Hyper Suprime-Cam (HSC) collaboration includes the astronomical
communities of Japan and Taiwan, and Princeton University. The HSC
instrumentation and software were developed by the National Astronomical
Observatory of Japan (NAOJ), the Kavli Institute for the Physics and
Mathematics of the Universe (Kavli IPMU), the University of Tokyo, the
High Energy Accelerator Research Organization (KEK), the Academia Sinica
Institute for Astronomy and Astrophysics in Taiwan (ASIAA), and
Princeton University. Funding was contributed by the FIRST program from
Japanese Cabinet Office, the Ministry of Education, Culture, Sports,
Science and Technology (MEXT), the Japan Society for the Promotion of
Science (JSPS), Japan Science and Technology Agency (JST), the Toray
Science Foundation, NAOJ, Kavli IPMU, KEK, ASIAA, and Princeton
University.
This paper makes use of software developed for the Large Synoptic Survey
Telescope. We thank the LSST Project for making their code available as
free software at \url{http://dm.lsst.org}

The Pan-STARRS1 Surveys (PS1) have been made possible through
contributions of the Institute for Astronomy, the University of Hawaii,
the Pan-STARRS Project Office, the Max-Planck Society and its
participating institutes, the Max Planck Institute for Astronomy,
Heidelberg and the Max Planck Institute for Extraterrestrial Physics,
Garching, The Johns Hopkins University, Durham University, the
University of Edinburgh, Queen's University Belfast, the
Harvard-Smithsonian Center for Astrophysics, the Las Cumbres Observatory
Global Telescope Network Incorporated, the National Central University
of Taiwan, the Space Telescope Science Institute, the National
Aeronautics and Space Administration under Grant No. NNX08AR22G issued
through the Planetary Science Division of the NASA Science Mission
Directorate, the National Science Foundation under Grant
No. AST-1238877, the University of Maryland, and Eotvos Lorand
University (ELTE) and the Los Alamos National Laboratory.

Based in part on data collected at the Subaru Telescope and retrieved
from the HSC data archive system, which is operated by Subaru Telescope
and Astronomy Data Center at National Astronomical Observatory of Japan.
\end{ack}

\vspace{1cm}

\appendix

\section{Pseudo-$C_\ell$ method in the flat-sky approximation and the
  test with mock shear catalogs}
\label{sec:app1}

\subsection{Formalism}
Following the methodology developed in \citet{Hikage11}, here we
provide a detailed description of the pseudo-$C_\ell$ method that we
adopt for unbiased measurements of cosmic shear power spectra from the
HSC first-year shear catalog. Throughout the paper we use a flat-sky
approximation because the curvature effect on each disjoint patch of
the current HSC survey data is negligible compared to the statistical
errors.  We define the observed shear field $\bm{\gamma}^{\rm (obs)}$
as a sum of weighted ellipticities at sky pixel position $\bm{\theta}$,
after correcting for the shape bias and responsivity
[equation~(\ref{eq:obsshear})]. This shear field is related to the true
shear field as
\begin{equation}
\bm{\gamma}^{\rm (obs)}(\bm{\theta})=W(\bm{\theta})\bm{\gamma}^{\rm (true)}(\bm{\theta}).
\end{equation}
The mask (weight) field $W({\bm\theta})$ is computed as a sum of
source weights $w_i$ within each pixel using a
nearest neighbor assignment scheme. The observed shear field is
decomposed into an E-mode (even parity) and B-mode (odd parity) in
Fourier space as
\begin{equation}
\tilde{E}_{\bm\ell}\pm i\tilde{B}_{\bm\ell}
=\int \bm{d\theta} \bm{\gamma}(\bm\theta)e^{i(\bm{\ell\cdot\theta}\pm 2\varphi_{\bm{\ell}})},
\end{equation}
where $\bm\ell$ denotes the two-dimensional harmonic vector,
$\varphi_{\bm{\ell}}$ is the angle of $\bm\ell$, and the
and the multipole $\ell$ corresponds to the angular scale
$\theta=\pi/\ell$. In practice, we consider a square boundary to cover
each of the separate fields of HSC data to perform the Fourier
transform. The angular scale of the survey boundary is set to be 720
arcmin for XMM, WIDE12H, and HECTOMAP, 900 arcmin for GAMA09H and
VVDS, and 1080 arcmin for GAMA15H. The number of pixels $N_{\rm pix}$
is set to be $1024^2$ for all of the fields. After
Fourier-transforming the observed (weighted) shear field, we obtain
the shear field convolved with the mask field as
\begin{eqnarray}
(\tilde{E}_{\bm\ell}\pm i\tilde{B}_{\bm\ell})^{\rm (obs)}
=\int \frac{\bm{d\ell}'}{(2\pi)^2}
(\tilde{E}_{\bm\ell'}\pm i\tilde{B}_{\bm\ell'})^{\rm (true)}
\tilde{W}_{\bm{\ell-\ell}'}e^{\pm 2i(\varphi_{\bm{\ell'\ell}})}, \nonumber \\
\end{eqnarray}
where $\varphi_{\bm{\ell\ell'}}$ denotes the angle between vectors $\bm\ell$ and $\bm\ell'$
and $\tilde{W}_{\bm\Delta\ell}$ is the Fourier transform of $W(\bm\theta)$
\begin{equation}
\label{eq:mask_fourier}
\tilde{W}_{\bm\Delta\ell}=\int \bm{d\theta} W(\bm{\theta})
e^{i\bm{\Delta\ell\cdot\theta}}.
\end{equation}
We define the auto and cross spectra of E- and B-mode cosmic shear as
\begin{eqnarray}
\langle X_{\bm\ell}Y^{*}_{\bm{\ell'}}\rangle&=&\Omega_{\rm sky}
\delta^{\rm K}_{\bm{\ell-\ell'}}C_{\bm\ell}^{XY},
\end{eqnarray}
where $X_{\bm\ell}$ and $Y_{\bm\ell}$ denote the Fourier transform of
E- and B-mode shear, respectively, $\delta^{\rm K}_{\bm\ell}$ is the
Kronecker delta, and $\Omega_{\rm sky}$ is the sky area with non-zero
$W(\bm{\theta})$. The power spectrum for the weighted field has mode
coupling because of the presence of the weight field in real
space. This coupling can expressed as
\begin{equation}
{\cal C}_{\bm\ell}^{\rm
  (obs)}=\sum_{\bm\ell'}\bm{M}_{\bm{\ell\ell'}}F_\ell^2{\cal
  C}_{\bm\ell'}^{\rm (true)}+\bm{N}_{\bm\ell}^{\rm (obs)},
\end{equation}
where ${\cal C}_{\bm\ell}=(C_{\bm\ell}^{EE}, C_{\bm\ell}^{BB},
C_{\bm\ell}^{EB})$, $\bm{M}$ is convolution matrix,
$F_\ell$ is the pixel window function as we use a pixelized map in
the $\bm\theta$ space, and
$\bm{N}_{\bm\ell}^{\rm (obs)}$ is the noise spectrum that is also
convolved with the weight field.
Non-zero components of the convolution matrix are given by
\begin{eqnarray}
\label{eq:convmat}
\bm{M}_{\bm{\ell\ell'}}^{EE,EE}=\bm{M}_{\bm{\ell\ell'}}^{BB,BB}=
\frac{{\cal W}_{\bm{\ell-\ell'}}^{\gamma\gamma}}{\Omega_{\rm sky}}
\cos^2(2\varphi_{\bm{\ell\ell'}}), \\
\bm{M}_{\bm{\ell\ell'}}^{EE,BB}=\bm{M}_{\bm{\ell\ell'}}^{BB,EE}=
\frac{{\cal W}_{\bm{\ell-\ell'}}^{\gamma\gamma}}{\Omega_{\rm sky}}
\sin^2(2\varphi_{\bm{\ell\ell'}}), \\
\bm{M}_{\bm{\ell\ell'}}^{EB,EB}=
\frac{{\cal W}_{\bm{\ell-\ell'}}^{\gamma\gamma}}{\Omega_{\rm sky}}
[\cos^2(2\varphi_{\bm{\ell\ell'}})-\sin^2(2\varphi_{\bm{\ell\ell'}})],
\end{eqnarray}
where ${\cal W}^{\gamma\gamma}_{\bm\ell}$ is the power spectrum of the
weight field $W(\bm\theta)$.
The convolution with the mask generates apparent B-mode
cosmic shear power spectrum which is leakage from the E-mode power
spectrum, even if there is no intrinsic power in the B-mode as is the
case for cosmic shear. We also account for the effect of the finite
square boundary of each field to compute the full mode coupling matrix
\citep{Hikage11}. The mode coupling matrix is inverted after binning.
To do so, we compute the binned dimensionless power spectrum as
\begin{equation}
{\cal C}_b\equiv \frac{1}{N_{{\rm mode},b}}\sum_{\bm\ell}^{\ell \in
  \ell_b}P_{b\ell}C_{\bm\ell},
\label{eq:cb_calc}
\end{equation}
where $b$ is the label for the $\ell$ bin, $P_{b\ell}=\ell^2/2\pi$ and
$N_{{\rm mode},b}$ is the number of modes in the bin, approximately given by
\begin{equation}
N_{{\rm mode},b}\simeq\frac{\Omega_{\rm sky}}{4\pi}(\ell_{b,{\rm max}}^2-\ell_{b,{\rm min}}^2).
\end{equation}
The sum in the equation~(\ref{eq:cb_calc}) runs over the $\ell$ modes
in the given bin $b$.

The binned power spectrum with the mask correction is obtained by
multiplying the inverse of the mode coupling matrix by the pseudo-spectrum
\begin{equation}
\label{eq:cb}
{\cal C}_b^{\rm (true)}=\bm{M}_{bb'}^{-1}\sum_{\bm\ell}^{|{\bm\ell}|\in \ell_b'}P_{b'\ell}
(\bm{C}_{\bm\ell}^{\rm (obs)}-\langle \bm{N}_{\bm\ell}\rangle_{\rm MC}),
\end{equation}
where
\begin{equation}
\bm{M}_{bb'}=\sum_{\bm\ell}^{{\bm\ell}\in \ell_b}\sum_{\bm{\ell'}}^{\bm{\ell'}\in\ell_{b'}}P_{b\ell}M_{\bm{\ell\ell'}}Q_{\ell'b'},
\end{equation}
with $Q_{\ell b}=2\pi/\ell^2$. In order to remove the shot noise
effect, we randomly rotate orientations of individual galaxies
to estimate the noise power spectrum $\bm{N}_\ell$. For
accurate estimates of the noise power spectrum, we repeat this
procedure 10000 times and use average noise spectrum
$\langle \bm{N}_\ell\rangle_{\rm MC}$ over these
realizations. This averaging allows us to subtract the shot noise
contribution to the cosmic shear power spectrum accurately.

\begin{figure*}
\begin{center}
\includegraphics[width=14cm]{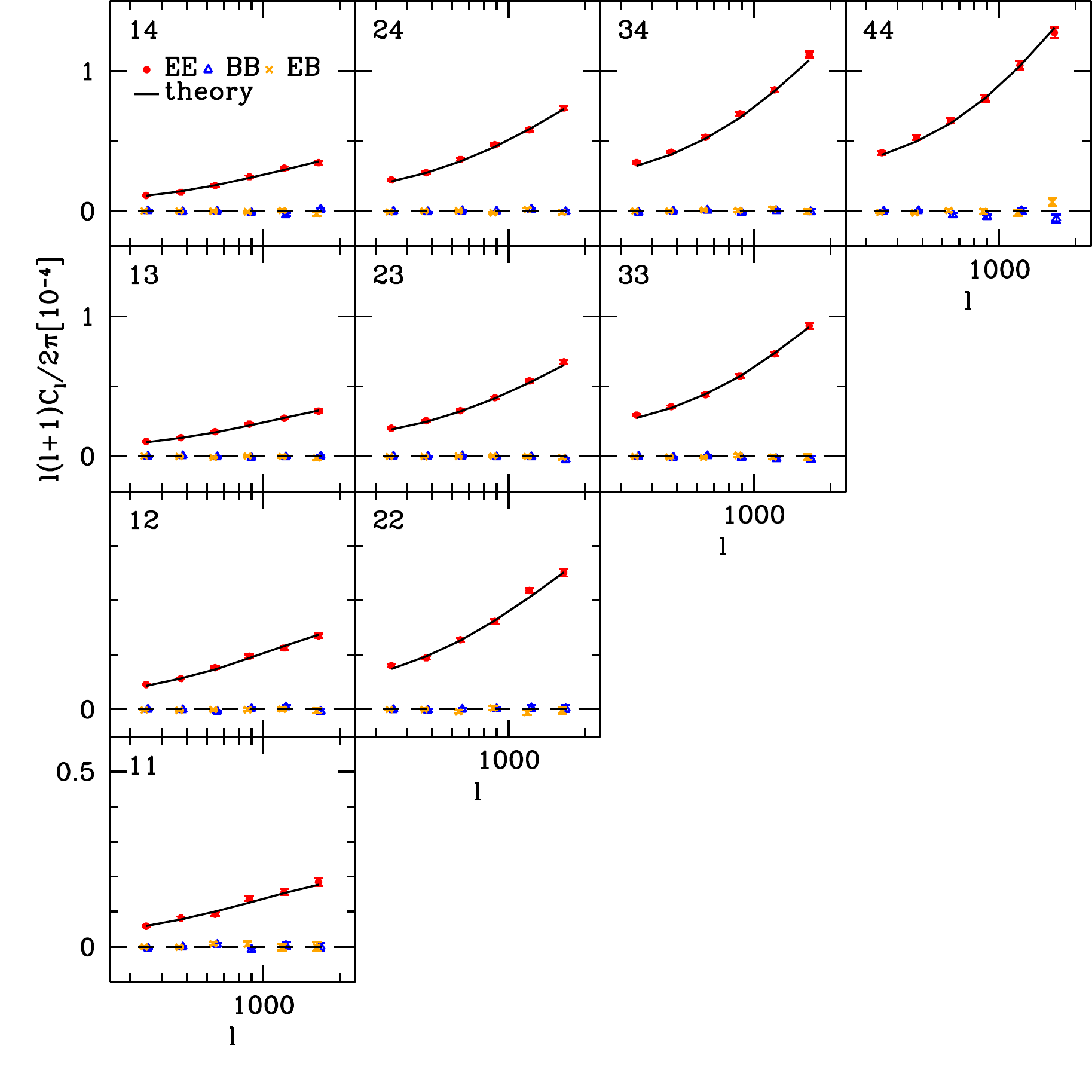}
\end{center}
\caption{Tomographic cosmic shear power spectra ({\it filled circles})
  measured on the HSC mock shear catalogs. The spectra are the
  weighted average of spectra for six disjoint fields (GAMA09H,
  GAMA15H, VVDS, WIDE12H, XMM, and HECTOMAP). The error-bars represent
  the 1$\sigma$ dispersion of the power spectra from the mock shear
  catalogs, divided by the square root of the number of mock
  realizations, $\sqrt{100}$. Input power spectra computed by
  equation~(\ref{eq:cl_model}) are shown as solid lines. We also
  plot BB ({\it open triangles}) and EB ({\it crosses}) spectra, which
  are consistent with zero.}
\label{fig:cl_mock}
\end{figure*}

\subsection{Test with HSC mock shear catalogs}

While the pseudo-$C_\ell$ method has been shown to recover input power
spectra accurately \citep{Hikage11,HikageOguri16}, we explicitly
check the validity and accuracy of the pseudo-$C_\ell$ method for our
analysis of the HSC first-year data by applying it to the HSC mock
shear catalogs \citep{Oguri18}. The mocks use realizations of cosmic
shear from all-sky ray-tracing simulations of \citet{Takahashi17}.
The cosmological model is flat $\Lambda$CDM with best-fit values
as measured by {\it WMAP9} \citep{Hinshaw13}: $\Omega_{\rm m}=0.279$,
$\Omega_{\rm b}=0.046$, $\Omega_\Lambda=0.721$, $h=0.7$, $n_{\rm
  s}=0.97$, and $\sigma_8=0.82$.
In order to construct realistic mock shear catalogs, we start with the
real HSC first-year shear catalog, and replace ellipticities of
individual galaxies with mock ellipticity values while retaining the
positions of these galaxies. Therefore the mock catalogs have exactly
the same spatial inhomogeneity and mask patterns as do the real HSC
first-year catalogs, and therefore are well suited for careful
tests of the accuracy of the correction for masking effects implemented
by the pseudo-$C_\ell$ method.

The mock ellipticity values include realistic noise properties as
well as cosmic shear from the all-sky ray-tracing simulations. We take
account of the redshift distribution of source galaxies by randomly
assigning redshifts of individual galaxies from their photo-$z$ PDFs,
using the {\tt MLZ} code \citep[see][]{Oguri18}. In the mock catalogs,
we also include the multiplicative bias $m$ by shifting shear values
taken from the ray-tracing simulations by a factor of $1+m$. Since the
analysis of the mock catalogs is conducted while the true HSC shape
catalog is blinded (see Section~\ref{subsec:blind}), we artificially
shift the values of $m$ by a factor of 1.3; this value is also kept
blinded during the analysis of the mock catalogs, in order to maintain
the blind nature of our analysis. We use 100 mock samples in each of
the six disjoint fields.

Figure~\ref{fig:cl_mock} shows tomographic cosmic shear power spectra
of the mock catalogs, measured by the pseudo-$C_\ell$ method. The
spectra are averaged over 100 mock samples, and also are averaged over
the six disjoint fields weighted by the effective number of source
galaxies, $\sum w_i$, in each field. The error represents the
1$\sigma$ scatter of measured power spectra from the mock catalogs
divided by the square root of the number of mock realizations,
$\sqrt{100}$.  For comparison, we show the input cosmic shear power
spectra computed by equation~(\ref{eq:cl_model}). We find that the
power spectra measured using the pseudo-$C_\ell$ method agree with the
input spectra quite well. The small deficit at $\ell
\gtrsim 2000$ in measured power spectra relative to the model
calculation originates from the limited angular resolution of the
all-sky ray-tracing simulations that are used for constructing the
mock shear catalogs. We also find that both B-mode auto and EB cross
spectra are consistent with zero.

We compute the $\chi^2$ values to quantify the goodness of fit between
the input (zero for B- and EB-modes) and measured spectra averaged
over the mocks, and find 74 for EE-auto, 70 for BB-auto, 62 for EB
for 60 degrees of freedom. The corresponding $p$-values are
0.10, 0.18, and 0.41, respectively. Measured power spectra with our
pseudo-$C_\ell$ method are consistent with the input power spectra even
for $1/\sqrt{100}$ smaller statistical errors than those expected for
the HSC first-year shear catalog, indicating that the systematic error
originating from the inaccuracy of the pseudo-$C_\ell$ method is well
below 10\% of the statistical error and therefore is negligibly
small. Furthermore, we perform the nested sampling analysis on the
average cosmic shear power spectra from the 100 mock shear catalogs to
test whether the input values of key cosmological parameters are
recovered by this nested sampling analysis. We find that
$\Omega_{\rm m}=0.292\pm 0.014$, $\sigma_8=0.801\pm 0.020$, and $S_8
(\alpha=0.5)=0.791\pm 0.005$, where these are the errors in the mean
over the 100 mock realizations. These values agree with the input
values of $\Omega_{\rm m}=0.279$, $\sigma_8=0.82$, and $S_8=0.791$
within the statistical errors.

\section{Analytic model of the covariance matrix}
\label{sec:covariance}

\subsection{Justification for use of an analytic model}
There are several different ways to estimate the covariance
matrix. One class of methods are resampling techniques, including the
bootstrap and jackknife methods, which allow us to derive a covariance
matrix directly from the observational
data. However, the covariance matrix derived by the resampling
technique represents a noisy estimate of the true covariance
matrix. This noise in the covariance matrix may affect the analysis in
several ways, such as the unstable inversion of the covariance matrix
and noise bias in derived constraints
\citep[e.g.,][]{Hartlap2007,Norberg2009,Friedrich2016}.
Furthermore, a covariance matrix derived from the data does not
contain the so-called super-survey modes which affect both large- and
small-scale covariance
via mode-coupling \citep{TakadaHu13}. One can instead resort to
$N$-body (ray-tracing) simulations to estimate covariance
matrices \citep[e.g.,][]{Jee16}. However, a technical challenge is how
to include the dependence of covariance matrices on cosmological
parameters, as this approach requires a
large number of $N$-body simulations for each set of cosmological
parameters. In this paper, we use the covariance matrix calculated
using the so-called halo model \citep{CooraySheth02}, which has no
issue with noise and also can include the cosmological parameter
dependence. Below we describe our analytic model in detail, and then
compare our model with the covariance matrix derived from realistic
HSC mock shear catalogs.

\begin{figure*}
\begin{center}
\includegraphics[width=14cm]{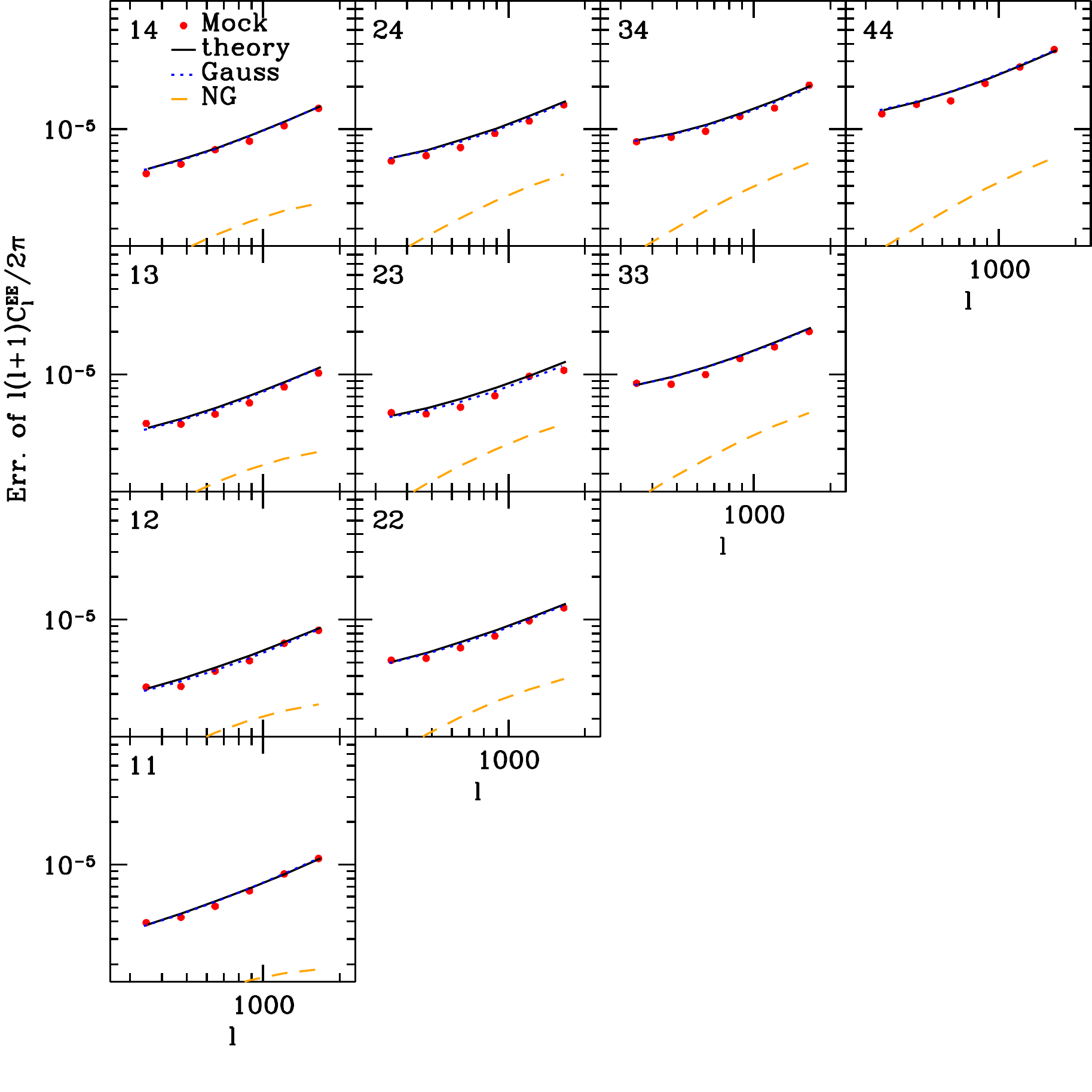}
\end{center}
\caption{Comparison of 1$\sigma$ errors of the tomographic cosmic
  shear power spectra computed from the HSC mock shear catalogs ({\it
    symbols}), which correspond to the diagonal part of the covariance
  matrix, with the analytic model of the covariance described in the
  text ({\it solid lines}).  The Gaussian and the non-Gaussian
  contributions including super-sample variance to the covariance
  matrix from the analytic model are shown by dotted and dashed lines,
  respectively. It is found that the contribution of the Gaussian term
  is dominant in the diagonal terms.}
\label{fig:cl_mock_err}
\end{figure*}

\subsection{Analytic Model}
As shown in equation~(\ref{eq:cov_tot}), we decompose the covariance
matrix of the cosmic shear power spectrum into three terms, the
Gaussian term ${\rm Cov}^{\rm (G)}$, the non-Gaussian term
${\rm Cov}^{\rm (NG)}$, and the super-sample covariance
${\rm Cov}^{\rm (SSC)}$. The diagonal part of the Gaussian term of the
covariance  ${\rm Cov}^{\rm (G)}$ is given by
\begin{eqnarray}
{\rm Cov}^{\rm (G)} (C_{b}^{(ij)},C_{b'}^{(i'j')})&=&
\frac{\delta_{bb'}^{K}}{N_{{\rm mode},b}^{\rm eff}}
\left[\hat{C}_{b}^{(ii')}\hat{C}_{b}^{(jj')}
\right. \nonumber \\ &&
\left.+\hat{C}_{b}^{(ij')}\hat{C}_{b}^{(i'j)}\right],
\label{eq:cov_gauss}
\end{eqnarray}
where $b$ is a binning number of the multipole $\ell$, and $i$ and $j$ refer
to tomographic bins. The effective number of Fourier modes
in the $b$-th bin $N_{{\rm mode},b}^{\rm eff}$ is given by
\begin{equation}
N_{{\rm mode},b}^{\rm eff}=N_{\rm mode,b}w_2^2/w_4.
\end{equation}
The factor $w_2^2/w_4$ represents the loss of modes due to
pixel weighting, where $w_k$ is the $k$-th moment of the weight
\citep{Hivon02} given by
\begin{equation}
w_k=\frac{1}{\Omega_{\rm sky}}\int\bm{d\theta} W^k(\bm\theta).
\end{equation}
The binned spectra $\hat{C}_{b}^{ij}$ includes the shot noise
\begin{equation}
\hat{C}_{b}^{(ij)}=C_{b}^{(ij)}+N_{b}^{(ij)},
\end{equation}
where $C_{b}^{(ij)}$ is the binned version of the model cosmic
shear power spectrum computed from equation~(\ref{eq:cl_model})
and $N_{b}^{(ij)}$ is the binned shot noise spectrum, which is zero
when $i\neq j$
\begin{equation}
N_{b}^{(ij)}=\delta_{ij}^{K}N_{b}^{(i)}.
\end{equation}
As mentioned in Appendix~\ref{sec:app1}, we measure the shot noise
spectrum $N_{b}^{(i)}$ directly from the data by randomly rotating
shapes of individual galaxies. We note that in tomographic cosmic
shear analysis the covariance between different tomographic bins is
significant, because source galaxies with different redshifts
share the same matter distribution along the line-of-sight.

The Gaussian covariance [equation~(\ref{eq:cov_gauss})] can be further
decomposed into the auto-term of cosmic shear power spectra, the
cross-term between cosmic shear and shape noise spectra, and the
auto-term of shape noise spectra. We write this explicitly as
\begin{eqnarray}
{\rm Cov}^{\rm (G)}_{SS} (C_{b}^{(ij)},C_{b}^{(i'j')})&=&
\frac{C_{b}^{(ii')}C_{b}^{(jj')}+C_{b}^{(ij')}C_{b}^{(i'j)}}{N_{{\rm mode},b}^{\rm eff}},
\label{eq:cov_gauss_ss}
\end{eqnarray}
\begin{eqnarray}
{\rm Cov}^{\rm (G)}_{SN} (C_{b}^{(ij)},C_{b}^{(i'j')})&=&
\frac{1}{N_{{\rm mode},b}^{\rm eff}}
\left[C_{b}^{(ii')}N_{b}^{(jj')}+N_{b}^{(ii')}C_{b}^{(jj')}
\right. \nonumber \\ &&
\left.+C_{b}^{(ij')}N_{b}^{(i'j)}+N_{b}^{(ij')}C_{b}^{(i'j)}
\right],
\label{eq:cov_gauss_sn}
\end{eqnarray}
\begin{eqnarray}
{\rm Cov}^{\rm (G)}_{NN} (C_{b}^{(ij)},C_{b}^{(i'j')})&=&
\frac{N_{b}^{(ii')}N_{b}^{(jj')}+N_{b}^{(ij')}N_{b}^{(i'j)}}{N_{{\rm mode},b}^{\rm eff}}.
\label{eq:cov_gauss_nn}
\end{eqnarray}
As discussed in Section~\ref{subsec:shearspectra}, the shape noise
covariance ${\rm Cov}^{\rm (G)}_{NN}$ estimated from the data shows
small off-diagonal components of the covariance, mostly between
neighboring multipole bins. In our analysis, we estimate ${\rm
  Cov}^{\rm (G)}_{NN}$ directly from the data using the shot noise
simulations described in Appendix~\ref{sec:app1}, with 10000 Monte
Carlo realizations of noise power spectra that are obtained by
randomly rotating the ellipticities of individual galaxies, whereas we
estimate ${\rm Cov}^{\rm (G)}_{SS}$ and ${\rm Cov}^{\rm (G)}_{SN}$
using equations~(\ref{eq:cov_gauss_ss}) and (\ref{eq:cov_gauss_sn}).

The non-Gaussian term of cosmic shear power spectra
${\rm Cov}^{\rm (NG)}$ originates from the mode-coupling due to
nonlinear gravitational evolution. This term was formulated in
previous work \citep{CoorayHu01,TakadaJain03,TakadaBridle07}, and
can be expressed as
\begin{eqnarray}
&{\rm Cov}&{}^{\rm (NG)} (C_{b}^{(ij)},C_{b'}^{(i'j')})
=\frac{1}{\Omega_{\rm sky}} \nonumber \\
&&\times\int_{|\mathbf{\ell}\in \ell_b|}\frac{\mathbf{d\ell}}{A(\ell_i)}
\int_{|\mathbf{\ell'}\in \ell_b'|}\frac{\mathbf{d\ell'}}{A(\ell_i')}
T^{(iji'j')}(\mathbf{\ell},-\mathbf{\ell},\mathbf{\ell'},-\mathbf{\ell'})
\nonumber \\
&\simeq & \frac{1}{\Omega_{\rm sky}}T^{\rm (iji'j')}
(\ell_b,\ell_b,\ell_{b'},\ell_{b'}),
\end{eqnarray}
where $\Omega_{\rm sky}$ is the observed sky area and $A(\ell_i)\equiv
\int_{|\mathbf{\ell}|\in \ell_b}\mathbf{d\ell}$. In the second line,
we assume that the trispectrum within bins of $\ell$ is represented
as
\begin{eqnarray}
T^{(iji'j')}(&\ell_b&,\ell_b,\ell_b',\ell_b') \nonumber \\
&=&\int_0^{\chi_H} d\chi \frac{q^{(i)}(\chi)q^{(j)}(\chi)
q^{(i')}(\chi)q^{(j')}(\chi)}{f_K^6(\chi)} \nonumber \\
&&\times T_\delta\left(\frac{\ell_b}{f_K(\chi)},\frac{\ell_b}{f_K(\chi)},
\frac{\ell_b'}{f_K(\chi)},\frac{\ell_b'}{f_K(\chi)}\right).
\end{eqnarray}
Since the one-halo term of the trispectrum dominates at $\ell\ge
100$  \citep{Sato09}, we include only the one-halo term of the
trispectrum given as
\begin{eqnarray}
T_\delta^{\rm 1h}(&&k_1,k_2,k_3,k_4)=\int dM\frac{dn}{dM}
\left(\frac{M}{\bar{\rho}_{\rm m}}\right)^4\tilde{u}_{\rm NFW}(k_1;M,z) \nonumber \\
&&\tilde{u}_{\rm NFW}(k_2;M,z)\tilde{u}_{\rm NFW}(k_3;M,z)
\tilde{u}_{\rm NFW}(k_4;M,z),
\end{eqnarray}
where $M$ is the halo mass defined as $M_{200m}$, the mass enclosed
in a sphere of 200 times the mean matter density, $\bar{\rho}_{\rm m}$ is the mean matter
density of the Universe, $dn/dM$ is the halo mass function
\citep{Tinker08}, and $\tilde{u}_{\rm NFW}$ is the Fourier transform
of the normalized \citet[][hereafter NFW]{Navarro97} density profile
for a halo with mass $M$. We adopt the concentration-mass relation by
\citet{DiemerKravtsov15}. These calculations are conducted making use
of the python package {\tt COLOSSUS} \citep{Diemer2018}.

In a finite sky area, mode fluctuations whose scales are larger than
the survey area generate excess covariance in the cosmic shear
power spectra \citep{TakadaBridle07,Sato09}. This is known as
super-sample covariance (SSC). It can be decomposed into the halo
sample variance (HSV), the beat coupling (BC) and the cross-term
BC-HSV \citep{TakadaHu13,LiHuTakada14}
\begin{equation}
{\rm Cov}^{\rm (SSC)}={\rm Cov}^{\rm (HSV)}+{\rm Cov}^{\rm (HSV-BC)}+{\rm Cov}^{\rm (BC)}.
\end{equation}
The halo sample variance is important on small scales (large $k$),
whereas the beat coupling is important on large scales (small $k$).
For the cosmic shear power spectrum, the HSV contribution to the
covariance matrix is written as \citep{TakadaHu13}
\begin{eqnarray}
\label{eq:err_hsv1}
&{\rm Cov}&^{\rm (HSV)} (C_b^{(ij)},C_{b'}^{(i'j')}) \nonumber \\
&=&
\int_0^{\chi_H} d\chi \frac{q^{(i)}(\chi)q^{(j)}(\chi)
q^{(i')}(\chi)q^{(j')}(\chi)}{f_K^6(\chi)} \nonumber \\
&&\times I_{\rm mm}(k_b,k_b)I_{\rm mm}(k_{b'},k_{b'})(\sigma_W^L(z))^2,
\end{eqnarray}
where $k_b=\ell_b/\chi$ and
\begin{eqnarray}
I_{\rm mm}(k,k')&\equiv &
\int dM\frac{dn}{dM}\left(\frac{M}{\bar{\rho}_{\rm m}}\right)^2 b(M,z) \nonumber \\
&& \times\tilde{u}_{\rm NFW}(k;M,z)\tilde{u}_{\rm NFW}(k';M,z).
\end{eqnarray}
We use a model of the halo bias $b(M,z)$ by \citet{Tinker10}. The
variance $(\sigma_W^L)^2$ represents the background fluctuation
convolved with the survey window function
\begin{eqnarray}
(\sigma_W^L(z))^2&=&\frac{1}{\Omega_{\rm sky}^2}
\int\frac{\mathbf{d\ell}}{(2\pi)^2}|\tilde{W}(\bm\ell)|^2
P^{\rm L}\left(k=\frac{\ell+1/2}{\chi};z\right), \nonumber \\
\end{eqnarray}
where $P^{\rm L}(k;z)$ is the three-dimensional linear power spectrum
at redshift $z$ and $\tilde{W}(\bm\ell)$ is the two-dimensional
Fourier transform of the square of the survey window function
[equation~(\ref{eq:mask_fourier})].
For simplicity, we use a square survey
window function with the same sky area in each field.

The other two covariance terms respectively are given by
\begin{eqnarray}
&{\rm Cov}&^{\rm (BC)} (C_b^{(ij)},C_{b'}^{(i'j')}) \nonumber \\
&\simeq& \left(\frac{68}{21}\right)^2\int_0^{\chi_H} d\chi
\frac{q^{(i)}(\chi)q^{(j)}(\chi)
q^{i'}(\chi)q^{(j')}(\chi)}{f_K^6(\chi)} \nonumber \\
&\times& [I_{\rm m}(k_b)I_{\rm m}(k_{b'})]^2
P^{\rm L}(k_b;z)P^{\rm L}(k_{b'};z)(\sigma_W^L(z))^2,
\end{eqnarray}
and
\begin{eqnarray}
&{\rm Cov}&^{\rm (HSV-BC)} (C_b^{(ij)},C_{b'}^{(i'j')}) \nonumber \\
&\simeq &\frac{68}{21}
\int_0^{\chi_H} d\chi \frac{q^{(i)}(\chi)q^{(j)}(\chi)
q^{(i')}(\chi)q^{j'}(\chi)}{f_K^6(\chi)} \nonumber \\
&\times & \left[I_{\rm m}^2(k_b)I_{\rm mm}(k_{b'},k_{b'})
P^{\rm L}(k_b;z)+(b\leftrightarrow b')\right](\sigma_W^L(z))^2, \nonumber \\
\end{eqnarray}
where
\begin{equation}
I_{\rm m}(k) \equiv \int dM\frac{dn}{dM}
\left(\frac{M}{\bar{\rho}_{\rm m}}\right) b(M) \tilde{u}_{\rm NFW}(k;M,z).
\end{equation}

\begin{figure}
\begin{center}
\includegraphics[width=7.5cm]{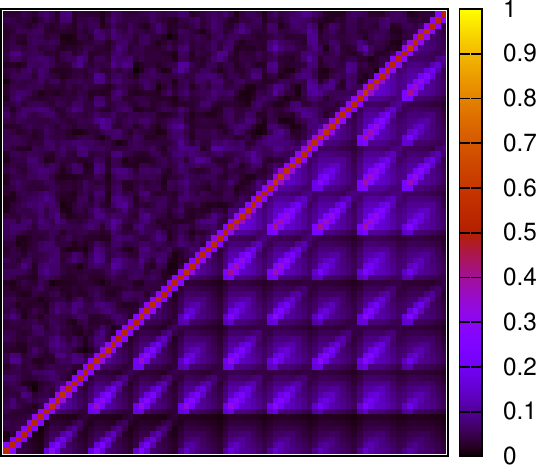}
\end{center}
\caption{Comparison between the analytic correlation matrix
  (covariance normalized by the diagonals, i.e.,
  $C_{ij}/\sqrt{C_{ii}C_{jj}}$) of four-bin tomographic cosmic shear
  power spectra and that derived from the HSC mock shear catalogs. The
  order of blocks is 11, 12, 13, 14, 22, 23, 24, 33, 34, 44 from left
  to right. We show the analytic correlation matrix in the lower
  right side of the triangle, whereas the absolute value of the difference
  between the analytic and mock correlation matrix is shown in the
  upper left side.}
\label{fig:cl_mock_cov}
\end{figure}

\subsection{Comparison with mock catalogs}

We compare the analytic covariance matrix with numerical estimates
using HSC mock shear catalogs presented in Appendix~\ref{sec:app1}. We note
that the HSC mock shear catalogs include cosmic shear from all-sky
tracing simulations, and hence the covariance matrix computed from the
mock catalogs naturally includes the effect of super-survey modes,
which we also include in the analytic model.
Figure~\ref{fig:cl_mock_err} shows the comparison of the diagonal
components of the covariance matrix of the tomographic cosmic shear
power spectra. The diagonal components are dominated by the Gaussian
term. Figure~\ref{fig:cl_mock_cov} compares the correlation
coefficient matrices of tomographic cosmic shear power spectra. We
find that our analytic model for the covariance well reproduces both
the diagonal and off-diagonal parts of the covariance matrix derived
numerically from the HSC shear mock catalogs. These comparisons
demonstrate the validity of the use of the analytic model of the
covariance matrix in our cosmological analysis.

\section{Convergence of our nested sampling results}
\label{sec:convergence}

The multimodal nested sampling algorithm called {\tt Multinest} is a
Bayesian inference tool to efficiently evaluate the Bayesian evidence
as well as posterior distributions. The nested sampling performs
Bayesian computations by maintaining a set of sample points (live
points) in parameter space inside the prior $\pi(\theta)$ and
repeatedly replacing the lowest likelihood point with another point
with higher likelihood. The remaining fraction of the prior volume
after the point with likelihood ${\cal L}_i$ is defined as
\begin{equation}
X({\cal L}_i)\equiv \int_{{\cal L}(\theta)>{\cal L}_i}\pi(\theta)d\theta.
\end{equation}
We note that $X$ exponentially shrinks as the likelihood ${\cal L}_i$
increases.

The nested sampling terminates when the remaining evidence
contribution estimated as ${\cal L}_{\rm max}X_i$, where ${\cal L}_{\rm max}$
is the maximum likelihood value of the current set of live points and
$X_i$ is the expected value of remaining prior volume, is less than a
user-defined tolerance.  We set the evidence tolerance factor {\tt
  tol}=0.1, a value that is expected to be appropriate for the
computation of Bayesian posterior and evidence.  We set sampling
efficiency ${\tt efr}=0.3$, which is the recommended value for evidence
calculation \citep{Feroz2009}. We adopt importance nested sampling
which obtains more accurate evidence values than does the standard
nested sampling \citep{Feroz2013}, but we do not adopt the
constant efficiency mode.

Another tuning parameter is the number of live points $N_{\rm live}$
at any given time. Larger $N_{\rm live}$ values increase the accuracy
of the Bayesian evidence, although the computation cost also
increases. Here we set $N_{\rm live}=2000$ to find that the standard
deviations of $S_8$ and
$\Omega_{\rm m}$ are $6.5\times 10^{-4}$ and $1.4\times 10^{-3}$
respectively estimated from 8 independent runs. These values are
negligible compared to the marginalized 1$\sigma$ errors of these
quantities (see Table~\ref{tab:S8_sys}). The standard deviation of
the $\log X$ value is 0.041, which is also sufficiently small to
enable the evaluation of the model preference from the evidence
ratio.

\begin{figure}
\begin{center}
\includegraphics[width=8cm]{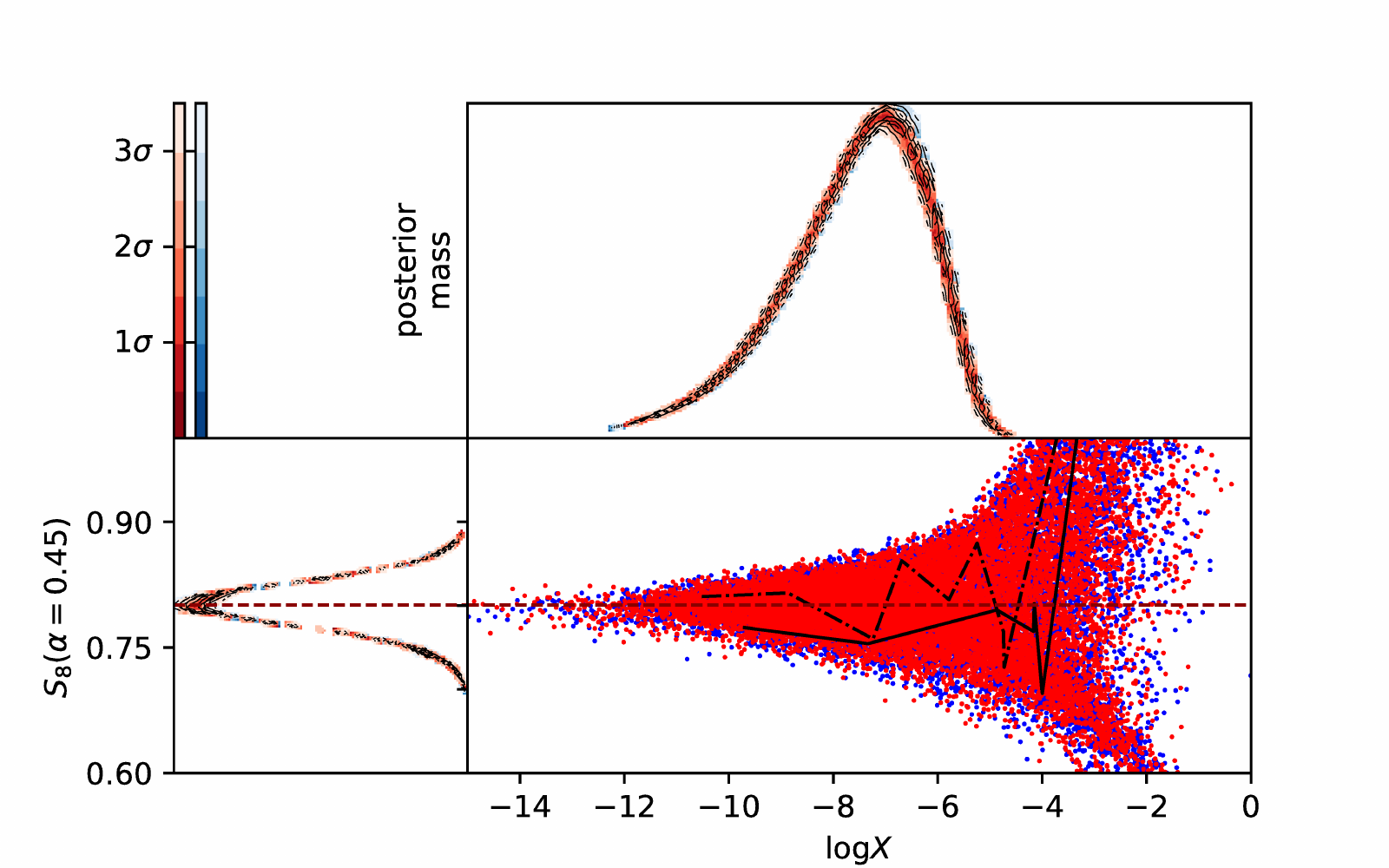}
\end{center}
\caption{Distributions in $\log X$ produced using {\tt nestcheck} for
  two nested sampling runs. The upper right panel shows the relative
  posterior mass at each $\log X$ value. The lower-right panel shows
  $S_8 (\alpha=0.45)$ values against $\log X$ and the lower-left
  panels shows the posterior distributions. The solid black lines show
  the evolution of an individual thread from each run chosen at
  random. The colored contours show iso-probability credible intervals
  on the marginalized posterior probability density function. The
  figure indicates that our nested sampling runs terminate at the
  point at which the remaining fraction of posterior mass is sufficiently
  small.}
\label{fig:logX_nestcheck}
\end{figure}

The convergence of the nested sampling result can be demonstrated
using the public diagnostic tool called {\tt nestcheck}
\citep{Higson2018}. Figure~\ref{fig:logX_nestcheck} is an output
generated from {\tt nestcheck} using the two nested sampling runs in
the fiducial case.  We show both the relative posterior mass at each
$\log X$ value, i.e., ${\cal L}(X)X$, where ${\cal L}(X)\equiv
X^{-1}({\cal L})$, and the distribution of $S_8$ at each $\log X$.
The Figure demonstrates that our nested sampling run terminates at the
point at which the relative posterior mass is very small and the $S_8$
value converges to the maximum likelihood point. The posterior
distributions of the two nested sampling runs agree within the error
estimated by bootstrap resampling.

\begin{figure*}
\begin{center}
\includegraphics[width=17cm]{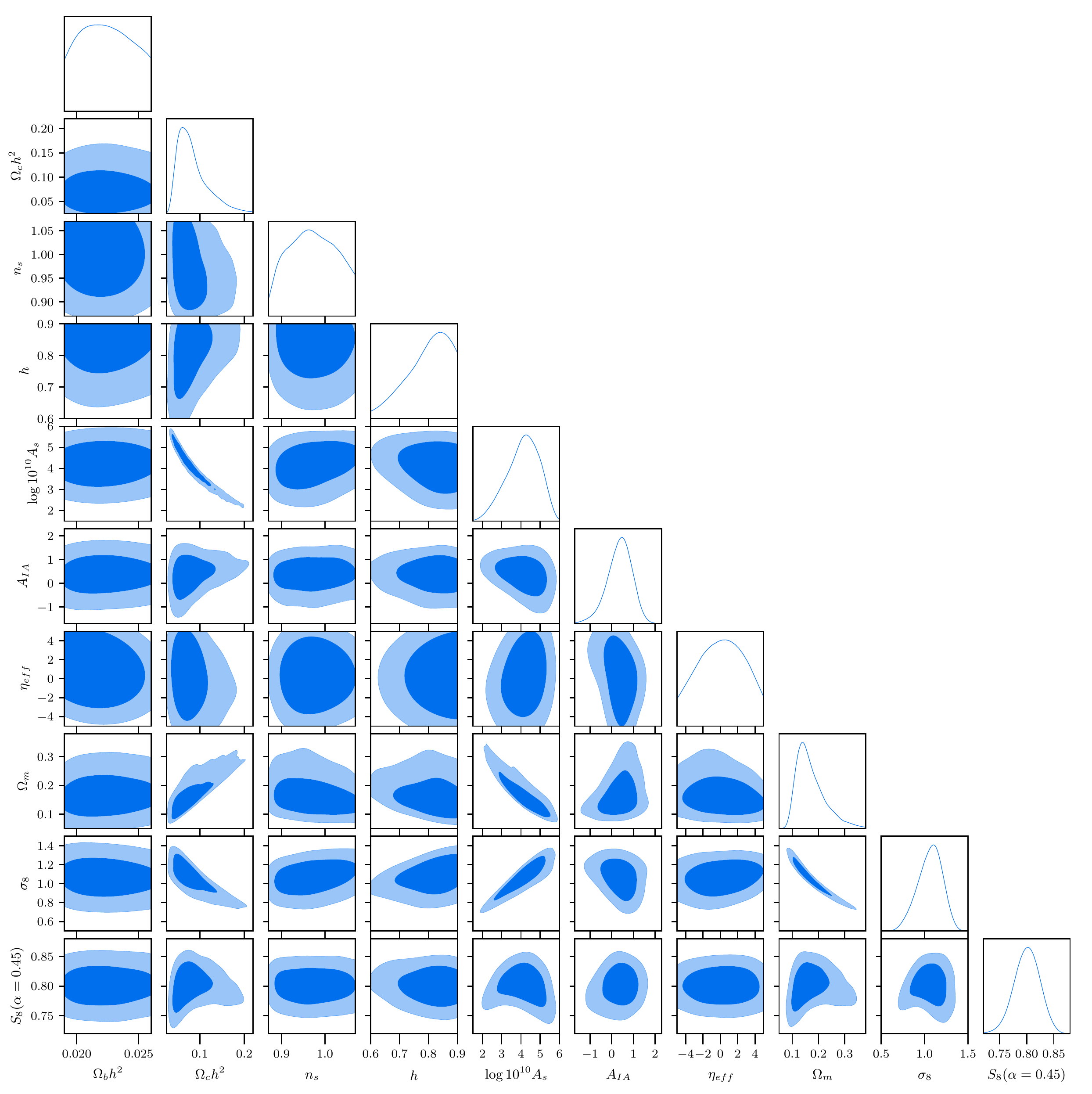}
\end{center}
\caption{Marginalized one-dimensional and two-dimensional posteriors
  of cosmological model ($\Omega_{\rm b}h^2$, $\Omega_{\rm c}h^2$,
  $n_s$, $h$, $\log(10^{10}A_s)$), and
  intrinsic alignment parameters ($A_{\rm IA}$, $\eta_{\rm eff}$) as
  well as derived parameters $\Omega_{\rm m}$, $\sigma_8$, and $S_8$ in the
  fiducial case. The contours in each two-dimensional constraint
  represent 68\% and 95\% credible intervals.}
\label{fig:triangle}
\end{figure*}

\section{Posterior distributions of the fiducial model parameters}
\label{sec:triangle}
Figure~\ref{fig:triangle} shows the marginalized one-dimensional and
two-dimensional posteriors of cosmological parameters and intrinsic
alignment parameters as well as derived parameters $\Omega_{\rm m}$,
$\sigma_8$, and $S_8$ in the fiducial setup. One can see that the
posteriors for $\Omega_{\rm b}h^2$, $n_s$, and $h$ among 5
cosmological parameters are strongly affected by the prior as listed
in Table~\ref{tab:params}. This is not surprising because the cosmic
shear is not very sensitive to these parameters. For intrinsic
alignment parameters, $\eta_{\rm eff}$ is prior-dominated whereas the
amplitude parameter $A_{\rm IA}$ is well constrained from the data. We
find that the prior of $h$ affects the quoted $\Omega_{\rm m}$ and
$\sigma_8$ values, but probably does not affect the $S_8$ value since
$h$ is not degenerate with $S_8$.

\vspace{1cm}

\end{document}